\documentclass[aps,groupedaddress,twocolumn,superscriptaddress,amsfonts,amssymb,amsmath,showpacs,floatfix,preprintnumbers]{revtex4-2}

\usepackage{dcolumn}
\usepackage{bm}
\usepackage{multirow}
\usepackage{verbatim}
\usepackage{graphicx}
\usepackage{ulem}
\usepackage{color}
\usepackage{placeins}
\usepackage{xcolor}
\usepackage{slashed}

\usepackage[dvipsnames]{xcolor}
\usepackage{graphicx}
\usepackage{placeins}
\usepackage{wrapfig}
\usepackage{amsmath}
\usepackage{amsfonts}
\usepackage{amssymb}
\usepackage{multirow}
\usepackage{slashed}
\usepackage{physics}
\usepackage{dsfont}
\usepackage{soul}
\usepackage{ulem}
\usepackage{verbatim}
\usepackage[colorlinks=true, linkcolor=blue, citecolor=blue, urlcolor=blue]{hyperref}

\begin{document}

\title{Reconstructing the full kinematic dependence of GPDs from pseudo-distributions}

\bibliographystyle{apsrev4-1}

\author{Herv\'e Dutrieux}
\email[e-mail: ]{herve.dutrieux@cnrs.fr}
\affiliation{Aix Marseille Univ, Universit\'e de Toulon, CNRS, CPT, Marseille, France.}

\author{ Robert G. Edwards}
\email[e-mail: ]{edwards@jlab.org}
\affiliation{Thomas Jefferson National Accelerator Facility, Newport News, VA 23606, USA}

\author{ Joe Karpie}
\email[e-mail: ]{jkarpie@jlab.org}
\affiliation{Thomas Jefferson National Accelerator Facility, Newport News, VA 23606, USA}

\author{ C\'edric Mezrag}
\email[e-mail: ]{cedric.mezrag@cea.fr}
\affiliation{Irfu, CEA, Université Paris-Saclay, 91191 Gif-sur-Yvette, France}
\affiliation{AIDAS, CEA, 91191, Gif-Sur-Yvette, France}

\author{ Christopher Monahan}
\email[e-mail: ]{cmonahan2024@coloradocollege.edu}
\affiliation{Department of Physics, Colorado College, Colorado Springs, CO 80903, USA}

\author{ Kostas Orginos}
\email[e-mail: ]{kostas@wm.edu}
\affiliation{Physics Department, William \& Mary, Williamsburg, VA 23187, USA}

\author{ Anatoly Radyushkin}
\email[e-mail: ]{radyush@jlab.org}
\affiliation{Department of Physics, Old Dominion University, Norfolk, Virginia, USA}
\affiliation{Thomas Jefferson National Accelerator Facility, Newport News, VA 23606, USA}

\author{ David Richards}
\email[e-mail: ]{dgr@jlab.org}
\affiliation{Thomas Jefferson National Accelerator Facility, Newport News, VA 23606, USA}

\author{ Eloy Romero}
\email[e-mail: ]{eromero@jlab.org}
\affiliation{Thomas Jefferson National Accelerator Facility, Newport News, VA 23606, USA}

\author{ Savvas Zafeiropoulos}
\email[e-mail: ]{savvas.zafeiropoulos@cpt.univ-mrs.fr}
\affiliation{Aix Marseille Univ, Universit\'e de Toulon, CNRS, CPT, Marseille, France.}

\begin{abstract}
We propose a reconstruction of the full $(x, \xi, t)$ dependence of unpolarized isovector proton generalized parton distributions (GPDs) $H^{u-d}$ and $E^{u-d}$ from lattice QCD data in the pseudo-distribution formalism. For the first time, we extract double distributions (DDs) directly from lattice data, enforcing therefore an important property of GPDs linked to Lorentz symmetry. We use the flexible framework of multidimensional Gaussian process regression to regularize the inverse problem and present an assessment of the impact of model dependence on the systematic uncertainty. Our lattice ensemble corresponds to a pion mass $m_\pi = 358$~MeV and a lattice spacing $a = 0.094$~fm. We use larger hadron momenta, up to 2.7~GeV, and kinematic coverage compared to our previous computations and extract additional skewness-dependent moments of the GPD.
\end{abstract}

\date{\today}
\preprint{JLAB-THY-26-4683}

\maketitle

\tableofcontents

\section{Introduction\label{sec:intro}}

Charting the internal structure of hadrons is a long-standing goal for nuclear physics and the focus of experimental and theoretical efforts over many decades. Relatively recently, theoretical and computational developments have enabled the first extractions of hadron structure directly from lattice quantum chromodynamics (QCD)~\cite{Liu:1993cv,Braun:2007wv,Ji:2013dva,Ji:2020ect,Radyushkin:2016hsy,Radyushkin:2017cyf,Ma:2014jla,Ma:2017pxb}, the numerical solution of the strong nuclear force, with a particular focus on parton distribution functions (PDFs). PDFs capture the longitudinal momentum structure of hadron, playing a central role in the Large Hadron Collider physics program~\cite{PDF4LHCWorkingGroup:2022cjn}, and serving as central objects of study at fixed target experiments in the US and beyond~\cite{Arrington:2021alx,Lewitowicz:2025qlr} and at the future Electron-Ion Collider~\cite{AbdulKhalek:2021gbh}.

Moving beyond one-dimensional measures of hadron structure has proved extremely difficult, however. Generalized parton distributions (GPDs) \cite{Muller:1994ses,Ji:1996ek,Ji:1996nm,Radyushkin:1996nd,Radyushkin:1996ru,Radyushkin:1997ki} were introduced close to 30 years ago, offering three-dimensional information on the inner structure of hadrons. Generalizing both PDFs and elastic form factors (EFFs), which encode the low-momentum-transfer interactions of hadrons with electromagnetic probes, they give access to a tomographic picture of the radial distribution of longitudinal momentum in a fast-moving hadron \cite{Burkardt:2000za,Burkardt:2002hr}. Their link to the hadronic energy-momentum tensor sheds light on the emergence of hadronic spin and so-called mechanical properties from the degrees of freedom of the QCD Lagrangian \cite{Ji:1996ek,Polyakov:2002yz,Polyakov:2018zvc}. 

The experimental access to GPDs is complicated partly by the small cross-section of the exclusive processes in which they play a role and partly by the nature of the deconvolution problem to extract them from experimental measurements \cite{Bertone:2021yyz,Qiu:2023mrm,Moffat:2023svr,Dutrieux:2024bgc}. Lattice QCD offers a complementary approach to GPD physics, with its own advantages and limitations. The hadronic energy-momentum tensor is related to Mellin moments, for instance the average momentum fraction $\langle x\rangle$ for quark GPDs, which can be extracted from local operators on the lattice. Such computations have reached a mature stage for ordinary parton distributions and were pioneered for GPDs 20 years ago in \cite{Hagler:2003jd,LHPC:2007blg}. Recent results can be found in \cite{Bali:2018zgl,Alexandrou:2019ali,Hackett:2023rif}. For moments of higher order or the full kinematic dependence of GPDs, efforts in lattice QCD rely on more recent frameworks, such as non-local matrix elements \cite{Ji:2013dva,Ji:2014gla}. We will use in this document the pseudo-distribution approach initiated in \cite{Radyushkin:2017cyf} and applied to GPDs in \cite{Radyushkin:2019owq,Radyushkin:2023ref}. On the one hand, these techniques bypass some limitations of experimental measurements linked, for instance, to the small cross-section, the deconvolution problem, the difficulty of disentangling some flavor contributions or the issues with hadronic probes other than the proton. On the other hand, they are limited in the range of kinematics that they are able to constrain, confronted to traditional lattice systematic errors and deal with their own inverse problem of reconstructing functions from a limited number of noisy Fourier harmonics \cite{Karpie:2019eiq, Dutrieux:2024rem, Dutrieux:2025jed, Xiong:2025obq, Medrano:2025cmg, Chen:2026aae}. The question of the size of power corrections, which is fundamentally intertwined with the range of useful Fourier harmonics computed in lattice QCD, and therefore the relevance of the reconstruction in the inverse problem, remains at an early stage of understanding. While this represents a serious limitation of the lattice approach for ordinary PDFs to discover beyond the Standard-Model physics, we can note that the majority of experimental data used to extract GPDs exists at low virtuality scales, and therefore suffers from large power corrections too (see for instance \cite{Martinez-Fernandez:2025jvk, Martinez-Fernandez:2025rcg}). The literature on GPDs from non-local matrix elements has grown quickly in recent years, as attested in \cite{Chen:2019lcm, Lin:2020rxa, Alexandrou:2020zbe, Lin:2021brq, Alexandrou:2021bbo, Bhattacharya:2022aob, Bhattacharya:2023ays, Bhattacharya:2023jsc, Bhattacharya:2023nmv, Lin:2023gxz, Bhattacharya:2024qpp, Hannaford-Gunn:2024aix, HadStruc:2024rix, Ding:2024saz, Bhattacharya:2024wtg,Schoenleber:2024auy, Bhattacharya:2025yba, Holligan:2025baj, Chu:2025kew, Gao:2025inf}. For a recent review addressing the advances in experimental, theoretical, and lattice QCD aspects of GPDs, see~\cite{Boer:2025ixc}.

Using the technology presented in \cite{HadStruc:2024rix}, where we extracted moments of GPDs including a systematic skewness expansion for the first time, we extend the framework to compute the full dependence in the three kinematic variables $(x, \xi, t)$ from the pseudo-distribution formalism. Pseudo-distributions are rooted in a Lorentz invariant formulation, so we extract from the lattice data, for the first time in the literature, a double distribution representation (DD) \cite{Radyushkin:1998es}, which encodes an important property of GPDs linked to Lorentz symmetry. Our results are therefore not bound to a specific value of the skewness $\xi$, but reflect the general entanglement that the $x$ and $\xi$ variables must obey in a flexible way. To obtain a solution to the inverse problem with reduced and controllable bias, we represent the DD on a 3D mesh of finite elements and use a Gaussian process regularization. This method yields by construction GPDs fulfilling the polynomiality property.

Our presentation is organized in the following way. Section \ref{sec:method} is devoted to the methodology. We first remind the reader of the theoretical framework relating GPDs to the matrix elements computed in lattice QCD and to DDs. Then we describe the finite element method to represent the DD and the regularization of the inverse problem through Gaussian process regression. In Section \ref{sec:ldata}, we show the lattice data, our analysis of correlation functions, and discuss a first simple quantity of interest: elastic form factors. In Section \ref{sec:results}, we show the results of our Gaussian process reconstruction, study the impact of the hyperparameters on the regularization and perform the perturbative matching. Our final results are displayed in Fig.~\ref{fig:final} and available online in the public Zenodo database \cite{dutrieux_2026_19695472}. In Section \ref{sec:gff}, we update our moment extraction from \cite{HadStruc:2024rix} using the extended kinematics and improved formalism.

\begin{widetext}

\section{Methodology\label{sec:method}}

\subsection{Theoretical framework}

\subsubsection{Relating GPDs to space-like matrix elements \label{eq:th1} }

GPDs are Lorentz invariant quantities parametrizing non-local hadronic matrix elements with a light-like separation and momentum transfer between the initial and final hadron states. In the convention of~\cite{Ji:1998pc} for a quark unpolarized GPD:

\begin{align}
    \int\frac{{\rm d}z^-}{2\pi}e^{ixP^+z^-}  \bra{N\left(p_f,\lambda_f\right)}\bar{\psi}^q\left(-\frac{z}{2}\right)\gamma^+\hat{W}\left(-\frac{z}{2},\frac{z}{2}\right)\psi^q\left(\frac{z}{2}\right)\ket{N\left(p_i,\lambda_i\right)}\mid_{z^+=0,\mathbf{z}_\perp=\mathbf{0}_\perp} \nonumber \\
    =\frac{1}{P^+}\overline{u}\left(p_f,\lambda_f\right)\left[\gamma^+ H^q\left(x,\xi,t\right)+\frac{i\sigma^{+\nu}q_\nu}{2m} E^q\left(x,\xi,t\right)\right]u\left(p_i,\lambda_i\right)\,,\label{eq:def-GPDs-HE}
\end{align}
where $\hat{W}$ is the Wilson line in the fundamental representation, $m$ the nucleon mass, $\lambda_i, \lambda_f$ are the initial and final state helicities, and $p_i, p_f$ are the initial and final state momenta that define the variables
\begin{equation}
    P \equiv \frac{1}{2}(p_i+p_f)\,,\ \ q \equiv p_f-p_i\,,\ \ t \equiv q^2\,,\ \ \xi \equiv -\frac{q^+}{2P^+}\,.\label{eq:var_def}
\end{equation}
We use the spinor normalization $\bar{u}(p,\lambda)u(p,\lambda')=2m\delta_{\lambda\lambda'}$. 

Non-zero light-like separations are not accessible on a Euclidean lattice, so we compute non-local matrix elements with an equal-time space-like separation $z^2 < 0$:
\begin{equation}
M^\mu\left(p_f,p_i,z\right)=\bra{N\left(p_f,\lambda_f\right)}\bar{\psi}^q\left(-\frac{z}{2}\right)\gamma^\mu\hat{W}\left(-\frac{z}{2},\frac{z}{2}\right)\psi^q\left(\frac{z}{2}\right)\ket{N\left(p_i,\lambda_i\right)}\,,\label{eq:mat-to-match}
\end{equation}

\end{widetext}
and form the ratio~\cite{Radyushkin:2017cyf, Orginos:2017kos}
\begin{equation}
\mathcal{M}^\mu\left(p_f,p_i,z\right) = \frac{M^\mu\left(p_f,p_i,z\right)}{M^0\left(0,0,z\right)}\label{eq:reducedmatelem}
\end{equation}
to cancel multiplicative ultraviolet divergences~\cite{Ishikawa:2017faj} that only depend on the length of the Wilson line. $\mathcal{M}$ has therefore a well-defined continuum limit and is renormalization group invariant.

The non-local matrix element  
admits a  decomposition involving  eight amplitudes  \cite{Bhattacharya:2022aob}
 that depend on the  Lorentz invariants $(p_i \cdot z)$, $(p_f \cdot z)$, $t$ and $z^2$. 
Introducing the combinations 
\begin{equation}
\nu = P \cdot z\,,\ \ \bar \nu = -(q \cdot z)/2 ,\label{eq:genskew}
\end{equation}
related to the average momentum $P$ and the momentum transfer $q$,
   we write 
\begin{align} 
 \mathcal{M}&^\mu\left(p_f,p_i,z\right)=
\langle\langle\gamma^\mu\rangle\rangle\mathcal{A}_1+z^\mu\langle\langle\mathds{1}\rangle\rangle\mathcal{A}_2+i\langle\langle\sigma^{\mu z}\rangle\rangle\mathcal{A}_3\nonumber  \\ 
&\qquad+\frac{i}{2m}\langle\langle\sigma^{\mu q}\rangle\rangle\mathcal{A}_4+\frac{q^\mu}{2m}\langle\langle\mathds{1}\rangle\rangle\mathcal{A}_5\nonumber \\
&\qquad+\frac{i}{2m}\langle\langle\sigma^{zq}\rangle\rangle\left[P^\mu\mathcal{A}_6+q^\mu\mathcal{A}_7+z^\mu\mathcal{A}_8\right]
    \label{eq:decomp}\,,
\end{align}
where each amplitude $\mathcal{A}_k$ is a function of $(\nu, \bar\nu, t, z^2)$. We have used the abbreviations $\sigma^{\mu z}\equiv\sigma^{\mu\rho}z_\rho$, $\sigma^{\mu q}\equiv\sigma^{\mu\rho}q_\rho$, $\sigma^{zq}\equiv\sigma^{\rho\lambda}z_\rho q_\lambda$, and \mbox{$\langle\langle\Gamma\rangle\rangle\equiv\overline{u}\left(p_f,\lambda_f\right) \Gamma u\left(p_i,\lambda_i\right)$.} 
The variable $\nu$ is known as the  Ioffe time~\cite{Braun:1994jq}.  Writing 
\begin{equation}
    \bar\nu \equiv \nu\xi \ ,\label{eq:cdb}
\end{equation}
generalizes  the definition of the skewness $\xi$ beyond the light-like case. 
This definition coincides with that  in Eq.~\eqref{eq:var_def} when $z$ is reduced to its light-cone component $z^-$. As we noted already in \cite{HadStruc:2024rix}, the light-cone skewness is bound in $[-1, 1]$, and in fact even more tightly when kinematics are considered, but the generalized skewness of Eq.~\eqref{eq:cdb} can take any real value when $z$ is not on the light-cone, including being undefined when $\nu = 0$.

 Note that in the non-forward kinematics, the variables $\nu$ and $\bar \nu$ are essentially on
 the same footing. As    we will see later, the variable $\bar\nu $
appears naturally in the perturbative matching and the DD formalism.

The identification of the amplitudes which contribute in the limit $z^2 \rightarrow 0$ is explained in \cite{HadStruc:2024rix} following the work in \cite{Bhattacharya:2022aob,Radyushkin:2023ref}. If we introduce generalized Ioffe-time distributions (GITD) as
\begin{align}
    \begin{pmatrix} {\mathcal H}^q \\ {\mathcal E}^q \end{pmatrix} (\nu, \bar\nu, t) = \int_{-1}^1 \mathrm{d}x\,e^{-ix\nu} \begin{pmatrix} H^q \\ E^q \end{pmatrix}(x, \xi, t)\,, \label{eq:GITD}
\end{align}
a possible choice of Lorentz invariant definition that converges to the light-cone one when $z^2 \rightarrow 0$ is
\begin{align}
    {\mathcal H}^q(\nu, \bar\nu, t, z^2) &= \mathcal{A}_1 - \xi A_5\,, \label{eq:defH}\\
    {\mathcal E}^q(\nu, \bar\nu, t, z^2) &= \mathcal{A}_4 + \nu \mathcal{A}_6 - 2\bar\nu \mathcal{A}_7 + \xi A_5\,.\label{eq:defE}
\end{align}
It is clear that there is no unique way to construct a Lorentz invariant object at $z^2 < 0$ that converges analytically towards the correct light-cone limit. The accuracy of the formalism is therefore limited by power corrections varying as $\mathcal{O}(z^2 \Lambda_{QCD}^2)$, where $\Lambda_{QCD}$ is a hadronic scale of the order of a few hundred MeVs.

At one-loop in perturbation theory with renormalization group improvement, the $z^2$-dependent non-singlet generalized Ioffe-time distributions can be matched to the $\overline{\rm MS}$ scheme in two steps \cite{Radyushkin:2019owq,Dutrieux:2023zpy}:
\begin{widetext}
\begin{itemize}
    \item First a matching kernel is applied, yielding an $\overline{\rm MS}$ distribution approximately at the scale $\mu_0^2 \sim -1/z^2$. For clarity, objects in the $\overline{\rm MS}$ scheme are denoted with a bar.
\begin{equation}
\bar{\mathcal F}\left(\nu,\bar\nu,t,\mu_0^2\right)={\mathcal F}\left(\nu,\bar\nu,t,z^2\right)+\frac{\alpha_s(\mu_0^2)C_F}{2\pi}[L + \ln\left(\frac{e^{2\gamma_E+1}}{4}\right)B] \otimes { \mathcal F}{} +\mathcal{O}\left(z^2\Lambda_{\rm QCD}^2\right) + \mathcal{O}\left(z^2 t\right)
  \label{eq:gitd-pgitd-matching}\,,
\end{equation}
where \cite{Radyushkin:2019owq}
\begin{equation}
L\otimes {\mathcal F} =\int_0^1{\rm d}\alpha \left[4\left(\frac{\ln\left(\bar\alpha\right)}{\bar\alpha}\right)_+\cos\left(\bar{\alpha}\bar\nu\right)-2\frac{\sin\left(\bar{\alpha}\bar\nu\right)}{\bar\nu}+\delta\left(\bar\alpha\right)\right]{\mathcal F} \left(\alpha\nu,\alpha\bar\nu,t\right)\,,\label{eq:LCrossM}
\end{equation}
$\bar\alpha = 1-\alpha$, ${\mathcal  F}$ denotes a non-singlet component of either ${\mathcal H}$ or 
${\mathcal E}$, and $B$ is defined below. Not only do larger values of $z$ increase the size of power corrections $\mathcal{O}(z^2\Lambda_{QCD}^2)$ and $\mathcal{O}(z^2t)$, but they also degrade the confidence in the perturbative matching as $\alpha_s(\mu_0^2)$ increases.
\item Then a differential equation is solved to perform renormalization-group improved evolution up to the desired final scale:
\begin{align}
\hspace{-9mm} \frac{\mathrm{d}}{\mathrm{d} \ln \mu^2}\, \bar{\mathcal F}(\nu, \bar\nu, t, \mu^2) = \frac{\alpha_s(\mu^2) C_F}{2\pi} B \otimes \bar{\mathcal F}\,,
\label{eq:evolGPD}\end{align}
where \cite{Radyushkin:2019owq}
\begin{equation}
B \otimes {\mathcal F} = \int_0^1{\rm d}\alpha \left[\left(\frac{2\alpha}{\bar\alpha}\right)_+\cos\left(\bar{\alpha}\bar\nu\right)+\frac{\sin\left(\bar{\alpha}\bar\nu\right)}{\bar\nu}-\frac{\delta\left(\bar\alpha\right)}{2}\right]
{\mathcal F}\left(\alpha\nu,\alpha\bar\nu,t\right)\,.
\end{equation}
\end{itemize}
\end{widetext}
The singularity for $\alpha = 1$ is regulated by the standard {\it plus-prescription} defined here as\begin{equation}\int_0^1{\rm d}\alpha\ G\left(\alpha\right)_+f\left(\alpha x\right)=\int_0^1{\rm d}\alpha\ G\left(\alpha\right)\left[f\left(\alpha x\right)-f\left(x\right)\right]\,.\end{equation}
Let us reiterate the comment from \cite{HadStruc:2024rix} that the evolution and matching take a much friendlier form in Ioffe time representation rather than in $x$-space, where the integral kernels have a non analytical behavior in $x / \xi$.

\subsubsection{The DD representation \label{sec:DDrep}}

The amplitudes $\mathcal{A}_i\left(\nu,\bar \nu, t,z^2\right)$, as functions of the Lorentz invariants 
$\nu$ and $\bar \nu$,
may be written in the  form of a  {\it double distribution} (DD) representation
\begin{align}
     {\cal A}_i (\nu, \bar\nu, t,z^2) =
     \int_\Omega \dd \beta\dd \alpha\,\,e^{-i\beta \nu -i\alpha \bar \nu}  a_i (\beta,\alpha, t, z^2)\,,
      \label{eq:DD}
\end{align}
where the support region $\Omega$  for  DDs $ a_i (\beta,\alpha, t, z^2)$  is   a rhombus specified by
 $|\alpha|+|\beta| \leq 1$ \cite{Radyushkin:1998bz}.

In fact, the DDs were introduced at the same time as GPDs \cite{Muller:1994ses,Radyushkin:1996nd,Radyushkin:1996ru,Radyushkin:1997ki}
 as an alternative way to parametrize the same off-forward non-local matrix elements.     
 Moreover, their definition does not involve parameters, like $\xi$, specifying relative size of $P$ and $q$, and one can use the same  DD  to obtain GPDs for different chosen values of $\xi$
   (see Eq.~(\ref{eq:radon}) below). Thus  DDs  are, at least, {\it as fundamental} as GPDs. 
Comparing (\ref{eq:DD}) with the {\it GPD representation} \begin{align}
     {\cal A}_i (\nu, \xi \nu, t,z^2) =
     \int_{-1}^1 \dd x \,e^{-ix  \nu}  A_i (x,\xi, t, z^2)
      \label{eq:GPD}
\end{align}
of the type in Eqs.~\eqref{eq:def-GPDs-HE} and~\eqref{eq:GITD}, 
one arrives at the relation \cite{Radyushkin:1998bz} between the (pseudo-)GPD $ A_i (x,\xi, t, z^2)$  and the (pseudo-)DD $a_i (\beta,\alpha, t, z^2)$,
\begin{equation}
A_i (x, \xi, t,z^2) = \int_\Omega \dd \beta\dd \alpha\,\delta(x - \beta-\alpha\xi) \,  a_i (\beta,\alpha, t, z^2) \ .\label{eq:radon}
\end{equation}
As noticed  in  \cite{Teryaev:2001qm},  the $\{\beta,\alpha\}$ integral may be treated formally as  a Radon transformation.

\begin{figure}[h]
    \centering
    \includegraphics[width=0.5\linewidth]{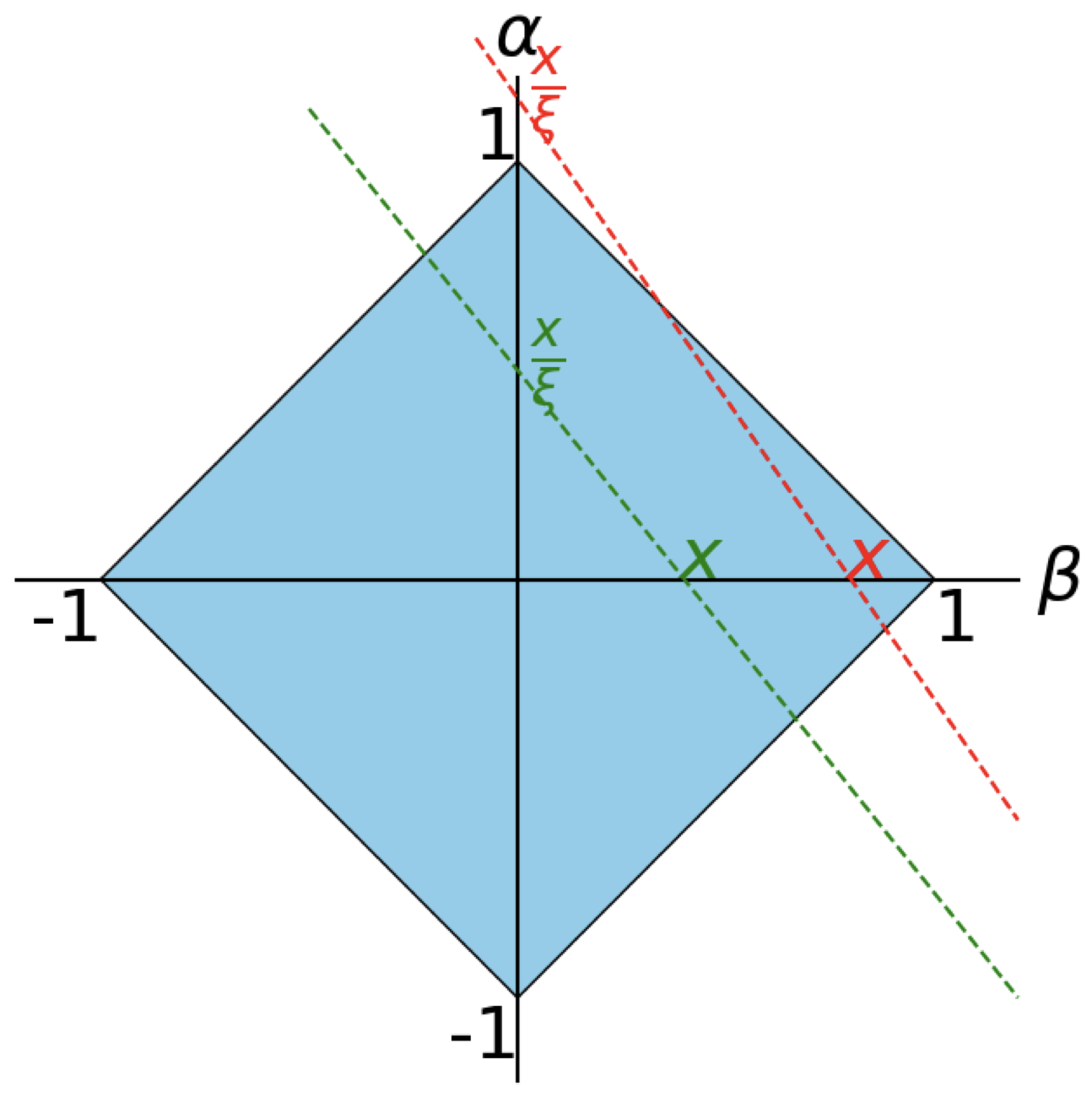}
    \caption{Illustration of the support in $(\beta, \alpha)$ of a DD. The associated GPD is obtained by integration along lines similar to the dotted ones, with green depicting the case $0 < x < \xi$ and red for $0 < \xi < x$. Figure borrowed from \cite{Mezrag:2022pqk} which presents a recent concise introduction to GPDs and DDs.}
    \label{fig:DD_Radon}
\end{figure}

It is useful to get a geometrical intuition of the connection between GPDs and DDs to understand their respective phenomenology. Fig.~\ref{fig:DD_Radon} shows a depiction of the domain of DDs and the Radon integration lines which define GPDs according to Eq.~\eqref{eq:radon}. 
 
The GPD-DD relation (\ref{eq:radon})
 leads to an   interplay between  the $x$  and $\xi$  dependences of GPDs: the 
$x^n$ Mellin moments of GPDs must be $n^{\rm th}$ order polynomials in~$\xi$ 
\begin{align}
    \int_{-1}^1 \dd x\,x^{n}  A_i (x, \xi, t,z^2)  = \sum_{k = 0}^{n} (A_i)_{n+1,k}(t,z^2) \xi^k \ .\label{eq:polrel}
\end{align}
This  result reflects  the fact that nonforward matrix elements 
of local operators involving  the $(zD)^n$  derivative are given by  a sum of  $(P\cdot z)^{n-k}(q\cdot z)^ka_{nk}$ 
terms. Hence,  the {\it polynomiality property}  (\ref{eq:polrel}) of GPDs is a simple consequence of Lorentz invariance.
Note that, due to  time reversal invariance, the DDs are either even or odd functions of $\alpha$.
Correspondingly, the summation in (\ref{eq:polrel}) involves even or odd values of $k$ only.

One advantage of working with DDs is that polynomiality is ``built-in'', so that the functional dependence of DDs is free from constraints with respect to that property, beyond simple notions of support and parity.
In particular, thanks to the polynomiality property, we were able to  extract skewness-related moments 
such as $A_{n,k}(t)$ with $k > 0$ in (\ref{eq:polrel}).

On the light cone, the  DD analog of the GPD representation (\ref{eq:def-GPDs-HE}) is 
\begin{align}
&\mathcal{M}^+\left(p_f,p_i,z_-\right)
  =
  \int_{\Omega} \dd \beta\dd \alpha\,e^{-i \beta  \nu -i\alpha\bar \nu } \, 
\bigg [\langle\langle\gamma^+ \rangle\rangle  \,  h(\beta, \alpha,t) \nonumber \\  &+\frac{i}{2M}   \langle\langle  \sigma^{+  q } \rangle\rangle    e(\beta, \alpha,t) 
  - \frac{q^+ }  {2M}  \,  \langle\langle\mathds{1}\rangle\rangle\, \, \delta (\beta)  \, D(\alpha,t)
  \bigg ] 
     \, ,\label{eq:def-DDs-HE}
\end{align}
where $h(\beta, \alpha,t)$ and $e(\beta, \alpha,t)$  are even functions of $\alpha$ and $D(\alpha,t)$ 
is an odd function of $\alpha$. This  form  corresponds to the {\it Polyakov-Weiss representation}
 for DDs \cite{Polyakov:1999gs} which contains  a  $\nu$-independent 
{\it $D$-term}, given by  the third contribution. 

Thus, in the DD representation,  we  deal with three independent functions:  $ h(\beta, \alpha,t),  e(\beta, \alpha,t)$
and $ D(\alpha,t)$. 
One may argue that, using the 
  Gordon decomposition
\begin{align}
 	&
 \frac{{\cal P}^\lambda }{M}   \bar u'   u =   \bar u'  \gamma^\lambda  u - 	 \frac1{2M}  \bar u' i \sigma^{\lambda q } u  \   
 \label{GD}
 	\end{align}
for $\lambda =+$ and the  definition $\xi = -q^+/2P^+$ , one  recovers (\ref{eq:def-GPDs-HE}) containing just   2  GPDs
\begin{widetext}
\begin{equation}
\begin{pmatrix}H^q \\ E^q\end{pmatrix} (x, \xi, t) = \int_\Omega \mathrm{d}\beta\mathrm{d}\alpha\,\delta(x - \beta-\alpha\xi) \left[\begin{pmatrix}h^q \\ e^q\end{pmatrix}(\beta, \alpha, t) \pm \xi \delta(\beta)D^q(\alpha, t)\right]\,,\label{eq:efwrb}
\end{equation}
where $\pm$ is  $+$ for $H$ and $-$ for $E$. Thus

\begin{align} 
 H^q\left(x,\xi,t\right) &= H^q_R\left(x,\xi,t\right) +{\rm sgn} (\xi)  D^q(x/\xi,t)  \ , 
 \nonumber \\ 
  E^q\left(x,\xi,t\right) &= E^q_R \left(x,\xi,t\right) - {\rm sgn} (\xi)  D^q(x/\xi,t) \ ,
  \label{HxiExi} 
\end{align} 
where $H^q_R\left(x,\xi,t\right)$ and $E^q_R\left(x,\xi,t\right)$ are the Radon transforms (\ref {eq:radon})
of $ h^q(\beta, \alpha,t) $ and $ e^q(\beta, \alpha,t) $.  Since the $D$-term $ D^q(x/\xi,t)$ is independent 
of $H^q_R\left(x,\xi,t\right)$ and $E^q_R\left(x,\xi,t\right)$, one deals anyway  with three independent functions.
For moments, one may write 
\begin{align}
    \int_{-1}^1 \dd x\,x^{n-1} \begin{pmatrix}H^q \\ E^q \end{pmatrix}(x, \xi, t) = \sum_{k = 0, \textrm{ even}}^{n-1} \begin{pmatrix}A_{n,k}(t) \\ B_{n,k} (t)\end{pmatrix} \xi^k \pm 
    \frac{1+(-1)^n}{2}  \xi^{n} C_n(t)\,,\label{eq:polrelHE}
\end{align}
 \end{widetext}
Note that the moments of the $D$-term, $C_n(t)$, contribute  for $n $ even only. One can  build the $D$-term without a reference to DDs, 
just from the highest powers of $\xi^k$ in the even-$n$ $x^{n-1}$ moments of $ H\left(x,\xi,t\right)$ or  $E\left(x,\xi,t\right)$.

A few observations can guide us when imposing proper modelling constraints on DDs. We assume in the following that both $x$ and $\xi$ are positive quantities to remove the need of constantly using absolute values. 
\begin{itemize}
\item \textbf{\textit{Small $\beta$ behavior}} If $x$ is small and $\xi$ of the order of $x$ or less, Radon integration lines are mostly vertical and only explore the region of $\beta$ of the order of $x$ or less. We can draw the rule ``the small $x$ behavior of the GPD at small $\xi$ is dictated by the small $\beta$ behavior of the DD". We know from phenomenology and the effect of evolution kernels that the small $x$ behavior of GPDs at small $\xi$ is divergent, so we expect a similar divergence of DDs at small $\beta$. At this stage however, we cannot infer that the divergence occurs at small $\beta$ for all values of $\alpha$. That fact is demonstrated using the convenient evolution of GPDs moments in Appendix \ref{sec:appA}.
\item \textbf{\textit{Large $\beta$ behavior}} If $x$ is close to 1 and $\xi$ smaller than $x$, Radon integration lines mostly probe the corner $(\beta, \alpha) \approx (1, 0)$ of the rhombus. Perturbation theory gives us insights on the behavior of the GPD at large $x$ \cite{Yuan:2003fs}, for instance:
\begin{equation}
H^q(x, \xi, t) \sim \frac{(1-x)^3}{(1-\xi)^2(1+\xi)^2}\label{eq:largexGPD}\,.
\end{equation}
The $\beta$-dependence at large $\beta$, which reflects the $x$-dependence at large $x$ and $\xi$ smaller than $x$, should therefore vanish as a power of $(1-\beta)$.
\item \textbf{\textit{Behavior on the edges}} Notice that in Eq.~\eqref{eq:largexGPD}, the limit $(x, \xi) \rightarrow (1, 1)$ is 0 along any path that approaches this limit in a linear fashion in $x$ and $\xi$ (\textit{i.e.} $\lim_{\varepsilon \rightarrow 0} H^q(1-\lambda\epsilon, 1-\epsilon) = 0$ for any $\lambda \geq 0$). This is due to the fact that the power of $(1-x)$ in the numerator is larger than the power of $(1-\xi)$ in the denominator.

However, in terms of DDs, as was previously derived in \cite{Bertone:2021yyz} for $\lambda < 1$:
\begin{equation}
\lim_{\varepsilon\rightarrow 0} H^q(1-\lambda\varepsilon, 1-\varepsilon) = \int_0^\lambda \mathrm{d}\alpha\,h^q(1-\alpha, \alpha)\,.
\end{equation}
The limit $(x, \xi) \rightarrow (1, 1)$ then depends on the angle of approach characterized by $\lambda$, unless the DD vanishes on its edge $\beta=1-\alpha$. There remains a point of contention at $(\beta, \alpha) = (0, 1)$ where the divergence at small $\beta$ and the cancellation on the edges conflict. One will notice that the corner $(\beta, \alpha) = (0, 1)$ is probed by Radon integration lines corresponding to $x = \xi$, a well-known point of non-analyticity for GPDs, reflected in the non-analytical behavior of the DD.
\end{itemize}

One of the most successful models of DDs used in the literature is Radyushkin's DD Ansatz (RDDA) \cite{Radyushkin:1998es,Radyushkin:1998bz,Musatov:1999xp}. It combines the three criteria that we have identified: divergence at small $\beta$, power law in $(1-\beta)$ at large $\beta$ and cancellation on the edges of the domain. The first two criteria are obtained by factorizing in the Ansatz a term looking like an ordinary PDF $q(x)$ with divergence at small $x$ and power law in $(1-x)$ at large $x$, up to the addition of a dependence in $t$:
\begin{equation}
h^q(\beta, \alpha, t) = q(\beta, t) \frac{((1-|\beta|)^2-\alpha^2)^N}{(1-|\beta|)^{2N+1}} \frac{\Gamma(N+3/2)}{\sqrt{\pi}\Gamma(N+1)}\,.\label{eq:RDDA}
\end{equation}
The last factor is a normalizing term depending on the Gamma function, which ensures that $H^q(x, 0, t) = q(x, t)$. The fact that the $\alpha$-dependence (and therefore the $\xi$-dependence) is controlled by a single parameter $N$
produces however a rather inflexible modelling \cite{Mezrag:2013mya}.

Besides the three features that we have listed above, parity constraints are helpful. The GPDs $H$ and $E$ are real and $\xi$-even \cite{Belitsky:2005qn}, so the DDs $h$ and $e$ are also real and even in $\alpha$. The $x$-odd contribution of the GPD translates to the $\beta$-odd contribution of the DD, and likewise for the even parts.

\begin{widetext}
Switching to Ioffe-time distributions  in Eq.~\eqref{eq:efwrb}, 
we have
\begin{align}
\begin{pmatrix} {\mathcal H}^q \\ {\mathcal E}^q \end{pmatrix} (\nu, \bar\nu, t) = \int_\Omega \mathrm{d}\beta\mathrm{d}\alpha\,e^{-i(\beta\nu+\alpha\bar\nu)} \begin{pmatrix}h^q \\ e^q\end{pmatrix}(\beta, \alpha, t) \pm \xi \int_{-1}^1 \mathrm{d}\alpha\, e^{-i\alpha\bar\nu}D^q(\alpha, t)\,.
\end{align}
where $\xi$ is understood as $\xi = \bar \nu /\nu$. Following the discussion of \cite{Radyushkin:2023ref} and comparing with Eqs.~\eqref{eq:defH} and \eqref{eq:defE}, we find the relations between the DDs extracted in this work and the lattice data as:
\begin{align}
\mathcal{A}_1(\nu, \bar\nu, t, z^2) &= \int_\Omega \mathrm{d}\beta\mathrm{d}\alpha\,e^{-i(\beta\nu+\alpha\bar\nu)}h^q(\beta, \alpha, t, z^2)\,, \\
[\mathcal{A}_4 + \nu \mathcal{A}_6 - 2\bar\nu \mathcal{A}_7](\nu, \bar\nu, t, z^2) &= \int_\Omega \mathrm{d}\beta\mathrm{d}\alpha\,e^{-i(\beta\nu+\alpha\bar\nu)}e^q(\beta, \alpha, t, z^2)\,, \\
\mathcal{A}_5(\nu, \bar\nu, t, z^2) &= - \int_{-1}^1 \mathrm{d}\alpha\, e^{-i\alpha\bar\nu}D^q(\alpha, t, z^2)\,.
\end{align}
Adding the parity constraints, we find for instance for the first relation:
\begin{align}
\textrm{Re}\, \mathcal{A}_1(\nu, \bar\nu, t, z^2) &= \int_\Omega \mathrm{d}\beta\mathrm{d}\alpha\,\cos(\beta\nu)\cos(\alpha\bar\nu)h^q(\beta, \alpha, t, z^2)\,, \\
\textrm{Im}\, \mathcal{A}_1(\nu, \bar\nu, t, z^2) &= -\int_\Omega \mathrm{d}\beta\mathrm{d}\alpha\,\sin(\beta\nu)\cos(\alpha\bar\nu)h^q(\beta, \alpha, t, z^2)\,,
\end{align}
whence it stems that $\textrm{Re}\, \mathcal{A}_1$ constrains the $\beta$-even part of the DD (and hence the $x$-even part of the GPD) and $\textrm{Im}\, \mathcal{A}_1$ the $\beta$-odd part. If we define
\begin{align}
h^{q(-)}(\beta, \alpha, t) &= h^q(\beta, \alpha, t) +  h^q(-\beta, \alpha, t) \,,\\
h^{q(+)}(\beta, \alpha, t) &= h^q(\beta, \alpha, t) -  h^q(-\beta, \alpha, t) \,,
\end{align} we finally obtain
\begin{align}
\textrm{Re}\, \mathcal{A}_1(\nu, \bar\nu, t, z^2) &= 2\int_{\Omega^+} \mathrm{d}\beta\mathrm{d}\alpha\,\cos(\beta\nu)\cos(\alpha\bar\nu)h^{q(-)}(\beta, \alpha, t, z^2)\,,\label{eq:frefvr} \\
\textrm{Im}\, \mathcal{A}_1(\nu, \bar\nu, t, z^2) &= -2 \int_{\Omega^+} \mathrm{d}\beta\mathrm{d}\alpha\,\sin(\beta\nu)\cos(\alpha\bar\nu)h^{q(+)}(\beta, \alpha, t, z^2)\,,
\end{align}
where the integrals are restricted to the triangular domain $\Omega^+ = \{(\beta, \alpha) \in [0,1]^2\,|\,\beta \leq 1-\alpha\}$. Since we will only compute a finite number of kinematics $(\nu, \bar\nu, t, z^2)$ dictated by the values of momenta $(p_f, p_i)$ and separations $z$ on the lattice, it is obvious that reconstructing the DDs amounts to solving an inverse problem, where we seek to implement in a flexible but meaningful way the physical features we have discussed.
\end{widetext}

\subsection{Gaussian process regression for DDs \label{sec:GPRDD}}

\subsubsection{Conceptual basis}

Gaussian process regression (GPR) is a non-parametric way to generate flexible models implementing constraints such as the smoothness of the reconstruction, asymptotic behavior or values at fixed points. GPR was explored in the context of parton distributions on the lattice in \cite{Alexandrou:2020tqq, Candido:2024hjt, Dutrieux:2024rem, Medrano:2025cmg}. Gaussian processes can be interpreted as a specific implementation of neural networks with a single extremely, if not infinitely, wide hidden layer. This geometry distinguishes it from other neural network applications to modelling parton distribution from lattice QCD calculations in~\cite{Karpie:2019eiq, Cichy:2019ebf, DelDebbio:2020rgv, Khan:2022vot, Chowdhury:2024ymm, Chu:2025jsi}.

Let us assume that we want to reconstruct a function $f^q$. We have information on $f^q$ through some linear transformation ${\mathcal B}$. For instance, we can rewrite Eq.~\eqref{eq:frefvr} as
\begin{equation}
{\mathcal M}(\nu, \bar\nu, t, z^2) = {\mathcal B}(\nu, \bar\nu) \otimes f^{q}(t, z^2)\,,\label{eq:3eferb}
\end{equation}
where 
${\mathcal B}(\nu, \bar\nu)$ is the operator of the 2D cosine transform on the domain $\Omega^+$ for the kinematics $(\nu, \bar\nu)$, and we have called the left-hand side ${\mathcal  M}$ to represent a generic observable. The information is obtained on a finite discrete set of kinematics, labelled by the index $k$ (here $(\nu_k, \bar\nu_k, t_k, z_k^2)$), and can be represented by a multivariate normal distribution $\mathcal{N}(M_k, \Sigma)$ with mean vector $(M_k)$ and covariance matrix $\Sigma$.

We represent the function $f^q$ by its values on the vertices of a fine mesh, labelled $f^q_i$. In the following $(f^q_i)$ will therefore denote a vector whose size is the number of vertices. For any point which is not a vertex of the mesh, the value is obtained by an interpolation method, chosen to be a linear operator applied to $(f^q_i)$, defined by the specific choice of finite elements. Evaluating the action of the operator ${\mathcal B}$ on the interpolation basis for the probed kinematics, the action of ${\mathcal B}$ on the function $f^q$ becomes the product of a matrix $(B_{k,i})$ with the vector $(f^q_i)$. The goal is therefore to determine the distribution of the random vector $(f^q_i)$ such that ${\mathcal B }\otimes f^q \equiv (B_{k,i}) \cdot (f^q_i)$ approximates the distribution of known data $\mathcal{N}(M_k, \Sigma)$.

As the number of free parameters in $(f^q_i)$ is considerably larger than the number of lattice data points, the linear system implied by the matrix $(B_{k,i})$ is severely under-constrained and we need a regularization method. We give ourselves a Bayesian prior on the random vector $(f^q_i)$ under the form of a multivariate normal distribution $\mathcal{N}(g_i, K)$. The diagonal terms of $K$ represent the \textit{a priori} uncertainty at each vertex. For instance, we will use them to enforce our prior expectation that the DD vanishes as a power of $(1-\beta)$ when $\beta$ is large. The off-diagonal terms of $K$ encode the correlation across different vertices. The chosen correlation length enforces requirements of smoothness on the solution. 

\begin{widetext}
Once the prior $I = \mathcal{N}(g_i, K)$ is chosen and the data $M = \mathcal{N}(M_k,\Sigma)$ computed, Bayes's theorem gives the posterior probability distribution of $(f^q_i)$ to be:
\begin{equation}
P(f^q_i | M, I) = \frac{P(M | f^q_i) P(f^q_i | I)}{P(M | I)}\,.
\end{equation}
$P(f^q_i|I)$ is simply given by the density of the prior distribution:
\begin{equation}
P(f^q_i | I) = \frac{1}{\sqrt{\det(2\pi K)}} \exp\left(-\frac{1}{2} (f^q_i - g_i)^T K^{-1} (f^q_i-g_i)\right)\,.
\end{equation}
The data likelihood term $P(M | f^q_i)$ is chosen as the correlated $\chi^2$ of ${B} \cdot f^q$ with the data:
\begin{equation}
P(M|f^q_i) = \frac{1}{\sqrt{\det(2\pi\Sigma)}} \exp\left(-\frac{1}{2} (M_k - { B} \cdot f^q)^T \Sigma^{-1} (M_k - {B} \cdot f^q)\right)\,.
\end{equation} 
\end{widetext}
The evidence term $P(M|I)$ collects the $f^q$-independent factors to normalize the probability distribution and it plays an important role in modelling strategies where the prior $I$ depends on hyperparameters that are integrated or optimized (see \cite{Medrano:2025cmg}). For this work, we choose to present a collection of posteriors using either fixed hyperparameters, or simple unweighted samples from sets of hyperparameters. Therefore, the evidence term is not central to our discussion.

With fixed hyperparameters, the posterior probability is normally distributed with mean:
\begin{equation}
\langle f^q_i \rangle = (g_i) + KB^T [\Sigma + BKB^T]^{-1}(M_k - B \cdot g) \,,
\end{equation}
and covariance matrix:
\begin{equation}
H = K - KB^T [\Sigma + BKB^T]^{-1} BK\,.
\end{equation}
The simplicity of the approach stems from the fact that the data is related in a linear fashion to the function to reconstruct, so that with fixed prior hyperparameters the distribution of the posterior can be computed with simple linear algebra.~\footnote{This is not a fundamental limitation. For processes such as Drell Yan, Semi-Inclusive Deep Inelastic Scattering, or Deeply Virtual Meson Production which contain two unknown functions, specifically parton distributions, fragmentation functions, GPDs, or distribution amplitudes, can be handled within GPR by applying a form of the Hubbard-Stratonovich transformation~\cite{1957SPhD....2..416S,Hubbard:1959ub}. This comes at the cost of introducing a new hyperparameter to be integrated over for each data point of that type.} On top of its flexibility compared to traditional non-linear fits with a small number of parameters, this approach is less prone to unphysical features of uncertainty tightening for our specific application \cite{Dutrieux:2024rem}.

\subsubsection{Choice of set-up}

To conduct a GPR as presented in the previous section, we need to arbitrate several choices: the mean $(g_i)$ and covariance $K$ of the prior, the mesh on which the function to reconstruct is discretized and the interpolation procedure. 

Let us make directly a remark on the $z^2$ variable, since the DDs we reconstruct from lattice data depend on $(\beta, \alpha, t, z^2)$. The $z^2$-dependence can be understood in terms of matching and evolution equations as we presented in Section~\ref{eq:th1}. Therefore, we do not include the $z^2$ dimension in our modelling of DDs. Instead, we perform a different reconstruction at each $z^2$, which we match to the same final $\overline{MS}$ scale. The difference obtained between various reconstructions will provide insights on the importance of the range of Ioffe time data, perturbative uncertainties, and power corrections. 

We are therefore reduced to a three-dimensional domain in $(\beta, \alpha, t)$. For the prior mean, owing to the complicated dependence of DDs on their three kinematic variables, we generally opt for a simple mean of zero. Tests with more sophisticated prior means did not result in a significant difference considering our wide prior covariance $K$. For the latter, we implement the three physical criteria that we discussed in Section \ref{sec:DDrep} in the following way:
\begin{widetext}
\begin{align}
K_{ij} &= K[(\beta_i, \alpha_i, t_i), (\beta_j, \alpha_j, t_j)] \,, \nonumber \\
&= \sigma^2 (\beta_i \beta_j)^{-\gamma} ((1-\beta_i)(1-\beta_j))^\delta ((1-\alpha_i-\beta_i)(1-\alpha_j-\beta_j))^\eta \nonumber \\
&\hspace{20pt}\times\exp\left(-\frac{(\ln(\beta_i)-\ln(\beta_j))^2}{2 l_\beta^2}\right)\exp\left(-\frac{(\beta_i \beta_j)^{-\gamma}(\alpha_i-\alpha_j)^2}{2 l_\alpha^2}\right)\exp\left(-\frac{(t_i-t_j)^2}{2 l_t^2}\right)\,.\label{eq:kernel}
\end{align}
\end{widetext}
The hyperparameters of the prior are therefore: $\sigma^2$, which controls the overall variance of the prior; $\gamma$, which controls the rate of divergence of the uncertainty at small $\beta$; $\delta$, which governs the convergence to zero at large $\beta$; $\eta$, which controls the convergence to zero on the edges of the domain; and the three correlation lengths $l_\beta$, $\l_\alpha$ and $l_t$, which control the smoothness of the reconstruction in the three kinematic variables. For the $\beta$ variable, we use a logarithmic scale to control the smoothness, allowing more flexibility at small $\beta$. Note that, for instance, we do not strictly force the reconstructed DD to diverge as some power $\beta^{-\gamma}$ at small $\beta$: we merely offer it the possibility to do so by allowing the prior uncertainty to diverge at that rate. If the data significantly disagreed with this \textit{a priori} expectation, the posterior would differ. However, we are well aware that our data, made of low Fourier modes of the DD, offer no significant sensitivity to the small $\beta$ region. Therefore it is likely that the posterior distribution will follow that of the prior at small $\beta$, corresponding to a mean of zero and power-divergent uncertainty given by $\beta^{-\delta}$. The non-parametric nature of the GPR is really put to use in the region where the data is informative.

The final ingredient is our choice of discretization mesh and the interpolation routine. For convenience of computation, we use a regular grid in $(\beta, \alpha, t)$ and a 3D cubic spline interpolation. The coarseness of the mesh in each direction is a crucial matter for the numerical efficiency of the framework in three dimensions. We address questions of performance in Appendix~\ref{app:perf}. In contrast, mesh coarseness was significantly less important in the one-dimensional applications we considered before, because it was computationally cheap to generate a finer-than-necessary mesh.

When performance is an issue, the coarseness of the mesh must be set in accordance with the smoothness imposed by the Bayesian prior. It is counter-productive to give a large number of degrees of freedom on a region of the mesh if the prior imposes very large correlations between neighboring vertices. Since DDs are slowly varying functions of $\alpha$ and $t$, a large number of points in those dimensions is not necessary. We will rather focus on the question of the number of points necessary along the $\beta$ dimension, whose phenomenological behavior is akin to the $x$-dependence of one-dimensional PDFs.

\begin{figure}
    \centering
    \includegraphics[width=0.8\linewidth]{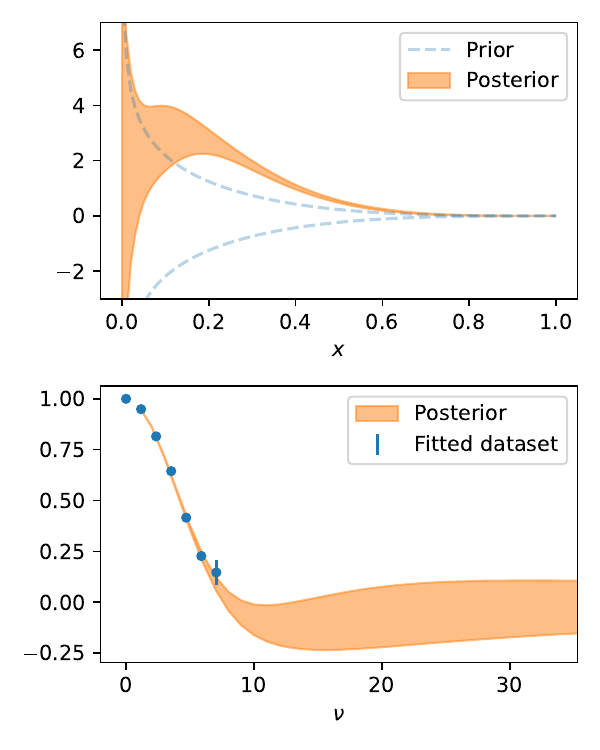}
    \caption{(top) Prior and posterior in $x$-space when using a regular grid of 100 points in $x$ -- (bottom) The fitted dataset and the posterior reconstruction in Ioffe time space.}
    \label{fig:100pts}
\end{figure}

To obtain insights on this question, we use the same dataset of one-dimensional unpolarized isovector PDFs that we used in \cite{Dutrieux:2024rem}, and a simple prior kernel:
\begin{align}
K(x, x') = 1.5^2 &(xx')^{-0.3} ((1-x)(1-x'))^3\nonumber\\
& \times \exp\left(-\frac{(\ln(x)-\ln(x'))^2}{2\ln(2)^2}\right)\,.
\end{align}
The result with a regular grid of 100 points in $x$ is shown in Fig.~\ref{fig:100pts}. Notice that the posterior can stray significantly from the prior in the region where data is informative (essentially $x \sim [0.2, 0.7]$), and follows the prior trend on the other parts of the domain. The agreement between the fitted dataset and the posterior in Ioffe time is excellent.

\begin{figure}
    \centering
    \includegraphics[width=0.85\linewidth]{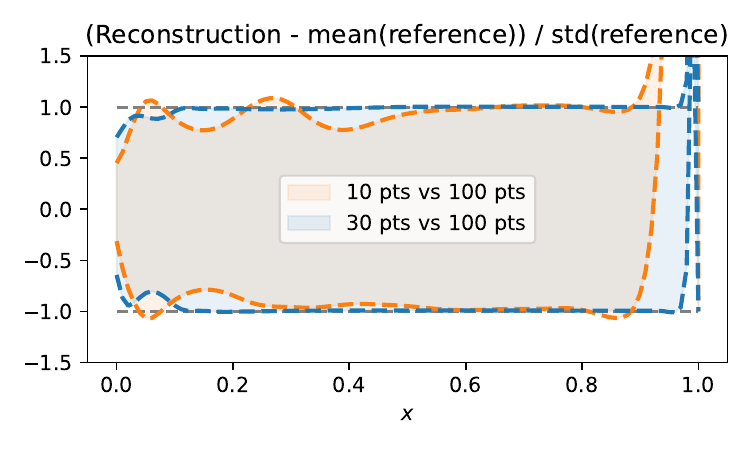}
    \caption{Comparison of the reconstruction in $x$-space using 30 and 10 points in the grid with respect to the reference of 100 points shown in Fig.~\ref{fig:100pts}.}
    \label{fig:compmeshnumb}
\end{figure}

The ratio of posteriors when reducing the number of points in the grid from 100 to 30 and to 10 is depicted in Fig.~\ref{fig:compmeshnumb}. The result with 30 points appears consistent compared to that with 100 points, except for a relatively small inconsistency below $x < 0.1$ and above $x > 0.97$. At small $x$, the grid lacks the resolution to reproduce exactly the divergence, while the relative error at large $x$ is mostly an artifact of dividing two very small numbers. With 10 points, more significant inaccuracies are observed up to the fairly large value of $x \sim 0.4$. Overall, this test using representative lattice data validates that, in the range of $x$ and $\beta$ where we expect the lattice data to be constraining, working with a grid of 30 points is sufficient. It is clear that phenomenological applications with sensitivity to GPDs at much lower values of $x$ cannot use the low resolution grid with which we work in this paper.

\begin{figure}
    \centering
    \includegraphics[width=0.9\linewidth]{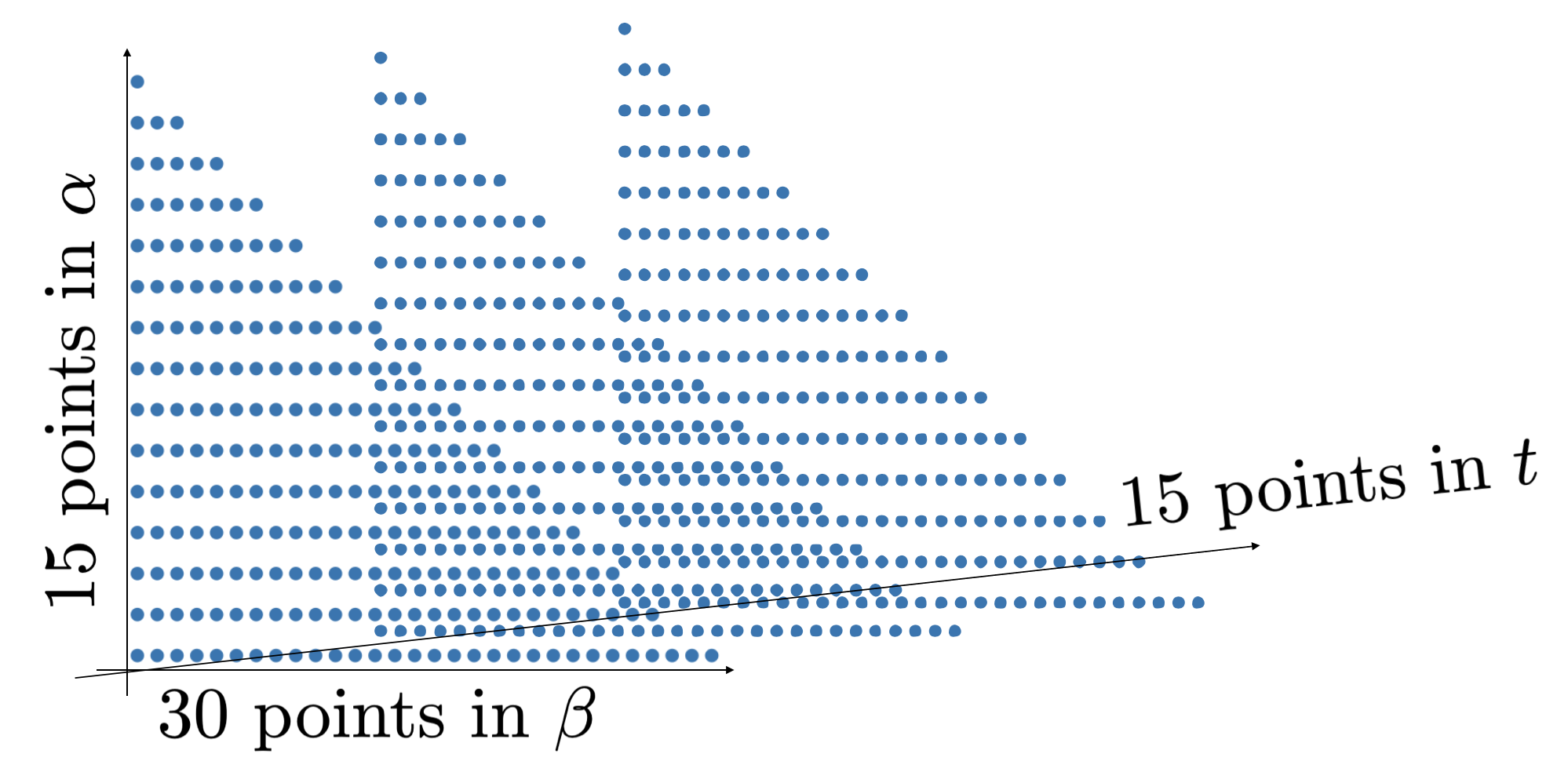}
    \caption{The discretization of the DD we use, with $30 \times 15 \times 15$ points in $(\beta, \alpha, t)$. Only three points in $t$ are shown for legibility.}
    \label{fig:mesh}
\end{figure}

Our final choice of interpolating mesh is therefore a regular 3D grid with 30 points along $\beta$, 15 along $\alpha$ and 15 along $t$, as shown in Fig.~\ref{fig:mesh}, where we have depicted three points in $t$. Because the domain $\Omega^+$ is triangular, we simply set to zero the degrees of freedom of the grid where $\beta > 1 - \alpha$. We have therefore effectively 3390 degrees of freedom in the grid. One should note that cubic splines still take non-zero values outside of the domain with this simple prescription, due to their long-distance behavior explained in more detail in Appendix~\ref{app:perf}. Instead, as described in Ref.~\cite{Medrano:2025cmg}, constraints could be enforced in the GPR to explicitly cause convergence to 0 in $(\alpha,\beta)$ space where desired. We simply ignore these (small) values outside of the domain. The range in $t$ covered by our mesh extends from 0 to -1.4 GeV$^2$.

After the DD reconstruction has been obtained, the information can be projected into different manners, each by evaluating appropriate integrals of $\alpha$ and $\beta$. In Sec.~\ref{sec:results} we produce the GITD in $\nu$ space and the GPD in $x$ space. In Sec.~\ref{sec:gff} we discuss the gravitational form factors in Mellin space. In App.~\ref{app:windows} we show the Gaussian windows proposed in Ref.~\cite{Karpie:2025ikc}.

\section{Lattice data\label{sec:ldata}}

\begin{table*}[t]
    \centering
    \begin{tabular}{ccccccccc}
    \hline\hline
    ID & $a$ (fm) & $m_\pi$ 3(MeV) & $\beta$ & $m_\pi L$ & $L^3\times N_T$ & $N_{\rm cfg}$ & $N_{\rm srcs}$ & ${\rm rk}\left(\mathcal{D}\right)$ \\
    \hline
    {\tt a094m358} & $0.094(1)$ & $358(3)$ & $6.3$ & $5.4$ & $32^3\times64$ & $348$ (\cite{HadStruc:2024rix}) & $4$ & $64$ \\
    & & & & & & $1490$ (new) & & \\
    \hline\hline
    \end{tabular}
    \caption{Lattice spacing, pion mass, inverse coupling, control parameter $m_\pi L$ for finite-volume effects, lattice volume. The total number of configurations $N_{\rm cfg}$, temporal sources $N_{\rm srcs}$ per configuration, and rank of the distillation space ${\rm rk}\left(\mathcal{D}\right)$ are also given.}
    \label{tab:ensem}
\end{table*}

We use the same gauge ensemble presented in various previous studies of parton distributions by the Hadstruc collaboration \cite{Joo:2019jct, Joo:2019bzr,Joo:2020spy,Egerer:2021ymv,HadStruc:2022nay,HadStruc:2021qdf,HadStruc:2021wmh,HadStruc:2022yaw}, with a pion mass $m_\pi=358\ {\rm MeV}$, $32^3\times64$ lattice volume and $a=0.094\ {\rm fm}$ lattice spacing~\footnote{It should be noted that the value of the lattice spacing cited in previous work by the HadStruc collaboration is larger than the higher statistics study with correctly tuned quark masses of Ref.~\cite{Yoo:2026uul}. For consistency with previous work we choose consistency with the previous HadStruc works and reserve use of the smaller lattice spacing for a future studies with continuum limit. }. This ensemble, denoted {\tt a094m358}, is an isotropic $2+1$ flavor Wilson clover fermion ensemble generated by the JLab/W\&M/LANL/MIT/Marseille collaborations~\cite{Yoo:2026uul}, with the strange quark fixed to its physical value and Wilson clover fermion sea quarks. While our previous GPD study in \cite{HadStruc:2024rix} used 348 configurations with four distillation ``sources" evenly separated in the time extent per configuration, our new results have been computed with approximately 4 times more statistics at 1490 configurations. This increase was motivated by the larger momentum of the new data, requiring better signal. Aspects of the {\tt a094m358} ensemble are summarized in Table~\ref{tab:ensem} -- further details can be found in Refs.~\cite{Yoon:2016dij,Yoon:2016jzj}. The construction of our correlation functions in the distillation framework \cite{HadronSpectrum:2009krc} is detailed in \cite{HadStruc:2024rix}. The use of distillation reduces computational resource costs by reusing the same expensive components of the calculation for many combinations of source and sink momenta. In this manner, it provides straight-forward access to what has been called ``symmetric'' and ``asymmetric'' matrix elements \cite{Bhattacharya:2022aob} without repeating
nearly identical inversions when varying the sources/sinks, which are otherwise required
for standard Gaussian-smeared correlation functions computed with the help of sequential inversion methods.

This work relies on calculating correlation functions of operators with large momentum. These correlation functions suffer from an exponentially decaying signal-to-noise ratio in the Euclidean time which must be as large as possible to obtain nucleon matrix elements. Worse, the rate of the exponential decay is proportional to the energy of the state. It was proposed in Ref~\cite{Bali:2016lva} to modify the interpolating operators such that the overlap with large momentum states is optimal by introducing a ``momentum smearing phase''. The dramatic improvement in signal is critical to the parton structure program. Within the distillation framework, one can choose the momentum smearing phase for the source and sink interpolator independently by rotating the eigenvectors at source and sink appropriately. This advantage allows for improved signal for transitions from low to high momenta. In this work, when the magnitude of the source or sink momentum in the $z$ direction is larger than $3 \frac{2\pi}L$, that interpolating operator obtains a momentum smearing phase factor of $\xi=\pm2 \frac{2\pi}L$, where the sign of the phase is the sign of the momenta.

\begin{figure}[t]
    \centering
    \includegraphics[width=0.8\linewidth]{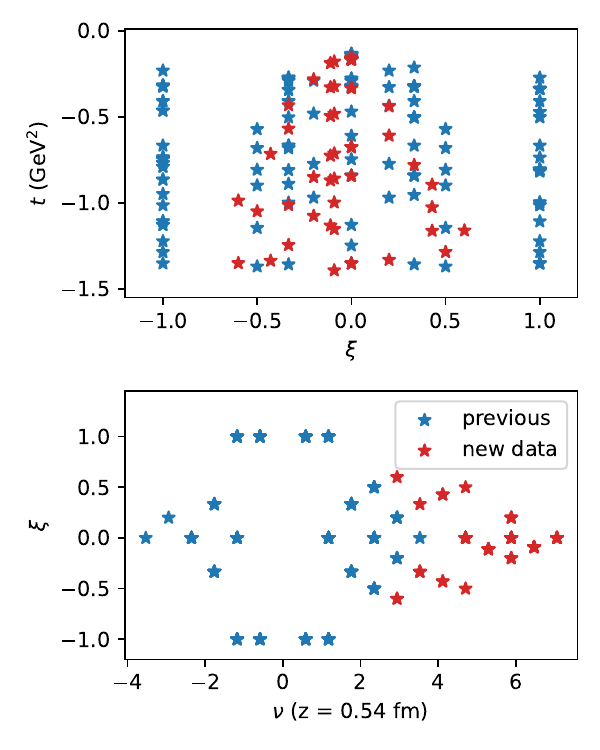}
    \caption{The kinematic coverage of this study, consisting of the 186 pairs $(\vec{p}_f, \vec{p}_i)$ of \cite{HadStruc:2024rix} in blue and 63 additional pairs computed for this study in red.}
    \label{fig:kincov}
\end{figure}

In our previous study \cite{HadStruc:2024rix}, we evaluated matrix elements corresponding to 186 pairs $(\vec{p}_f, \vec{p}_i)$ with momenta up to 1.4 GeV. In this work, we compute 63 additional pairs $(\vec{p}_f, \vec{p}_i)$ with up to double the largest momenta at 2.7 GeV. This allows us to double the maximal range in Ioffe time from $\nu \approx 0.6 z/a$ to $1.2 z/a$. For the largest value of $z$ that we will consider, namely $z = 6a = 0.56$ fm, the maximal Ioffe time reaches 7 as seen on Fig.~\ref{fig:kincov}. Our coverage in skewness is also denser thanks to the new kinematics.

\subsection{Analysis of correlation functions}

We analyze the correlation functions in a similar way to the analysis we presented in Ref.~\cite{HadStruc:2024rix}. The spectral decomposition of the two-point and three-point functions reads:
\begin{widetext} \begin{align}
    C_{2{\rm pt}}\left(\vec{p},T\right)&=\langle\mathcal{N}\left(-\vec{p},T\right)\overline{\mathcal{N}}\left(\vec{p},0\right)\rangle=\sum_n\frac{\left|\mathcal{Z}_n\left(\vec{p}\right)\right|^2}{2E_n\left(\vec{p}\right)}e^{-E_n\left(\vec{p}\right)T}\,,\label{eq:2pt} \\
    C_{3{\rm pt}}^{\left[\gamma^\mu\right]}\left(\vec{p}_f,\vec{p}_i,T;z,\tau\right)&=\langle\mathcal{N}\left(-\vec{p}_f,T\right)\overline{\psi}\left(-\frac z2,\tau\right)\gamma^\mu W^{(f)}\left(-\frac z2,\frac z2;A\right)\psi\left(\frac z2,\tau\right)\overline{\mathcal{N}}\left(\vec{p}_i,0\right)\rangle \nonumber\\
    &=\sum_{n',n}\frac{\mathcal{Z}_{n'}\left(\vec{p}_f\right)\mathcal{Z}_n^\dagger\left(\vec{p}_i\right)}{4E_{n'}\left(\vec{p}_f\right)E_n\left(\vec{p}_i\right)}\bra{n'}\mathring{\mathcal{O}}^\mu\left(z;A\right)\ket{n}e^{-E_{n'}\left(\vec{p}_f\right)\left(T-\tau\right)}e^{-E_n\left(\vec{p}_i\right)\tau}\,,\label{eq:3pt}
\end{align}
\end{widetext}
where the source and sink nucleon interpolators $\mathcal{N}$ are separated by a Euclidean time $T$. The bare quark bilinear, abbreviated by $\mathring{\mathcal{O}}^\mu\left(z;A\right)$, is introduced between the source and sink interpolators for Euclidean times $1\leq\tau\leq T-1$.

\begin{figure}
    \centering
    \includegraphics[width=0.8\linewidth]{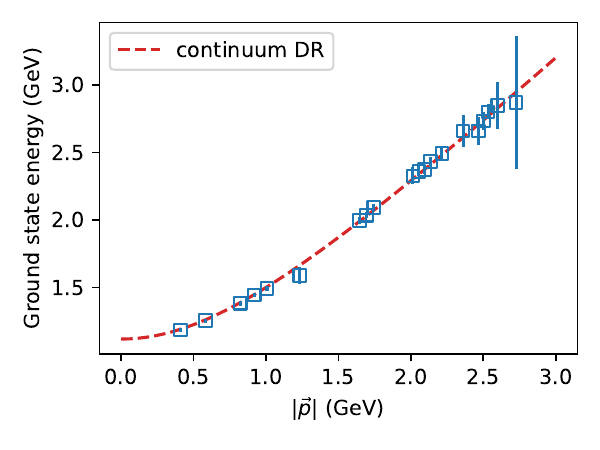}
    \includegraphics[width=0.8\linewidth]{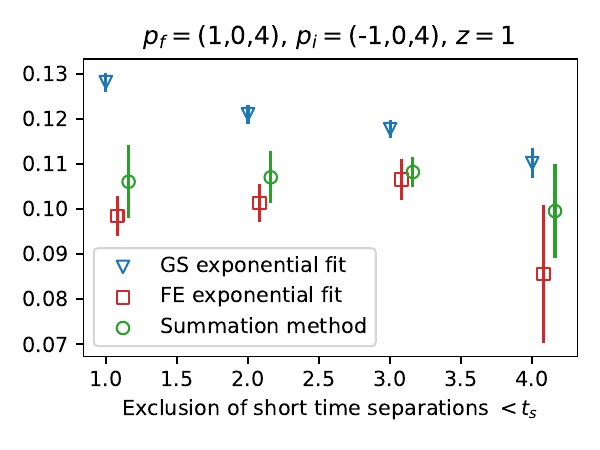}
    \caption{(top) Dispersion relation $E_0(|\vec{p}|)$ obtained from non-linear fits of the ground state and first excited state on the two-point correlation functions. The dotted line is the continuum dispersion relation. -- (bottom) Impact of the exclusion of short-time separations below $t_s$ on the extraction of the same bare matrix element as in Fig.~\ref{fig:rainbow}. We show the result of a ground state exponential fit (blue), exponential fit with one excited state (red), and summation method (green). The standard result in this analysis will use both the red point at $t_s = 3$ and at $t_s = 4$ to evaluate excited-state contamination.}
    \label{fig:DR}
\end{figure}

The energies $E_n(\vec{p})$ and overlap factors $\mathcal{Z}_n(\vec{p})$ of the ground state and first excited state are extracted from a non-linear exponential fit of the two-point functions for every jack-knife sample. We perform the fit on data with source-sink separations ranging from 4 to 16 lattice units. The resulting dispersion relation is depicted on the top panel of Fig.~\ref{fig:DR}. The proton mass is approximately 1.12 GeV on our lattice. Then the four matrix elements involving the ground state and first excited state are obtained from a linear fit of the three-point functions, independently for the real and imaginary part. We only consider data such that the operator insertion is at least $t_s$ lattice units away from the source and sink. The dependence on $t_s$ can be observed on the bottom panel of Fig.~\ref{fig:DR}, where we compare the fit with one excited state (red) to a fit with only the ground state (blue) and with the summation method (green) that we describe now.

A ratio of the three-point function with two-point functions can be designed so that it converges to the ground-state matrix element in the limit of infinite Euclidean separation between the source, sink and operator insertion:
\begin{align}
R^\mu(\vec{p}_f, \vec{p}_i, z; T, \tau) &=\frac{C^\mu_{3{\rm pt}}\left(\vec{p}_f,\vec{p}_i,z;T,\tau\right)}{C_{2{\rm pt}}\left(\vec{p}_f;T\right)} \times \nonumber \\
&\hspace{-70pt}\sqrt{\frac{C_{2{\rm pt}}\left(\vec{p}_i;T-\tau\right)C_{2{\rm pt}}\left(\vec{p}_f;\tau\right)C_{2{\rm pt}}\left(\vec{p}_f;T\right)}{C_{2{\rm pt}}\left(\vec{p}_f;T-\tau\right)C_{2{\rm pt}}\left(\vec{p}_i;\tau\right)C_{2{\rm pt}}\left(\vec{p}_i;T\right)}}\,,
\label{eq:ratio} \\
&\hspace{-40pt}\overset{\tau,T\rightarrow\infty}{\longrightarrow}\frac{1}{\sqrt{2}}\times\frac{1}{\sqrt{4E_0\left(\vec{p}_f\right)E_0\left(\vec{p}_i\right)}}M^\mu\left(p_f,p_i,z\right)\,.
\end{align}
The additional factor of $\sqrt{2}$ is due to the normalization of the isovector vector current~\footnote{
The electromagnetic current in the light-quark sector reads $\mathcal{J}^\mu=\frac{2}{3}\overline{u}\gamma^\mu u-\frac{1}{3}\overline{d}\gamma^\mu d$, which can be expressed in terms of isovector $\rho^\mu$ and isoscalar $\omega^\mu$ components according to $\mathcal{J}^\mu=\frac{\rho^\mu}{\sqrt{2}}+\frac{\omega^\mu}{3\sqrt{2}}$, where $\rho^\mu=\left(\overline{u}\gamma^\mu u-\overline{d}\gamma^\mu d\right)/\sqrt{2}$ and $\omega^\mu=\left(\overline{u}\gamma^\mu u+\overline{d}\gamma^\mu d\right)/\sqrt{2}$. We use the current $\rho^\mu$ to probe the isovector flavor structure of the off-forward nucleon matrix elements.}.

Summing the ratios $R^\mu\left(\vec{p}_f,\vec{p}_i,z;T,\tau\right)$ over the operator insertion time $\tau$, excluding all time separations below a threshold $t_s$, produces an alternative method to extract the ground-state matrix element \cite{Maiani:1987by,Capitani:2012gj}:
\begin{equation}
\sum_{\tau/a=t_s}^{T-t_s}R^\mu\left(\vec{p}_f,\vec{p}_i,z;T,\tau\right) \approx C+M^\mu\left(p_f,p_i,z\right)T + \mathcal{O}\left(e^{-\Delta ET}\right)\,,
\label{eq:SRfit}
\end{equation}
where $C$ is an irrelevant constant and $\Delta E$ is the energy gap between the lowest-lying effective excited state and the ground state. We use this method as a cross-validation of the results of exponential fits.

\begin{figure}
    \centering
    \includegraphics[width=0.99\linewidth]{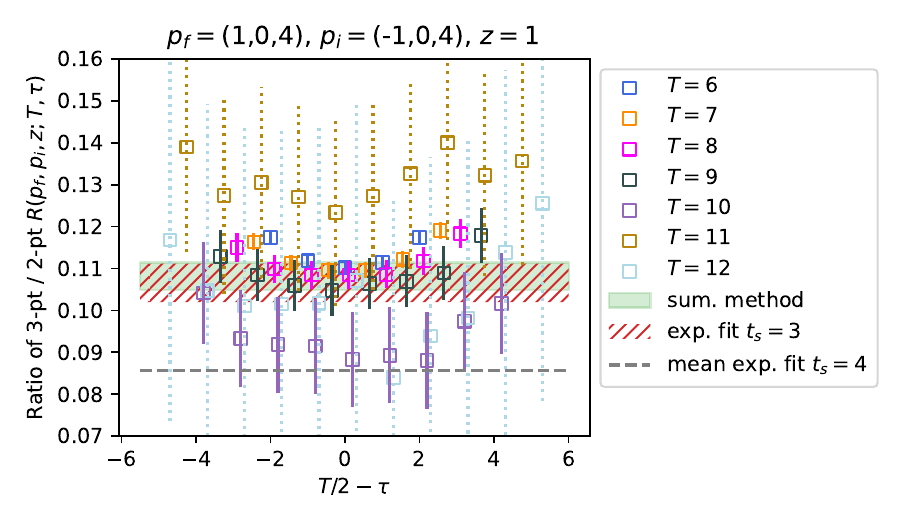}
    \caption{Ratio $R^0(\vec{p}_f,\vec{p}_i, z= a; T,\tau)$ for $\vec{p}_f = (1,0,4)$ and $\vec{p}_i = (-1,0,4)$ (lattice units) and a specific combination of helicities and operator insertion, for various values of the source-sink separation $T$ and operator insertion time $\tau$. We show the result of the matrix element fit using either the summation method (green band) or an exponential fit with one excited state (red hatches), with the exclusion of all $\tau$ and $T-\tau <3$. The mean of the exponential fit with one excited state and an exclusion of 4 is shown as the dashed grey line. The data for $T = 11$ and 12 is dotted for legibility purposes.}
    \label{fig:rainbow}
\end{figure}

The three-point functions are computed for Euclidean source-sink separations $T$ between 6 and 12. An example of ratio of three-point function with two-point functions is shown in Fig.~\ref{fig:rainbow}. We also plot the result of our extraction of the ground-state matrix element using the one-excited state fit and the summation method, both with $t_s = 3$. The mean of the extraction with the one-excited state fit and $t_s = 4$ is also depicted. As can also be seen in Fig.~\ref{fig:DR}, the statistical uncertainty explodes for $t_s = 4$, which can be linked to the quickly growing statistical uncertainty of the raw data at larger $T$. This is a manifestation of the classical exponential signal-to-noise problem of baryon correlators at large Euclidean time, which is worsened by the large hadronic momentum.

For our final extraction of the ground-state matrix element, we choose the one-excited state fit with $t_s = 3$, whose uncertainty will be quoted as statistical error. The mean difference between the extraction with $t_s = 3$ and 4 is added in quadrature as a measure of systematic uncertainty linked to excited-state contamination.

\subsection{Elastic form factors \label{sec:effs}}

When the operator separation $z$ is 0, the local operator gives access to EFFs following the simplification of Eq.~\eqref{eq:decomp} into:
\begin{align}
\mathcal{M}^\mu&\left(p_f,p_i,0\right)=\langle\langle\gamma^\mu\rangle\rangle\mathcal{A}_1\left(0,\xi,t,0\right)
\nonumber \\ & 
+\frac{i}{2m}\langle\langle\sigma^{\mu q}\rangle\rangle\mathcal{A}_4\left(0,\xi,t,0\right)+\frac{q^\mu}{2m}\langle\langle\mathds{1}\rangle\rangle\mathcal{A}_5\left(0,\xi,t,0\right)
\nonumber  \\
&\hspace{1.5cm} =\langle\langle\gamma^\mu\rangle\rangle F_1\left(t\right)+\frac{i}{2m}\langle\langle\sigma^{\mu q}\rangle\rangle F_2\left(t\right)
\nonumber \\ & \hspace{1.5cm} +\frac{q^\mu}{2m}\langle\langle\mathds{1}\rangle\rangle\mathcal{A}_5\left(0,\xi,t,0\right)\label{eq:dirac-pauli-FFs}\,,
\end{align}
where $F_1\left(t\right)$ and $F_2\left(t\right)$ are the Dirac and Pauli form factors of the nucleon. $\mathcal{A}_5(z=0)$ vanishes in principle due to the Ward identity $q_\mu M^\mu=0$. Departures of $\mathcal{A}_5$ from zero in the local limit can be interpreted as lattice artifacts since the local vector current is not the conserved vector current on the lattice.

The amplitudes $\mathcal{A}_k$ are related to the matrix elements $\mathcal{M}^\mu$ through a linear combination whose coefficients are constructed from kinematic factors and spinor products. For a given pair $(\vec{p}_f, \vec{p}_i)$ and a separation $z$, 16 matrix elements are obtained by varying the hadron helicity and Lorentz structure of the inserted operator. On the other hand, there are up to 8 amplitudes $\mathcal{A}_k$. The linear system relating matrix elements to amplitudes is both redundant and, for some kinematics, not of full column rank. We extract the amplitudes from the matrix elements using the pseudo-inverse of the matrix of kinematic factors, which amounts to finding the least-squares solution to the linear system. 

\begin{figure}
    \centering
    \includegraphics[width=0.99\linewidth]{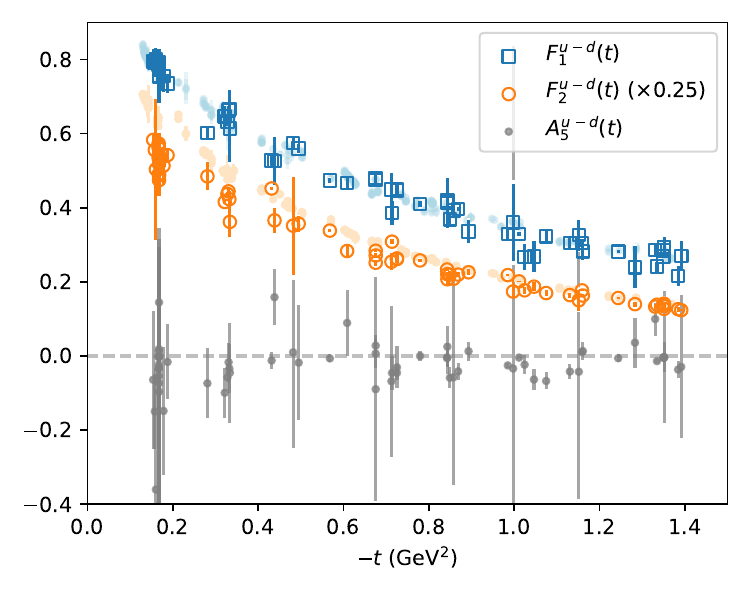}
    \caption{Elastic form factors of the proton for the isovector flavor combination $u-d$, extracted from the local matrix elements. The faint blue and orange points represent the extraction of $F_1(t)$ and $F_2(t)$ using lower hadron momentum in our previous work \cite{HadStruc:2024rix}. The points with more intense color represent the extraction performed using the new data with larger hadron momentum. Only the new data is shown for $\mathcal{A}_5$ for legibility.}
    \label{fig:EFFphasedisovec}
\end{figure}

In Fig.~\ref{fig:EFFphasedisovec}, we overlay the new 63 computations of the EFFs at $z = 0$ on top of the previous ones of \cite{HadStruc:2024rix}. The agreement is generally good, apart at small $|t|$ for the $F_2(t)$ form factor where the new data seem to exhibit a systematic downward shift. The uncertainty displayed includes the statistical uncertainty of the one-excited state exponential fit of the matrix elements with a cut $t_s = 3$ and the systematic uncertainty derived from the difference of central values between the cut $t_s = 3$ and $t_s = 4$. The EFF $A_5(t)$ is generally compatible with zero. Some clutter in the points is observed at lower momenta, and may result from unaccounted for discretization errors, or under-evaluated excited state contamination.

\begin{figure}
    \centering
    \includegraphics[width=0.99\linewidth]{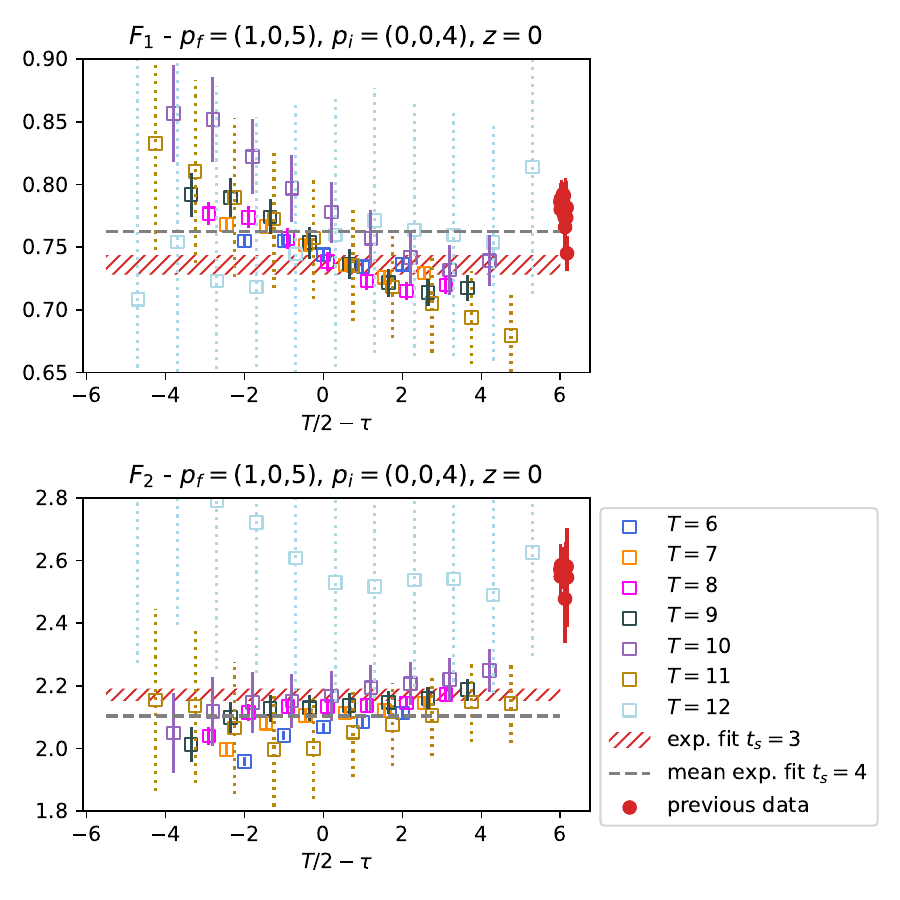}
    \caption{Applying the pseudo-inverse of the kinematic matrix directly on the three-point function data at $z = 0$ produces data for the EFFs $F_1$ (top) and $F_2$ (bottom) with excited state contamination, which depend on $(T, \tau)$. The red points represent extractions of $F_1$ and $F_2$ for nearby values of $t$ from the previous study of \cite{HadStruc:2024rix}. While the agreement is good for $F_1$, tension is visible in $F_2$ and seems difficult to explain solely by excited state contamination.}
    \label{fig:amp_cont}
\end{figure}

We investigate the possibility that poorly accounted for excited state contamination at larger hadron momentum causes the downward shift of $F_2(t)$ at small $|t|$ in the new data. We pick the specific kinematic $\vec{p}_f = (1,0,5)$ and $\vec{p}_i = (0,0,4)$ corresponding to $t = -0.19$ GeV$^2$ which stands noticeably below the previous extraction of $F_2$. In our current analysis, we first extract the ground-state matrix elements $\mathcal{M}^\mu$ from the three-point function data, and then obtain the Lorentz invariant amplitudes through the pseudo-inverse of the kinematic matrix. Instead, we propose to apply the pseudo-inverse of the kinematic matrix directly to the three-point function data. We obtain therefore amplitudes with excited-state contamination on various $(T, \tau)$ kinematics. Only then do we extract the ground-state contribution. This alternative method offers the possibility to gauge how excited state contamination is distributed among the different Lorentz amplitudes. The result is shown on Fig.~\ref{fig:amp_cont}, illustrating the good agreement of our new dataset with the previous data in red for $F_1$, and the tension for $F_2$. It seems that the data itself, irrespective of the specific procedure to extract the ground state contribution, is in tension with the previous value of $F_2$, unless one only considers the extremely noisy $T = 12$ data. The cause of this discrepancy remains to be clarified, and could for instance stem for discretization errors that we cannot assess at this stage.

\section{Results}\label{sec:results}

\subsection{Baseline reconstruction}

\begin{figure}
    \centering
    \includegraphics[width=0.8\linewidth]{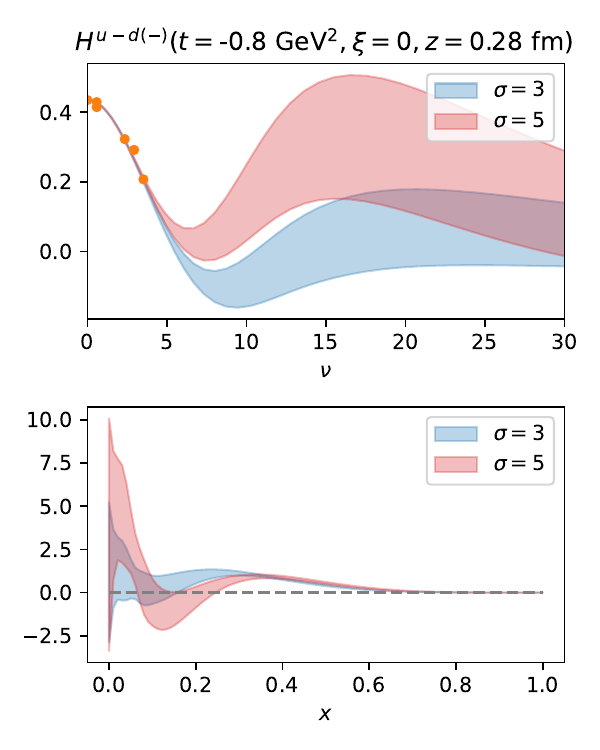}
    \caption{Demonstration of the effect of increasing the prior width from $\sigma = 3$ to 5, all other hyperparameters kept equal in Ioffe time (top) and $x$-space (bottom). The fitted dataset is shown in orange on the left panel.}
    \label{fig:issue}
\end{figure}

We present a result of the GPR procedure where the seven hyperparameters of the Gaussian process kernel of Eq.~\eqref{eq:kernel} have been fixed to physically-motivated values:
\begin{center}
    \begin{tabular}{ccccccc}
    \hline\hline
    $\sigma$ & $\gamma$ & $\delta$ & $\eta$ & $l_\beta$ & $l_\alpha$ & $l_t$ \\
    \hline
    variable & 0.3 & 2 & 1 & $\ln(2)$ & 0.2 & 0.3 \\
    \hline\hline
    \end{tabular}
\end{center}
Setting the overall magnitude of the prior, controlled by the $\sigma$ parameter, is the most delicate task. If $\sigma$ is too small, the reconstruction will be biased towards zero, the mean of our prior. Maybe less intuitively, $\sigma$ too large is also an issue beyond a simple overestimation of uncertainty, as depicted in Fig.~\ref{fig:issue}. The looser prior in $x$-space (red curve) allows for a stark oscillation around $x = 0.1$ on the bottom panel, which is perfectly compatible with the fitted dataset as seen on the top panel, but results in an extravagant extrapolation at larger Ioffe time. This exemplifies the ill-definedness of the inverse problem we are facing, especially when $z$ is small. The value of $\sigma$ is chosen to be as large as possible for each GPD type before the onset of unphysical extrapolation behavior at larger Ioffe time. We use $\sigma = 3$ for $H^{u-d(-)}$, $\sigma = 4$ for $H^{u-d(+)}$, $\sigma = 10$ for $E^{u-d(-)}$ and $\sigma = 7$ for $E^{u-d(+)}$.

\begin{figure*}
    \centering
    \includegraphics[width=0.42\linewidth]{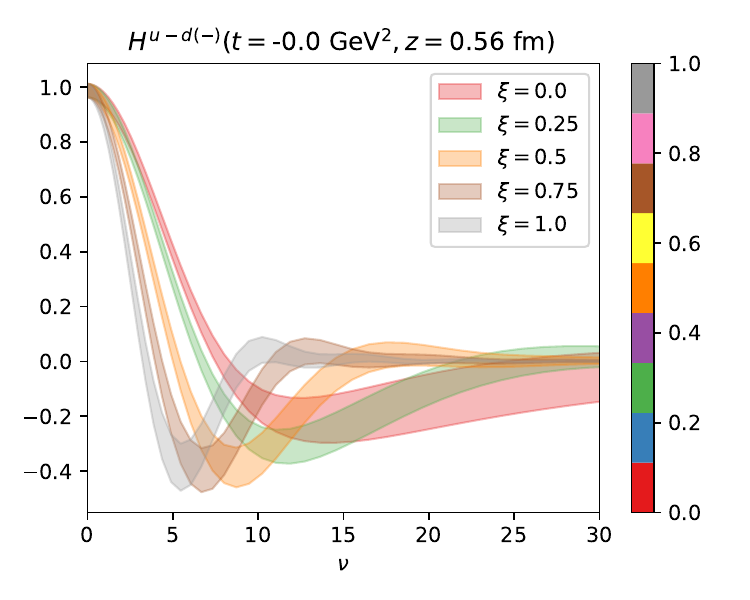}\includegraphics[width=0.42\linewidth]{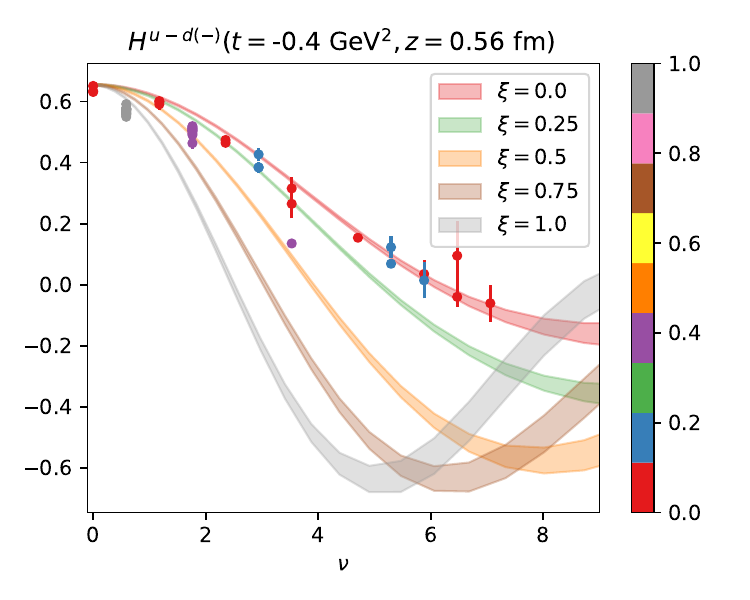}
    \includegraphics[width=0.42\linewidth]{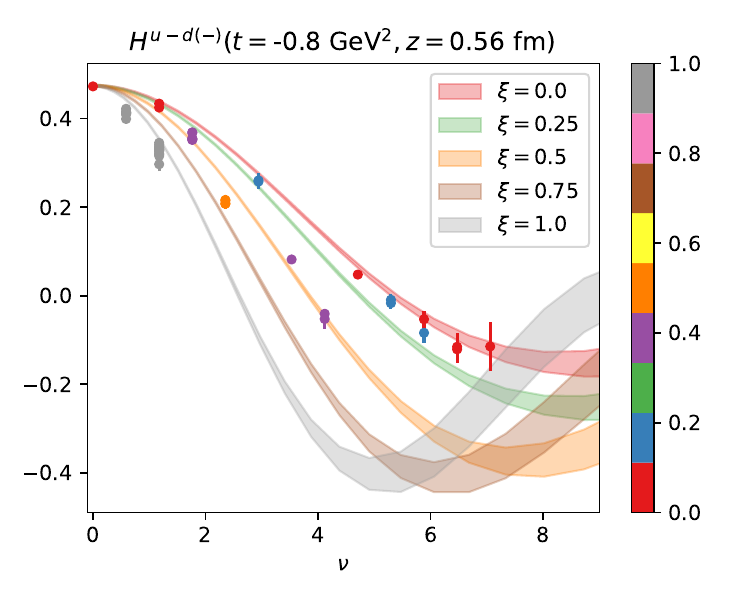}\includegraphics[width=0.42\linewidth]{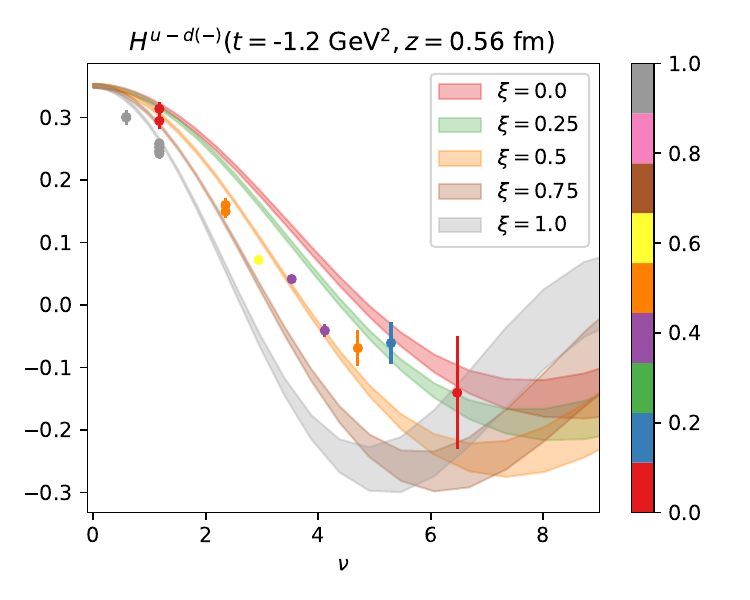}
    \caption{Baseline reconstruction of the real part of the Ioffe-time distribution corresponding to the unmatched $H^{u-d(-)}$ GPD fitted on the $z = 6a = 0.56$ fm data ($t_s=3$). The upper-left plot presents the result of the extrapolation to $t = 0$ and demonstrates the reconstruction behavior up to large values of $\nu$. The datapoints within 0.1 GeV$^2$ of the quoted value of $t$ have been depicted, with a  correction to account for the mismatch in $t$ learned from the $t$-dependence of the reconstruction.} \label{fig:GITD_res}
\end{figure*}

Presenting visually the results of the fit is a challenge, since our 249 pairs $(\vec{p}_f, \vec{p}_i)$ form a diffuse cloud of kinematics $(\nu, \bar\nu, t)$, as seen on Fig.~\ref{fig:kincov}. On Fig.~\ref{fig:GITD_res}, we show our results in Ioffe time for the real part corresponding to the $x$-even $H^{u-d(-)}$ GPD for 4 different values of $t$ and $z = 6a = 0.56$ fm, the largest value that we will allow ourselves to consider. The correlated $\chi^2$ divided by the number of fitted points is 1.7, showing a fair agreement with the full dataset.

Since our lowest values of $|t|$ in the dataset are about 0.15 GeV$^2$, the results at $t = 0$ in the upper-left plot are an extrapolation whose flexibility is notably controlled by the correlation length $l_t$. We see that at $\nu = 0$, the extrapolation reaches 1 within uncertainty, as expected from the construction of our matrix elements as a ratio. The fact that the value at $\nu = 0$ is systematically the same regardless of $\xi$ is a consequence of our exact respect of the
polynomiality property through the use of the DD formalism, and highlights that the integral over $x$ of the GPD is independent of $\xi$. The upper-left plot also allows us to appreciate the behavior of our reconstruction in the extrapolation region at larger $\nu$.

\begin{figure}[h!]
    \centering
    \includegraphics[width=0.8\linewidth]{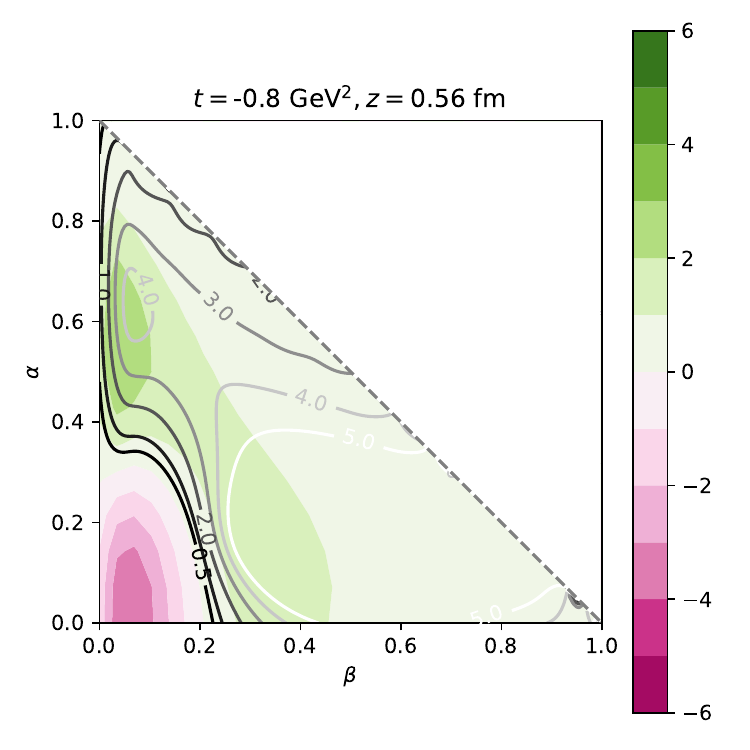}
    \caption{Baseline reconstruction of the unmatched DD corresponding to the $x$-even $H^{u-d(-)}$ GPD, fitted on the $z = 6a = 0.56$ fm data ($t_s=3$), displayed at $t = -0.8$ GeV$^2$. The color scheme depicts the central values of the DD, while the contour lines represent the number of standard deviations away from 0. The stark negative central values at small $\beta$ and $\alpha$ are therefore fully compatible with 0 within uncertainties.} \label{fig:DD_res}
\end{figure}

The three other plots correspond to values of $t$ which are at the heart of our dataset. To give a sense to the reader of the quality of our fit, we represent all data points which lie within less than 0.1 GeV$^2$ of the depicted value of $t$. The discrepancy caused by the varying value of $t$ is small. We reduce it further using the $t$-dependence of our own reconstruction at that value of $(\nu, \xi)$. The data depicted on the three plots represent just half of all the fitted dataset, the other points being too far from any of the depicted $t$ values to be shown. One can see precise values at $\xi = 0$ extending up to $\nu = 7$, and other precise data up to $\xi \approx 0.5$ and $\nu \approx 5$, which play a crucial role in the fit. On the other hand, data at $\xi = 1$ is restricted to very low values of $\nu$.

\begin{figure}
    \centering
    \includegraphics[width=0.8\linewidth]{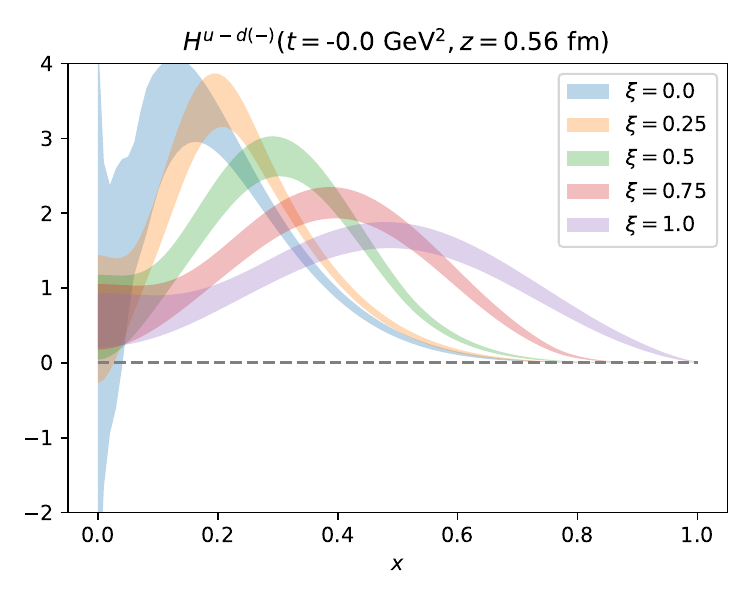}
    \includegraphics[width=0.8\linewidth]{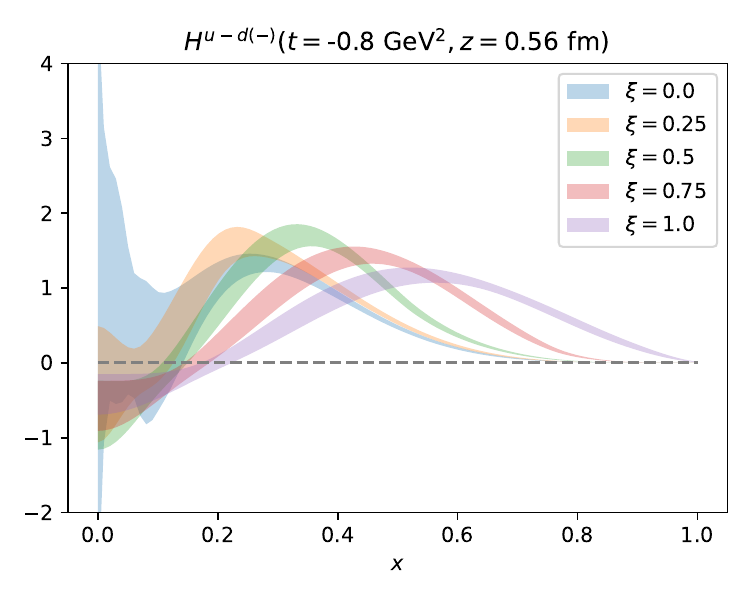}
    \caption{Baseline reconstruction of the unmatched $x$-even $H^{u-d(-)}$ GPD fitted on the $z = 6a = 0.56$ fm data ($t_s=3$).} \label{fig:GPD_res}
\end{figure}

On Fig.~\ref{fig:DD_res}, we show the corresponding DD. Its behavior at small $\alpha$ and $\beta$ is very poorly resolved. The functional freedom of GPR allows the central values and uncertainty of the DD to take complex profiles that we would have a lot of difficulty modelling with a simple parametric form. The associated GPD reconstruction, corresponding to the unmatched $x$-even $H^{u-d(-)}$ GPD is depicted on Fig.~\ref{fig:GPD_res} for various values of the skewness and $t$ parameter.

\begin{figure*}
    \centering
    \includegraphics[width=0.42\linewidth]{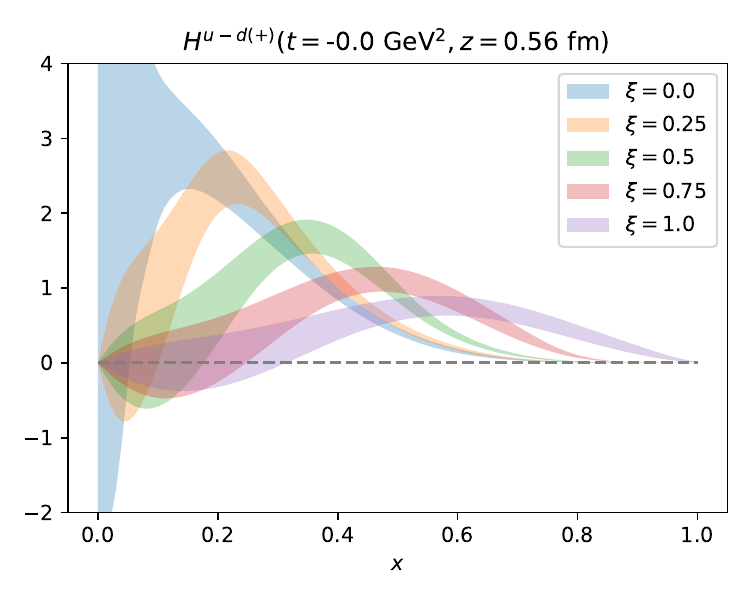}\includegraphics[width=0.42\linewidth]{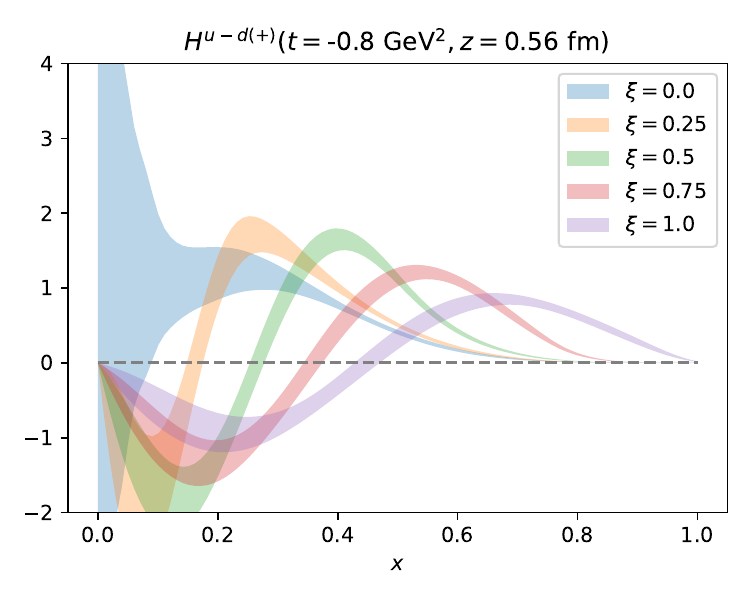}
    \includegraphics[width=0.42\linewidth]{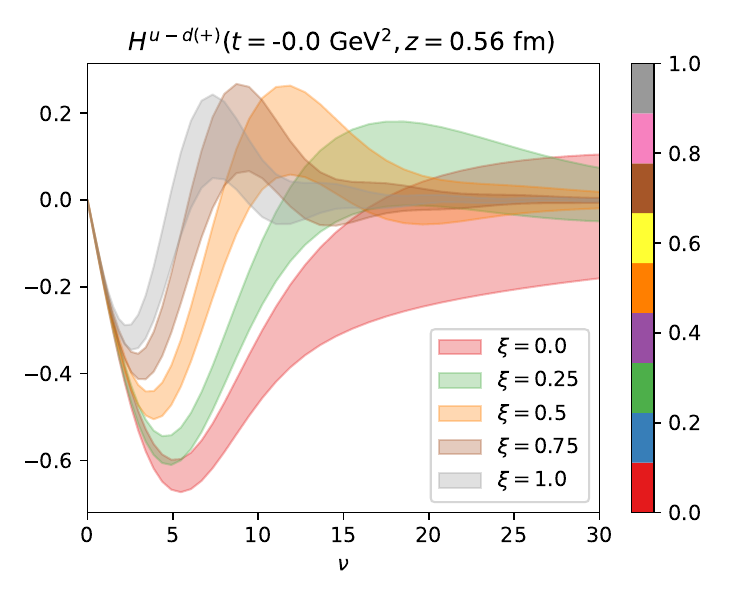}\includegraphics[width=0.42\linewidth]{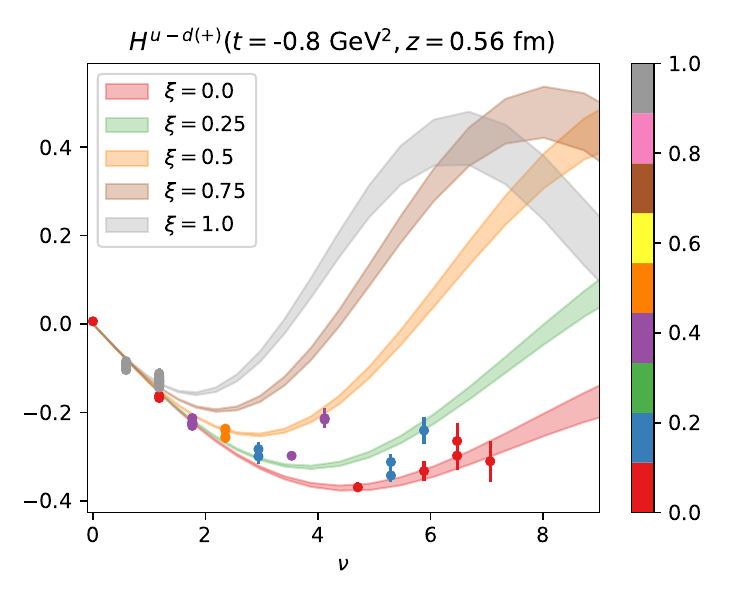}
    \caption{(top) Baseline reconstruction of the unmatched $x$-odd $H^{u-d(+)}$ GPD fitted on the $z = 6a = 0.56$ fm data ($t_s=3$) showcasing the extrapolation to $t = 0$ and the behavior at $t = -0.8$ GeV$^2$. (bottom) The imaginary part of the corresponding Ioffe-time distribution.} \label{fig:imHz6}
\end{figure*}

We depict further the results for the imaginary part corresponding to the $x$-odd $H^{u-d(+)}$ in Fig.~\ref{fig:imHz6}. The correlated $\chi^2$ divided by the number of fitted points is again 1.7. Notice that even though we show the data with $z = 0.56$ fm offering the largest extent of Fourier components in this study, the data of the imaginary part have not become compatible with zero yet. According to our extrapolation model provided by the GPR, one would need to reach approximately $\nu = 15$ for the data to be compatible with zero for $\xi = 0$. This underlines that our choice of regularization exposes us to a possible bias in the regularization of the inverse problem.

\begin{figure}
    \centering
    \includegraphics[width=0.49\linewidth]{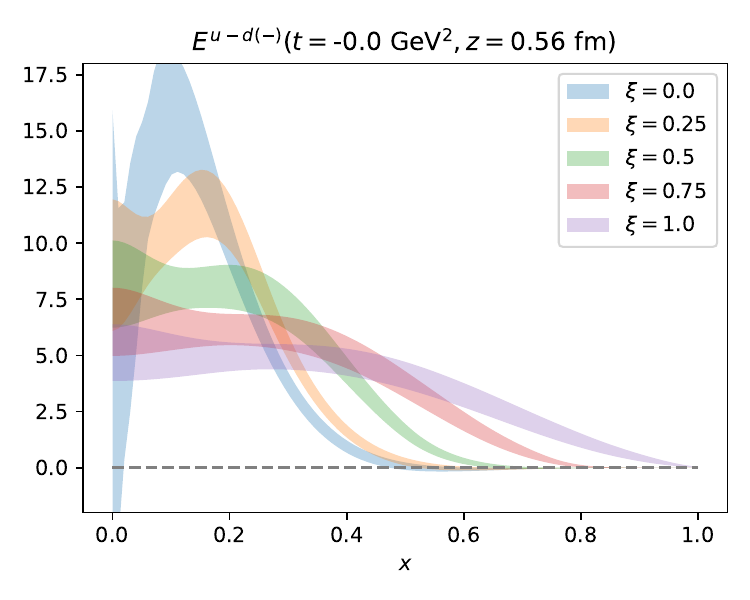}\includegraphics[width=0.49\linewidth]{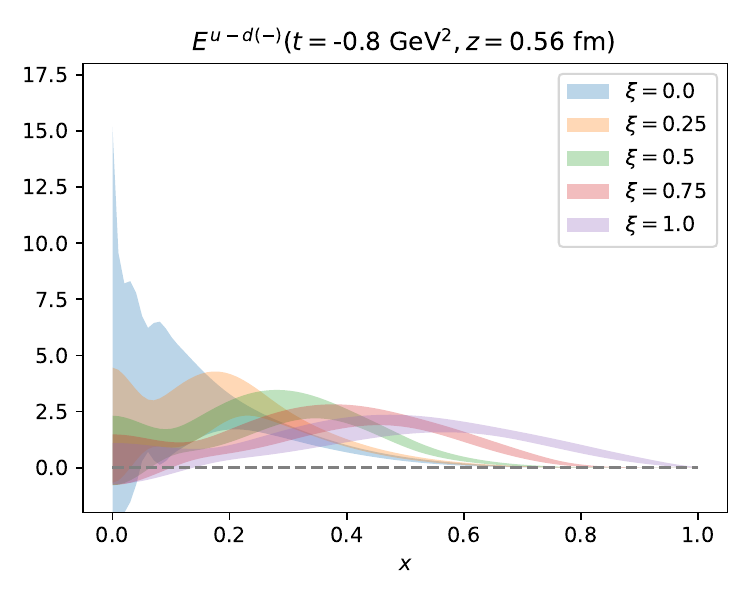}
    \includegraphics[width=0.49\linewidth]{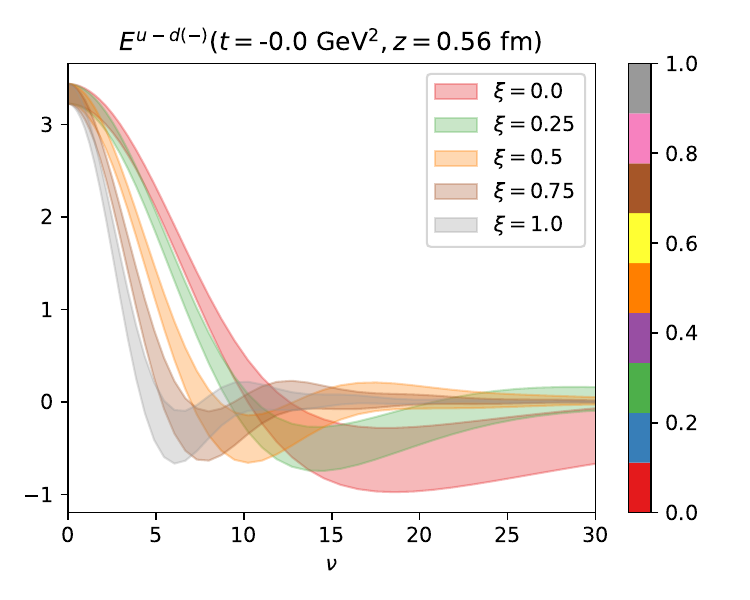}\includegraphics[width=0.49\linewidth]{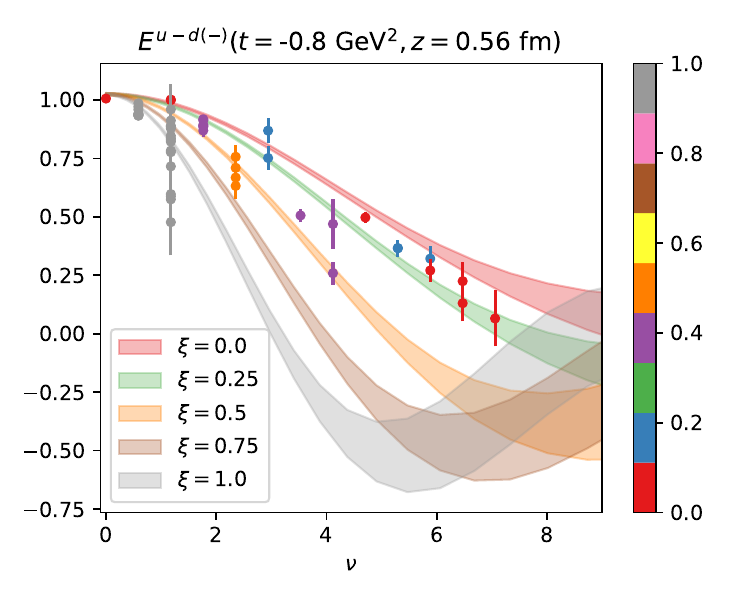}
    \caption{Same as Fig.~\ref{fig:imHz6} for the $x$-even $E^{u-d(-)}$ GPD.} \label{fig:reEz6}
\end{figure}

\begin{figure}
    \centering
    \includegraphics[width=0.49\linewidth]{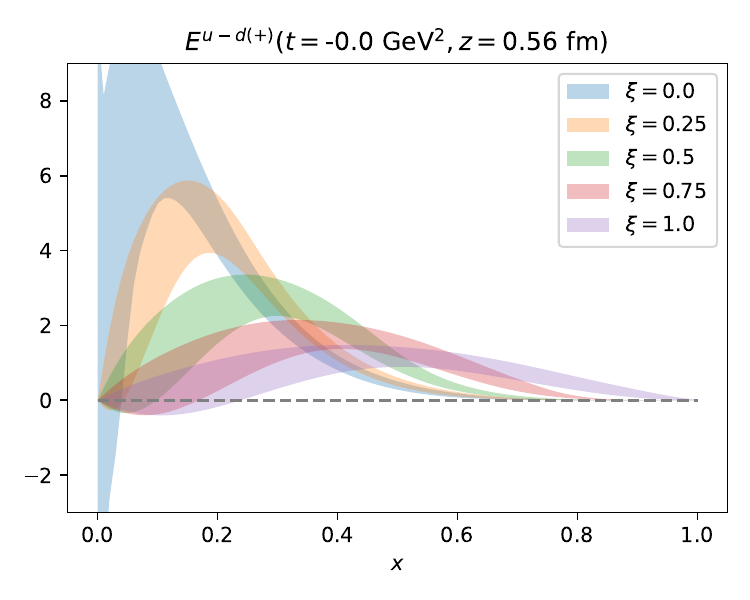}\includegraphics[width=0.49\linewidth]{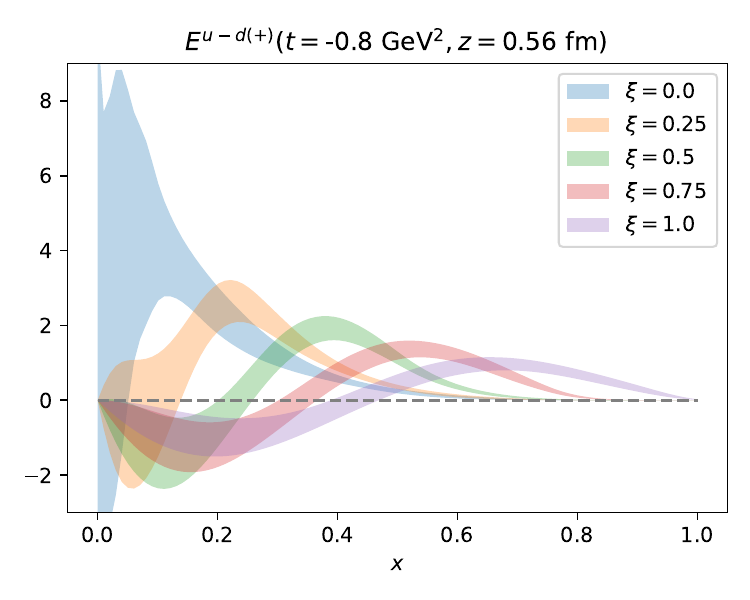}
    \includegraphics[width=0.49\linewidth]{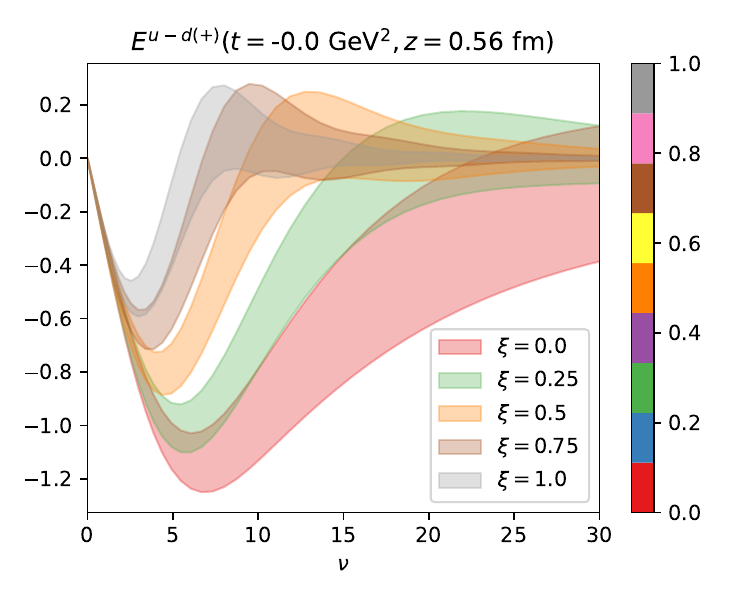}\includegraphics[width=0.49\linewidth]{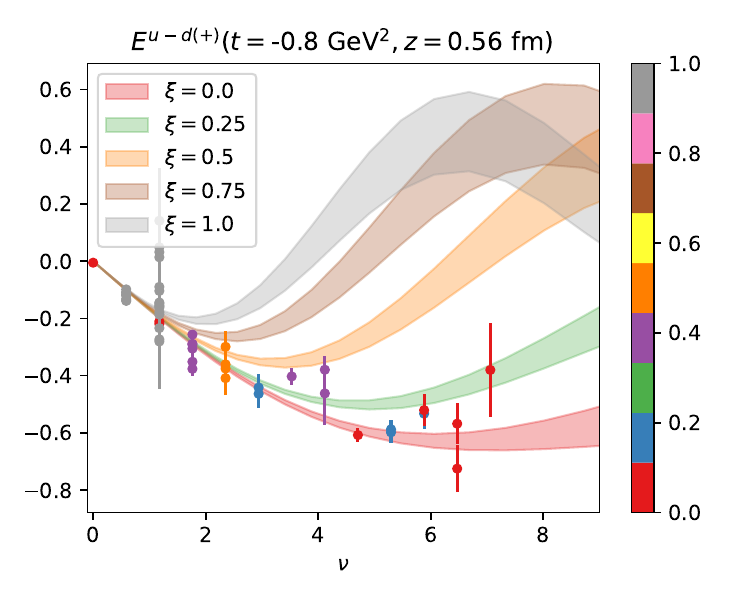}
    \caption{Same as Fig.~\ref{fig:imHz6} for the $x$-odd $E^{u-d(+)}$ GPD.} \label{fig:imEz6}
\end{figure}

The result for $E^{u-d(-)}$ is shown in Fig.~\ref{fig:reEz6}, with a correlated $\chi^2$ per fitted points of 1.8. Fig.~\ref{fig:imEz6} shows the imaginary part corresponding to the $x$-odd $E^{u-d(+)}$, with a correlated $\chi^2$ per fitted points of 1.0.

\subsection{Assessment of model dependence}

So far, all reconstructions that we have presented resulted from a fixed custom choice of hyperparameters. While we believe that it is a reasonable compromise that includes all of our prior knowledge on the physics of the system and offers a fairly conservative extrapolation uncertainty in the large Ioffe time region, different choices could be envisioned and would result in different reconstructions. This represents a model dependence that we should try to assess, within the limits of physical relevance of the reconstruction. Even then, defining a metric for model dependence relies on arbitrary decisions. 

We present a partial assessment of the model dependence by allowing the hyperparameters to be distributed in a uniform uncorrelated distribution spanning the following range: $\sigma$ between its current value and twice that value, $\gamma \in [0.1, 0.5]$, $\delta \in [1,3]$, $\eta \in [0,2]$, $l_\beta \in [0.3, 1]$, $l_\alpha \in [0.1, 0.4]$, $l_t \in [0.2, 0.5]$. $\eta = 0$ does not ensure a cancellation on the edges of the DD domain, in contradiction to what we argued in Section \ref{sec:DDrep}. However, since our main argument for the cancellation was a derivation in perturbation theory which used a number of approximations, we find it useful to allow flexibility on that criterion as part of the model dependence assessment. As we have already stressed, excessive freedom in the prior at small $x$ results in unphysical extrapolations for $\xi = 0$ at large Ioffe time. To exclude such sets of hyperparameters from our exploration, we discard those where the uncertainty at $\nu = 30$, $\xi = 0$ and $t = -0.8$ GeV$^2$ is 50\% larger than the one achieved with our fixed choice of hyperparameters. For all GPDs and values of $z$, this corresponds to discarding consistently between 50 and 55\% of the nominal volume of hyperparameters.

\begin{figure*}
    \centering
\includegraphics[width=0.42\linewidth]{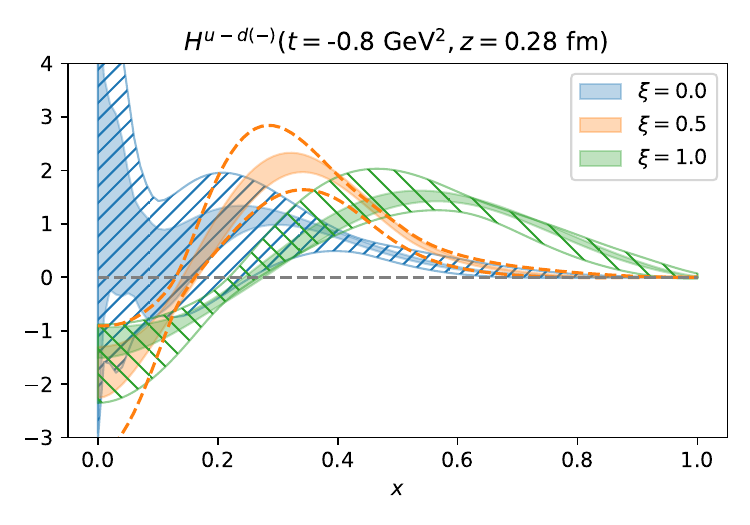}\includegraphics[width=0.42\linewidth]{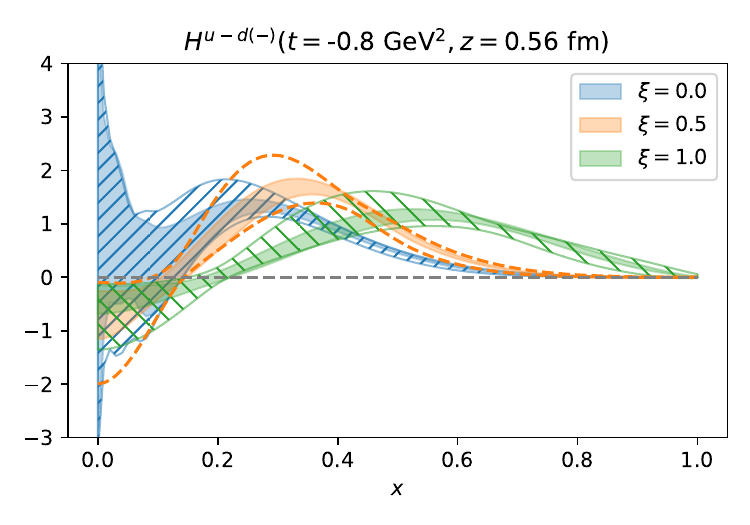}

\includegraphics[width=0.42\linewidth]{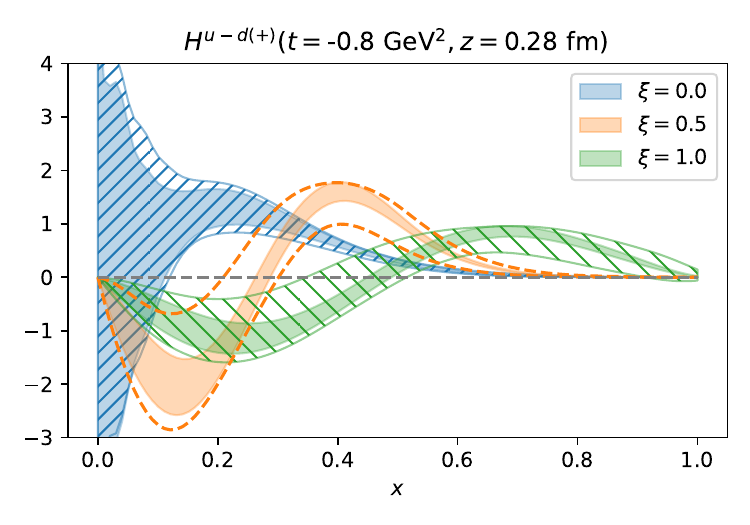}\includegraphics[width=0.42\linewidth]{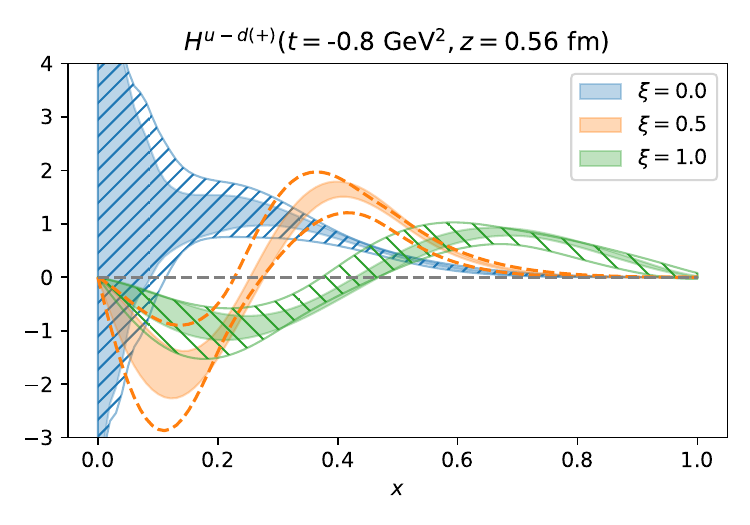}

\includegraphics[width=0.42\linewidth]{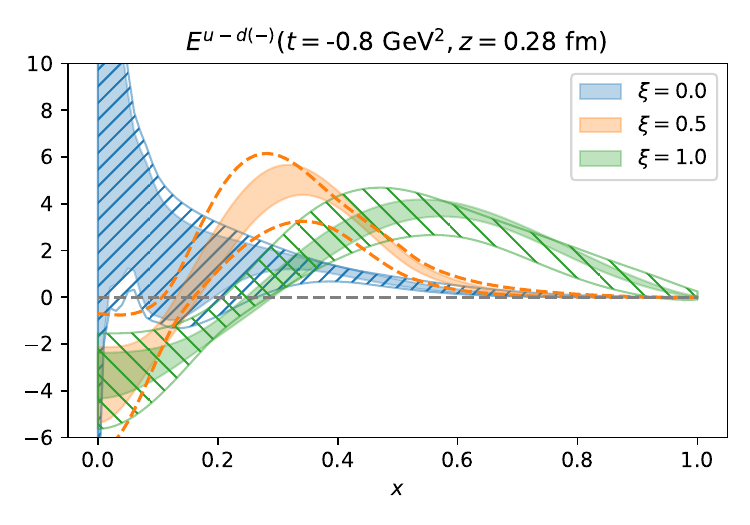}\includegraphics[width=0.42\linewidth]{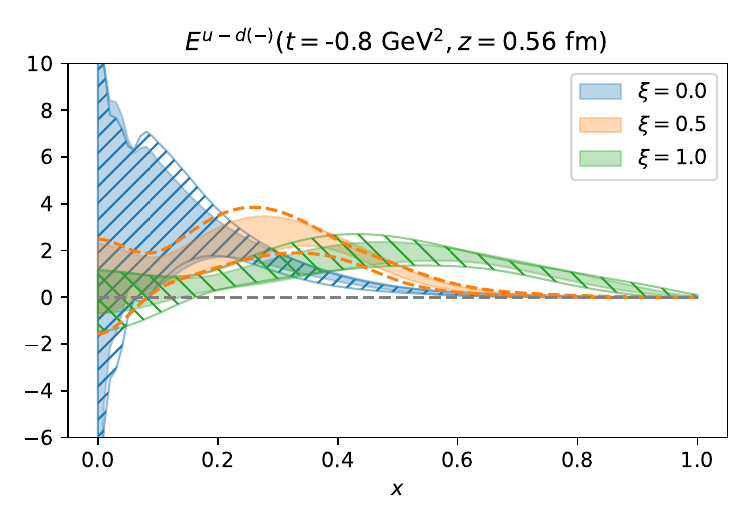}

\includegraphics[width=0.42\linewidth]{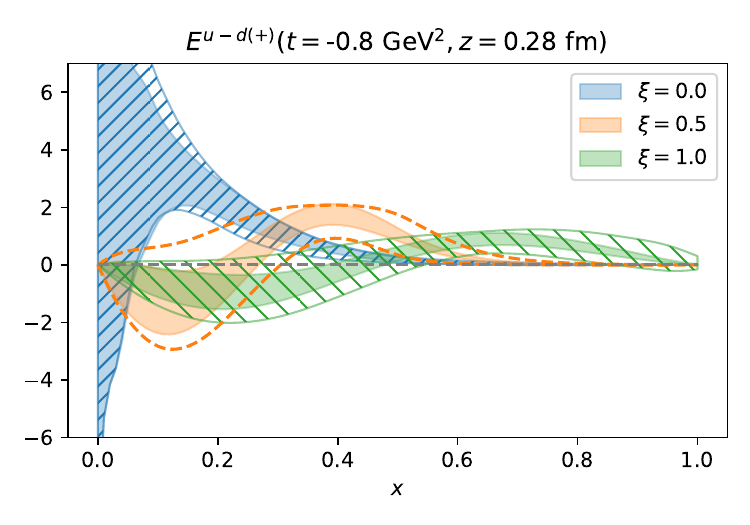}\includegraphics[width=0.42\linewidth]{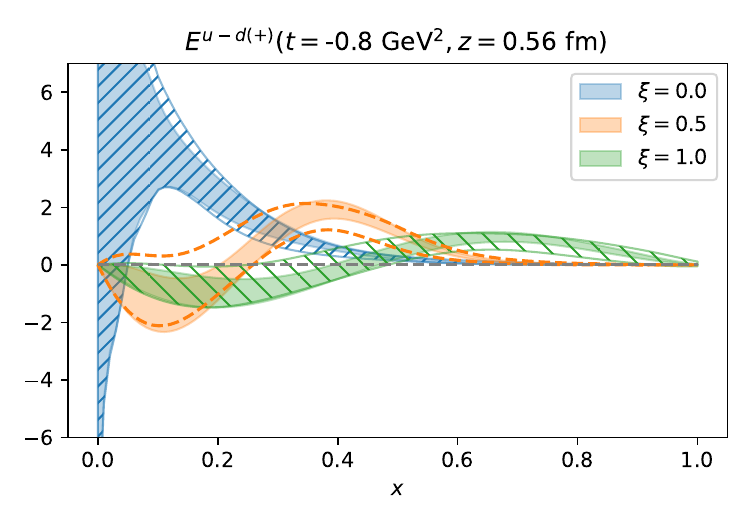}

\caption{Effect of varying hyperparameters to assess the model dependence. The solid bands represent the baseline reconstruction with fixed hyperparameters, the hatched or dotted bands represent the result with varying hyperparameters. Results are shown for $z = 3a = 0.28$ fm (left column) and $z = 6a = 0.56$ fm (right column) for $t_s = 3$ and the unmatched $H^{u-d(-)}$ (top row), $H^{u-d(+)}$ (2nd row), $E^{u-d(-)}$ (3rd row) and $E^{u-d(+)}$ (last row) at $t = -0.8$ GeV$^2$ and various values of $\xi$.} \label{fig:HYPvar}
\end{figure*}

Practically, we take 100 random samples of hyperparameters that fulfill our criterion on the uncertainty at large Ioffe time in the previously mentioned distribution. Then for each set of hyperparameters we pick 10 random samples of the posterior distribution. This gives us a collection of 1000 replicas which we compare to the posterior with fixed hyperparameters in Fig.~\ref{fig:HYPvar}. We also varied the prior mean using for half of the replicas an RDDA model of the form of Eq.~\eqref{eq:RDDA} with $N = 1$, $q(\beta, t) = C \beta^{-0.2}(1-\beta)^2 (1 - t)^{-2} / {\cal N}$, where ${\cal N} = 0.39$ normalizes the PDF contribution to unity, $C = 1$ for $H^{u-d}$, $C=3$ for $E^{u-d(-)}$ and $C=2$ for $E^{u-d(+)}$. We found that this choice produced no appreciable difference in the posterior. Our exploration of model dependence remains only partial as we have not deviated from the general parametric form of the prior covariance, but it helps pin-point where our baseline uncertainty assessment may be too optimistic.

From Fig.~\ref{fig:HYPvar}, it is clear that our baseline reconstruction is very optimistic in its uncertainty assessment as $z = 0.28$~fm, especially for the real parts and the imaginary parts at non-zero skewness. There is less difference with the variation of hyperparameters when $z = 0.56$~fm. As expected, the model dependence -- linked to the ill-definedness of the inverse problem -- decreases when the range in Ioffe time increases at larger $z$. 

The imaginary parts at zero skewness demonstrate smaller changes under hyperparameter variation and a rather similar final uncertainty assessment for $z = 0.28$ or 0.56 fm. The latter observation could result from the fact that the data is less informative on the imaginary part, and the reconstruction more prior-driven. Finally, we note that allowing $\eta$ close to 0 in the space of hyperparameters generates some non-vanishing contribution close to $(x, \xi) = (1, 1)$.

\subsection{Perturbative matching\label{pertmatch}}

We perform the perturbative matching in Ioffe time, first by applying the kernel of Eq.~\eqref{eq:gitd-pgitd-matching} to match to a scale $\mu_0 \sim 1/\sqrt{-z^2}$, and then by solving the integro-differential equation of Eq.~\eqref{eq:evolGPD} to evolve the result to a final scale of 2 GeV. To capture reliable features of the reconstruction over the whole range of $x$ of interest, we work with 200 points in Ioffe time spanning quadratically the range $\nu \in [0, 600]$. This means that we have a denser sampling of the function at small $\nu$, where it varies faster. The differential equation is solved by the Runge-Kutta method RK4. Notice that it is possible to apply matching and evolution to a function which does not vanish at $x = 1$, as it clearly represents no issue in Ioffe time space. The result has still a support in $[0,1]$ and is generally well-behaved.

To give a sense of the perturbative uncertainty associated to this matching, we use the heuristic of varying the initial scale $\mu_0$ in the interval $[1/\sqrt{-2z^2}, \sqrt{2/-z^2}]$. We fix the running of $\alpha_s$ by $\alpha_s(2 \textrm{ GeV}) = 0.28$, corresponding to $\Lambda_{QCD} = 165$ MeV.
\begin{figure}
    \centering
\includegraphics[width=0.8\linewidth]{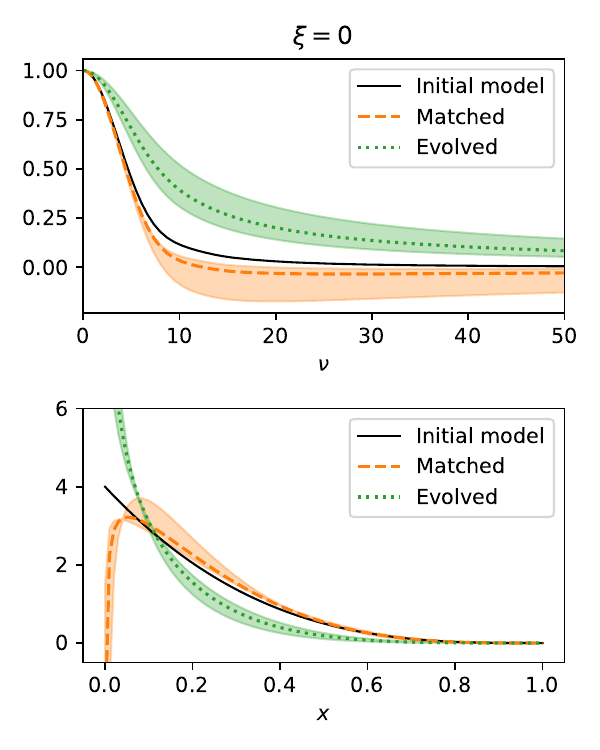}

\includegraphics[width=0.8\linewidth]{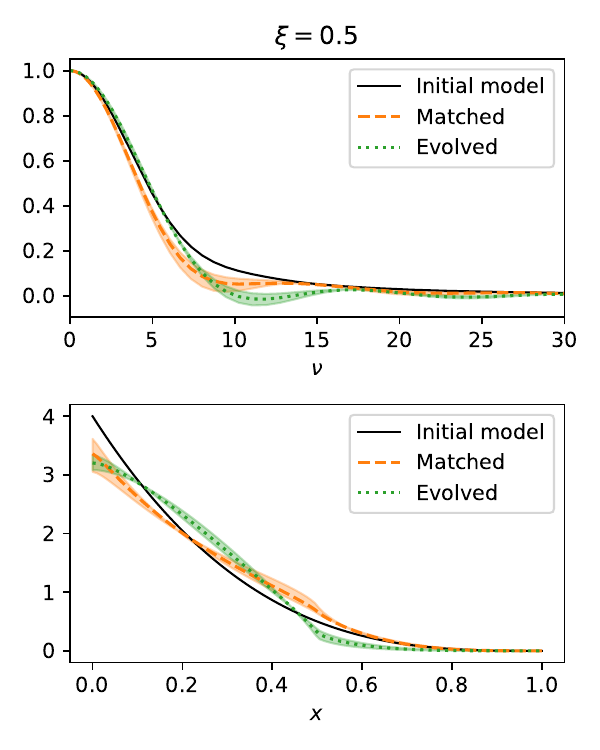}

\caption{Matching vs pure evolution of a toy model from $z = 0.56$ fm to 2 GeV, for $\xi = 0$ (top) and $\xi = 0.5$ (bottom). $z = 0.56$ fm corresponds to an initial scale of 0.35 GeV if $\mu_0 = 1/\sqrt{-z^2}$ (dotted and dashed lines), which we vary between 0.25 and 0.49 GeV to give an account of scale fixing uncertainty (error bands).} \label{fig:evol_effect}
\end{figure}

On Fig. \ref{fig:evol_effect}, we depict the effect of both the full matching, and pure evolution without matching on a simple GPD toy model $H(x, \xi) = 4 (1-x)^3$. The pure evolution part is benchmarked against the APFEL++ evolution code \cite{Bertone:2013vaa,Bertone:2017gds,Bertone:2022frx} and provides excellent numerical agreement. The absolute difference between our result and APFEL++ in $x$-space is typically of the order of $10^{-3}$, mostly because of uncertainty introduced by the inverse Fourier transform from Ioffe time to $x$-space, which is satisfactory considering the statistical and systematic uncertainty of our reconstruction.

\begin{figure*}
    \centering
\includegraphics[width=0.42\linewidth]{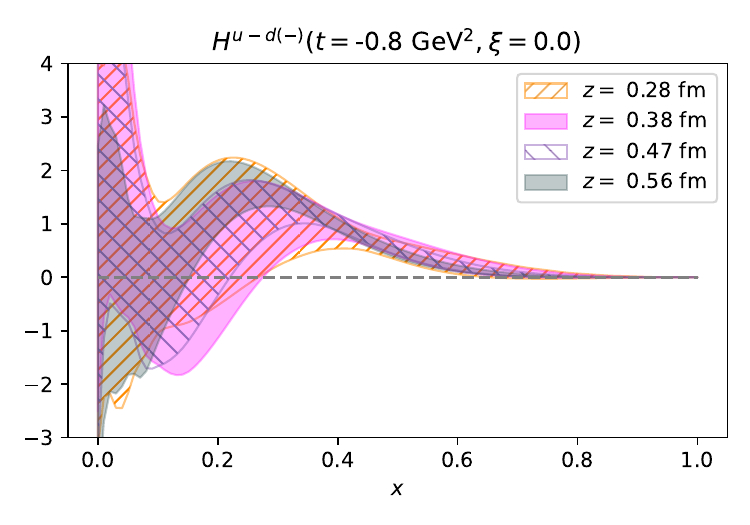}\includegraphics[width=0.42\linewidth]{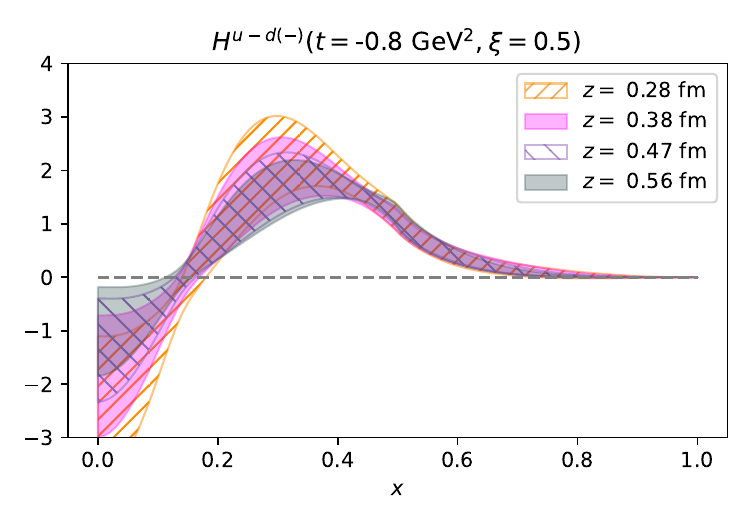}

\includegraphics[width=0.42\linewidth]{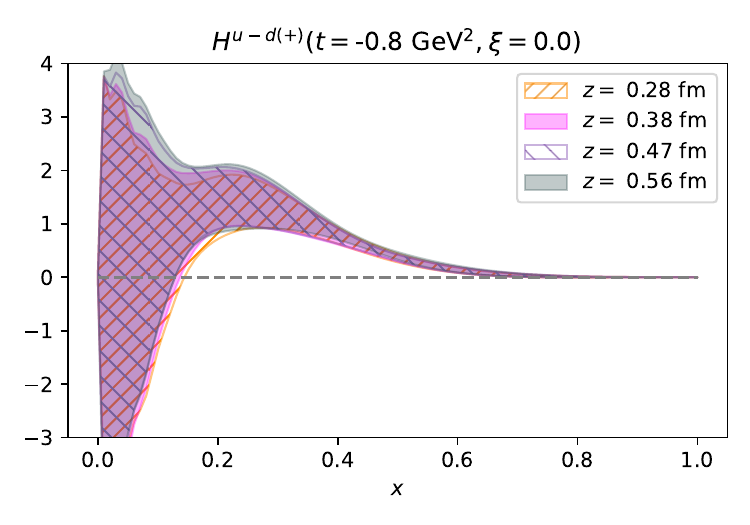}\includegraphics[width=0.42\linewidth]{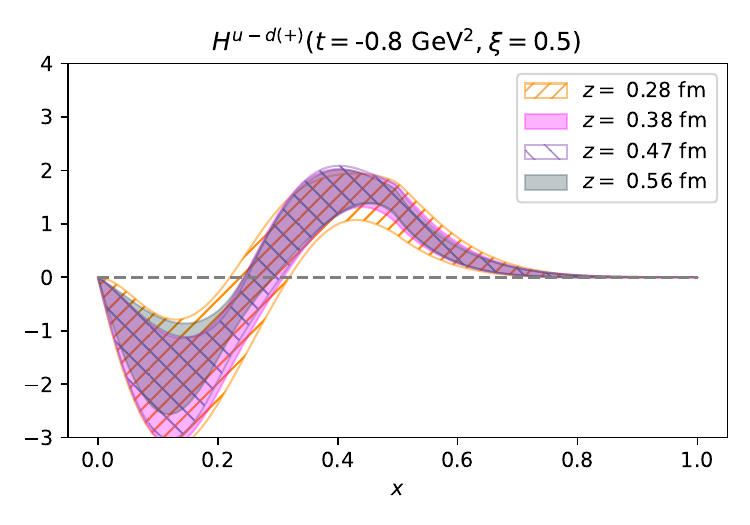}

\includegraphics[width=0.42\linewidth]{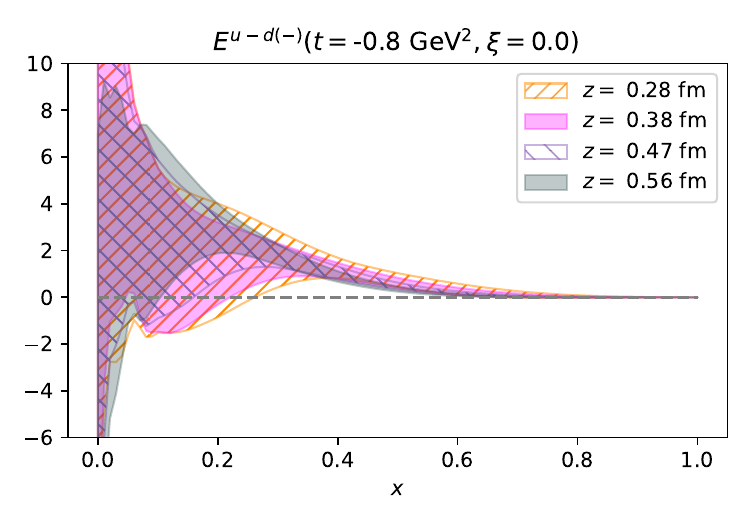}\includegraphics[width=0.42\linewidth]{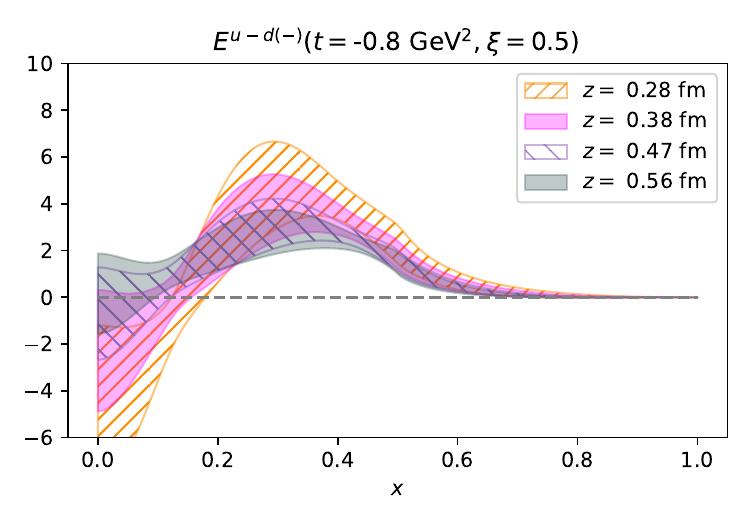}

\includegraphics[width=0.42\linewidth]{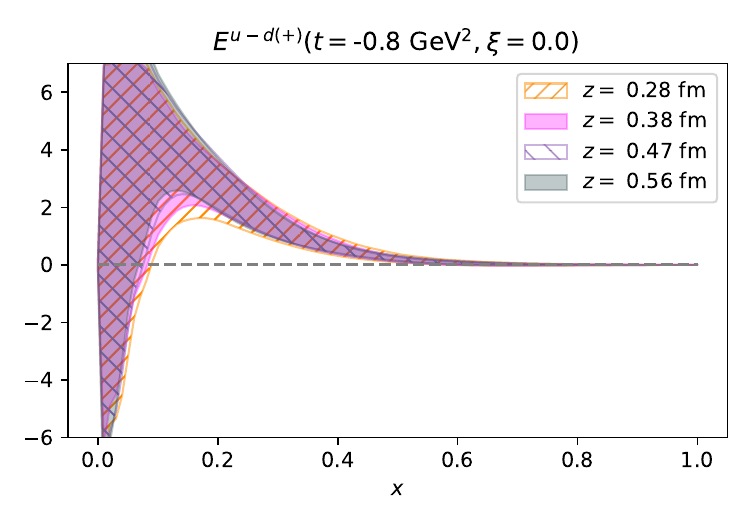}\includegraphics[width=0.42\linewidth]{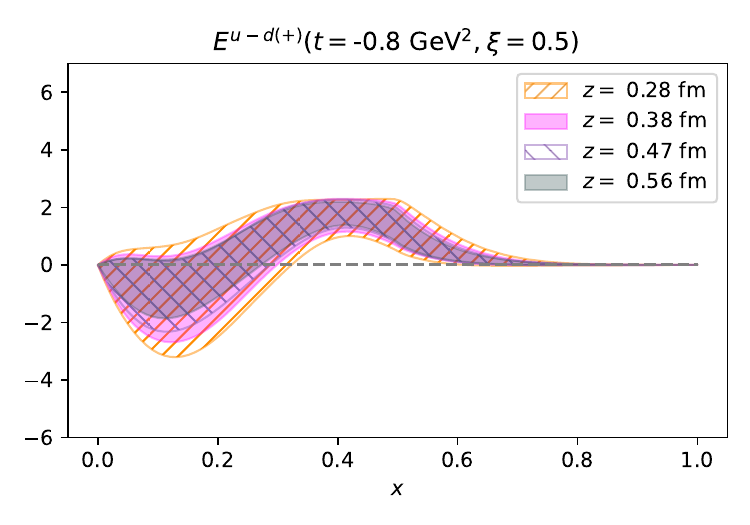}

\caption{Matched GPDs at $\mu = 2$ GeV, $\xi = 0$ (left) and $\xi = 0.5$ (right) for $t = -0.8$ GeV$^2$ depending on the value of $z$ ($t_s = 3)$, for $H^{u-d(-)}$ (top), $H^{u-d(+)}$ (second row), $E^{u-d(-)}$ (third row) and $E^{u-d(+)}$ (bottom).} \label{fig:matchedz}
\end{figure*}

For $\xi = 0$, we observe the expected effect of pure evolution to deplete the large $x$ region and generate a divergence at small $x$. For $\xi = 0.5$, evolution still depletes the large $x$ region, but changes behavior at the $x = \xi$ point. When considering the full matching however, much of the effect of evolution is cancelled out. This feature has been consistently observed in the pseudo-distribution formalism. As a result, the matched distribution is quite close to the initial model. In the end, the uncertainty associated with scale variation is essentially negligible compared to the statistical and model dependence uncertainty, even for the largest value $z = 0.56$ fm. We had made a similar observation at the level of moments in \cite{HadStruc:2024rix}. Of course, the heuristic of scale variation is only a very partial assessment of the real perturbative uncertainty and performing matching at higher order would be a useful task, left for a further study.

The result of the matching is shown on Fig.~\ref{fig:matchedz}, applied to the reconstruction with hyperparameter variation. It demonstrates globally a good agreement between all $z$ values, with more precise results at larger $z$ especially for the $x$-even GPDs and non-zero skewness for $x$-odd GPDs. We have already observed that at zero skewness, the $x$-odd GPD reconstructions were very similar regardless of the value of $z$. As is usually the case in practical computations of pseudo-distributions using the ratio method of Eq.~\eqref{eq:reducedmatelem}, there is no strong evidence of deviations linked to the use of fairly large values of $z$ where one would expect perturbation theory to become unreliable and higher-twist contaminations to grow. Placing a clear bound on the size of power corrections is, however, an important and difficult task for computations of parton distributions on the lattice and our observations are constrained by the context of this study on a single lattice spacing.

\subsection{Final results}

For the final results, we choose the reconstruction fitted on the data $z = 5a = 0.47$ fm, with a cut in Euclidean time of the three-point functions $t_s = 3$ and hyperparameter variation for uncertainty assessment. We add in quadrature the difference between the means of that reconstruction and the one with $z = 6a$ ($t_s = 3$) as well as $z = 5a$ ($t_s = 4$) to include more systematic effects. Our final uncertainty assessment reflects therefore the statistical uncertainty of the data, a systematic uncertainty linked to the identification of the ground-state matrix elements and efforts to address the inverse problem and offer a reduced model-dependence. We found scale variation to be negligible compared to the previous uncertainties, while we acknowledge that it is a very partial view on the question of perturbative uncertainty. We are not able to quantify lattice artefacts such as discretization errors or finite volume effects, nor can we address the challenging question of higher-twist contamination.

\begin{figure*}
    \centering
\includegraphics[width=0.42\linewidth]{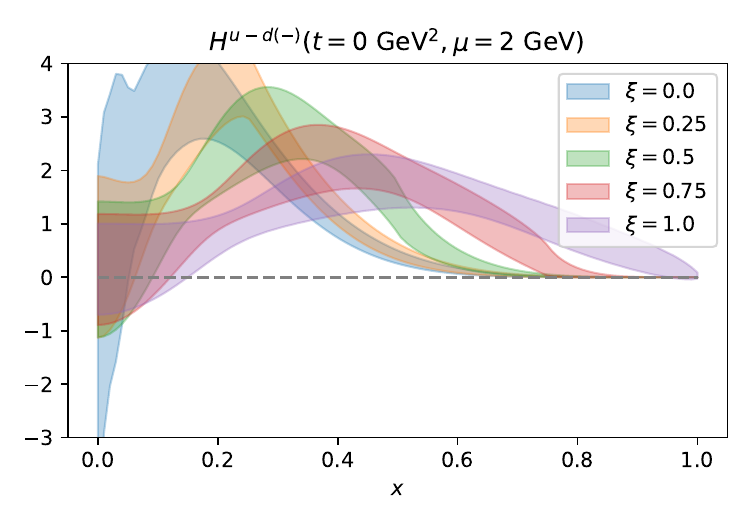}\includegraphics[width=0.42\linewidth]{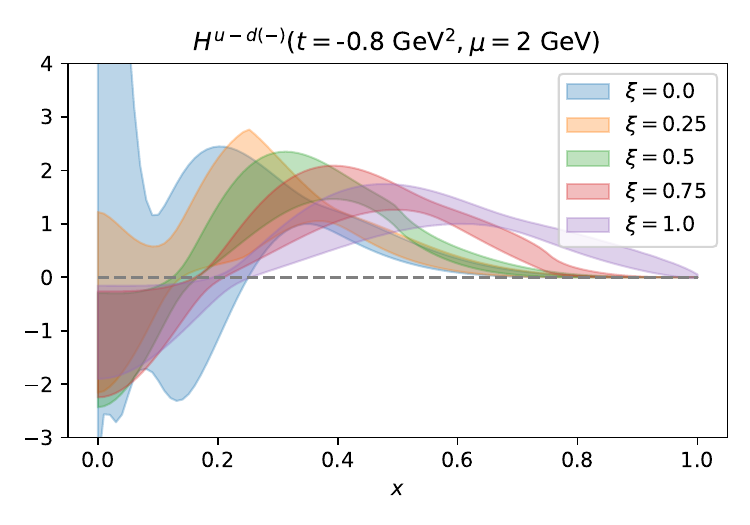}

\includegraphics[width=0.42\linewidth]{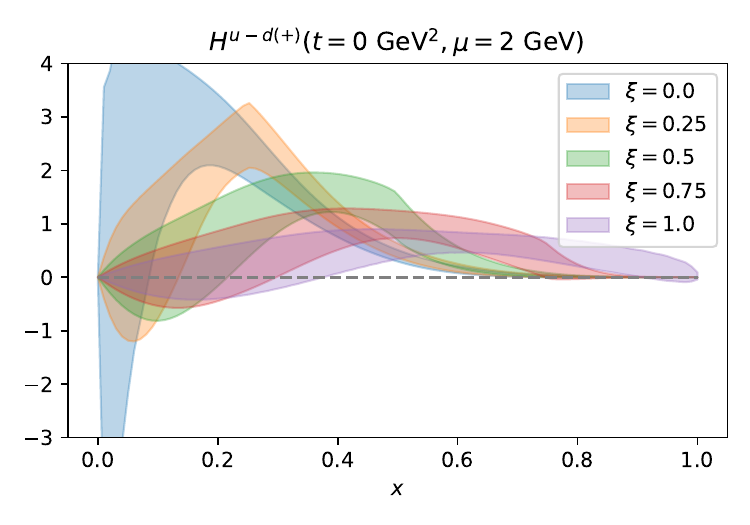}\includegraphics[width=0.42\linewidth]{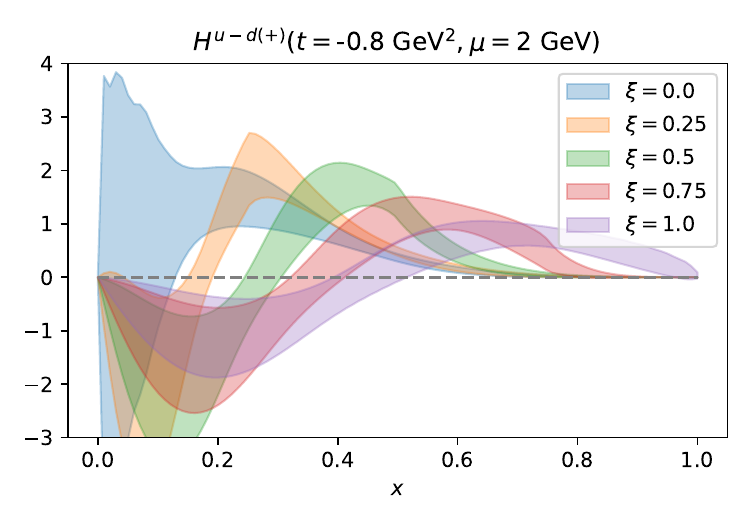}

\includegraphics[width=0.42\linewidth]{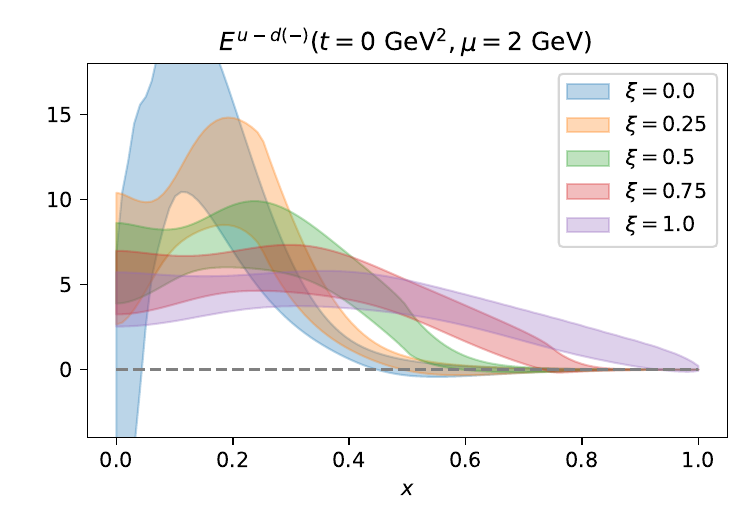}\includegraphics[width=0.42\linewidth]{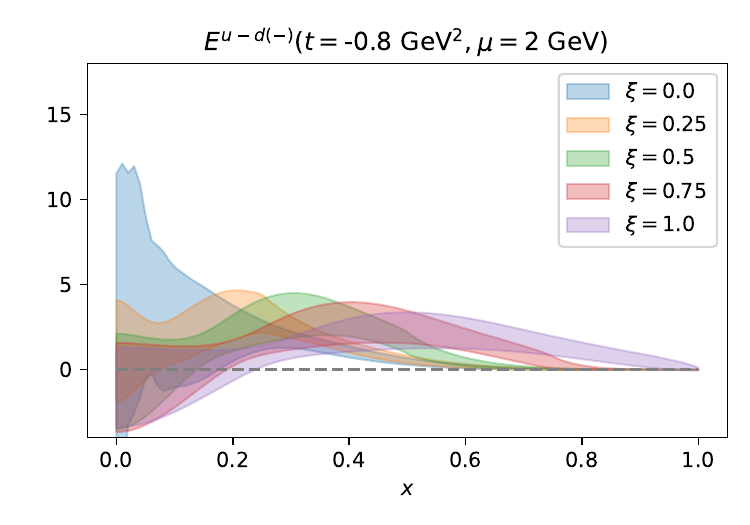}

\includegraphics[width=0.42\linewidth]{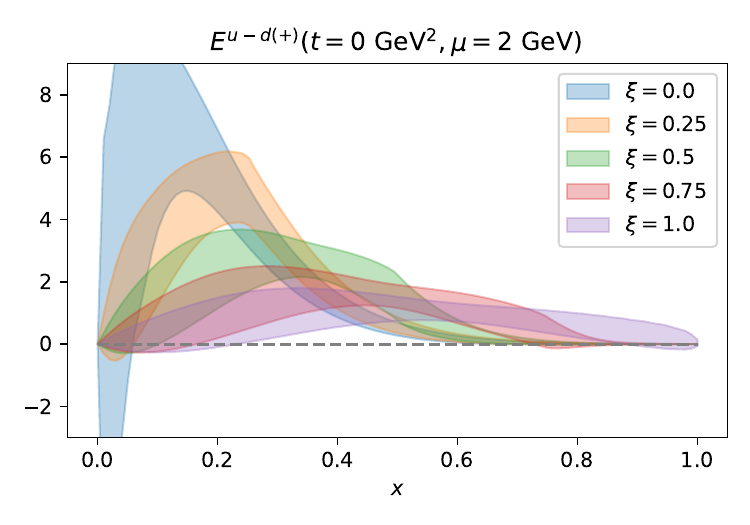}\includegraphics[width=0.42\linewidth]{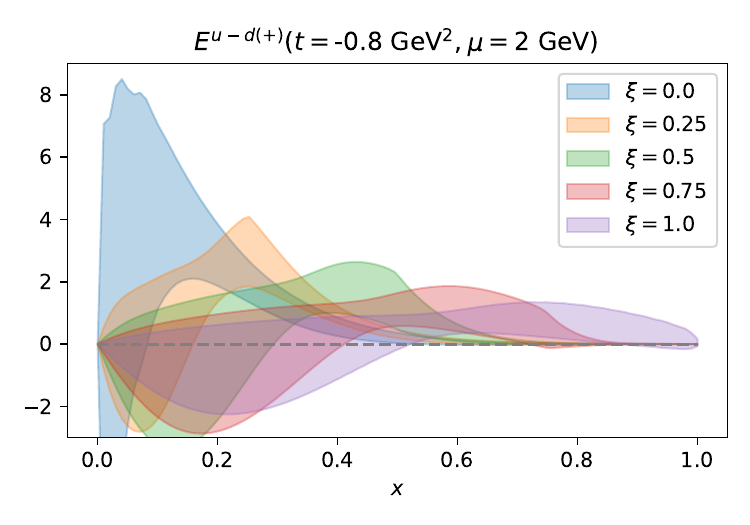}

\caption{Final extraction of isovector unpolarized GPDs at $\mu = 2$ GeV.} \label{fig:final}
\end{figure*}

Our final results are depicted on Fig. \ref{fig:final} for a range of values of $(\xi, t)$. At $(x,\xi) = (1,1)$, possible non-zero values result from our loose enforcement of the cancellation of the DD on its edges. These results are available publicly in the online Zenodo database \cite{dutrieux_2026_19695472}. A notice on how to read those datafiles is provided in Appendix~\ref{app:notice}.

Concerning the isovector $D$-term contained in the $\mathcal{A}_5$ Lorentz invariant, there is virtually no signal in the data to improve on the extraction of \cite{HadStruc:2024rix}. We do not attempt a functional reconstruction on data that lack any statistically resolved features and do not extend further than $\nu = 2.5$.

\subsection{Phenomenological observations}
\label{sec:pheno}

\begin{figure*}
    \centering
\includegraphics[width=0.4\linewidth]{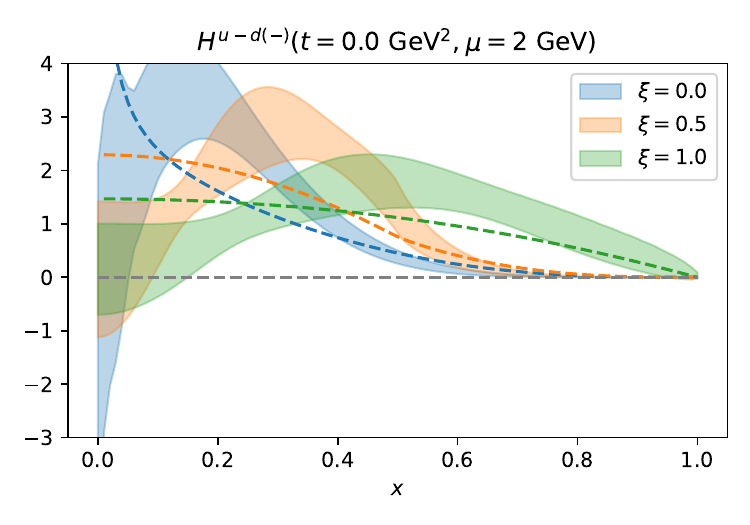}\includegraphics[width=0.4\linewidth]{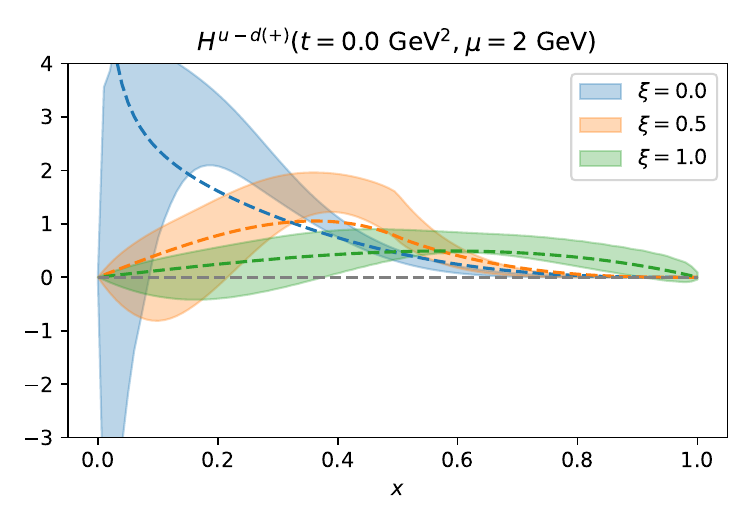}

\includegraphics[width=0.4\linewidth]{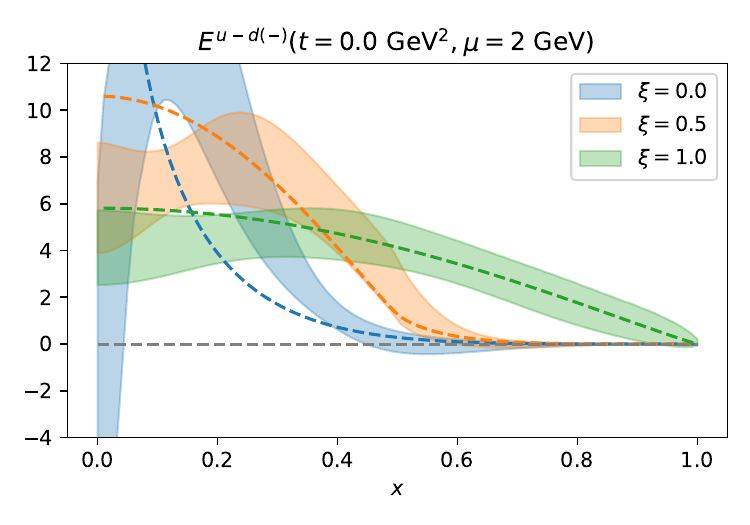}\includegraphics[width=0.4\linewidth]{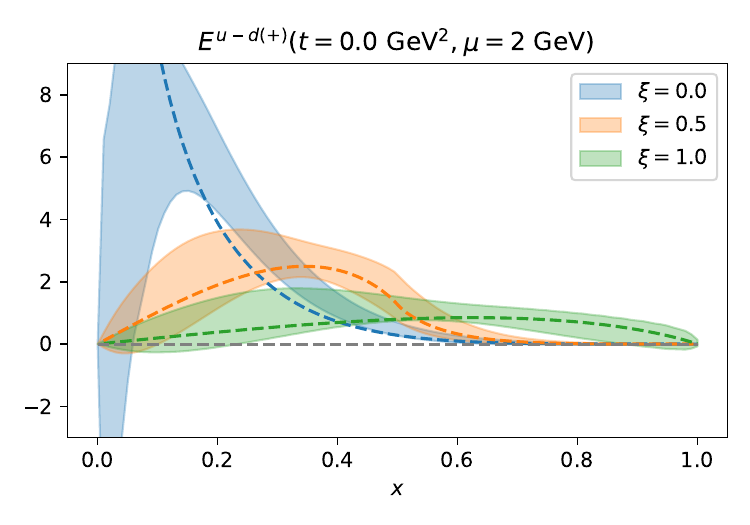}

\caption{Comparison of our extraction (colored bands) with the Goloskokov-Kroll (GK) model (dashed lines) at $t = 0$. At larger $t$, the phenomenological model decreases significantly faster than our extraction, making comparisons less informative.} \label{fig:compGK}
\end{figure*}

We compare our extraction to the popular Goloskokov-Kroll (GK) model \cite{Goloskokov:2005sd,Goloskokov:2007nt,Goloskokov:2009ia} that we evaluate using the PARTONS software \cite{Berthou:2015oaw} \footnote{See the website \url{https://3d-partons.github.io/partons/}. We used the PARTONS::GPDGK16 class to evaluate the GK model.}. The GK model is fundamentally based on the RDDA of Eq.~\eqref{eq:RDDA}. Considering that we work with a larger-than-physical pion mass of about 360 MeV, the $t$-dependence of our GPDs is milder than the one fitted on experimental data. Therefore, phenomenological GPDs decrease in magnitude more quickly than ours at larger $t$. We limit ourselves to a comparison at $t = 0$ on Fig.~\ref{fig:compGK}. The agreement is generally fair in the range where we expect our extraction to be most reliable at intermediate values of $x$, but shows significant differences for our most precise extraction of $H^{u-d(-)}$.

As an example of well-known effect of the pion mass discrepancy, the average quark momentum fraction $\langle x \rangle_{u-d}$ increases with the pion mass (see for instance Fig. 17 of \cite{LHPC:2007blg}). While the NNPDF40NLO global fit \cite{NNPDF:2021njg} gives $\langle x \rangle_{u-d} = 0.150 (5)$, the GK model stands at $0.17$, and our own extraction at non-physical pion mass at $0.21(1)$ (see next section). This systematic effect translates in our reconstruction being systematically larger than the GK model for $H^{u-d(+)}$ (upper-right plot of Fig.~\ref{fig:compGK}).

An interesting aspect of GPD phenomenology is \textit{positivity bounds} \cite{Pire:1998nw, Radyushkin:1998es}. Using the overlap representation for GPDs in terms of light-front wave functions establishes potential bounds \cite{Diehl:2000xz,Pobylitsa:2001nt,Pobylitsa:2002gw} on the GPD for $|x| > |\xi|$ as a function of the PDF. It should be noted that renormalization includes subtractions that can violate positivity bounds at low scale \cite{Altarelli:1998gn,Collins:2021vke}, so it is interesting to see whether our extraction satisfies this property. Positivity bounds are in fact a family of inequalities, involving kinematic combinations of several GPDs and with various constraining power. 

\begin{figure}
    \centering
    \includegraphics[width=0.85\linewidth]{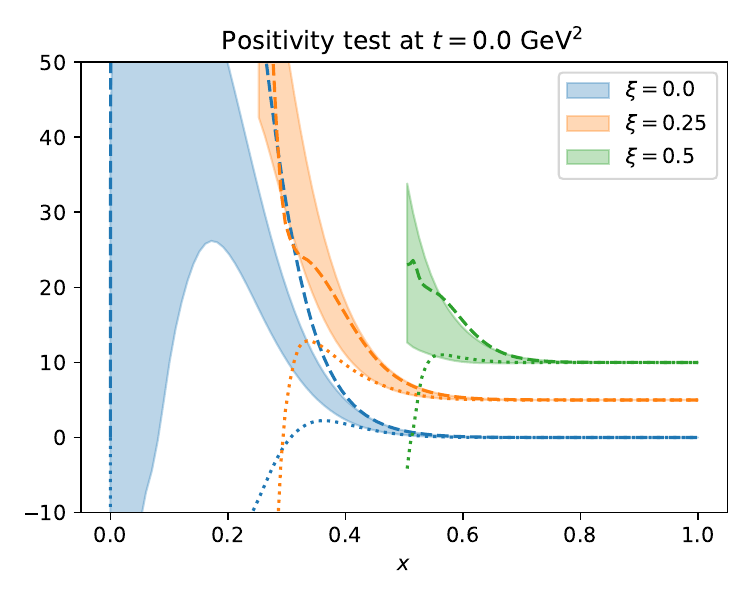}
    \includegraphics[width=0.85\linewidth]{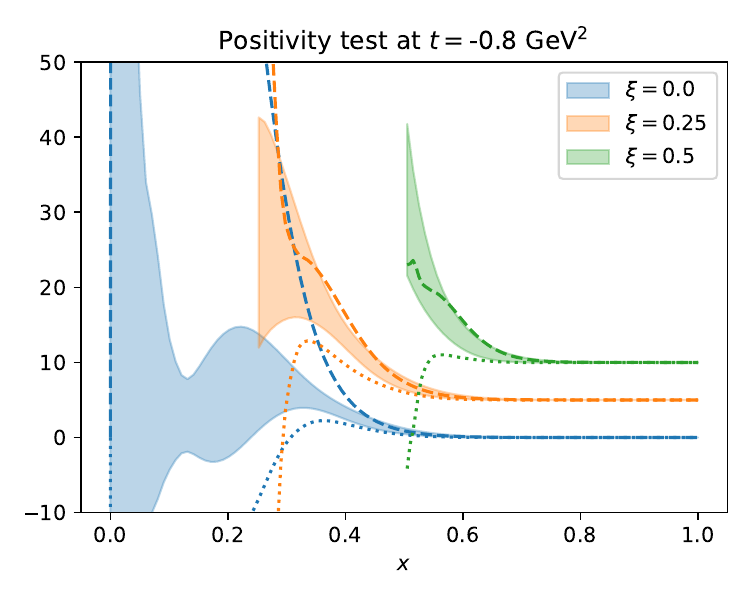}
    \caption{The colored bands represent the left-hand side of Eq.~\eqref{eq:pos} at $t = 0$ (top) and $-0.8$ GeV$^2$ (bottom). The dashed and dotted lines represent an estimate of the upper and lower uncertainty of the right-hand side. Because of our choice of adding a systematic uncertainty in quadrature using different $t_s$ and $z$ values, the plotted standard deviation can exceed the value of the mean of the distribution, giving rise to negative values when the left-hand side of Eq.~\eqref{eq:pos} is clearly a positive quantity. For $\xi = 0.25$, all values have been multiplied by a factor 2 and offset by 5 for legibility of the plot. For $\xi = 0.5$, the factor was 10 and the offset 10.}
    \label{fig:positivity}
\end{figure}

\begin{widetext}
We focus on the following simple relation that is applicable at any value of $t$:
\begin{equation}
    \bigg| H^q(x, \xi, t) - \frac{\xi^2}{1-\xi^2} E^q(x, \xi, t) \bigg|^2 \leq q\left(\frac{x+\xi}{1+\xi}\right)q\left(\frac{x-\xi}{1-\xi}\right)\frac{1}{1-\xi^2}\,,\label{eq:pos}
\end{equation}
\end{widetext}
where $q$ designates the usual PDF $q(x) = H(x, 0, 0)$. The comparison of both sides of this inequality is given on Fig.~\ref{fig:positivity}. 

It appears that positivity is generally satisfied, and that the right-hand side of Eq.~\eqref{eq:pos} provides in fact a qualitatively relevant model of the behavior of this GPD combination at $x > \xi$.

\section{Revisiting the generalized form factors}  \label{sec:gff}

Our previous GPD publication \cite{HadStruc:2024rix} was devoted to the extraction of generalized form factors (GFFs) defined from GPD Mellin moments using the polynomiality property of Eq.~\eqref{eq:polrelHE}. Using a subset of the kinematics presented in this study with hadron momentum limited to 1.4 GeV, we had extracted the GFFs at each value of $z$ using a binning in $t$ and a polynomial fits of the small $(\nu, \bar\nu)$ expansion. After matching, we had fitted the highly correlated GFFs at various values of $z$ by a constant to obtain the final result, which was made public in the Zenodo database \cite{dutrieux_2024_13504271}.

\begin{figure*}
    \centering
\includegraphics[width=0.25\linewidth]{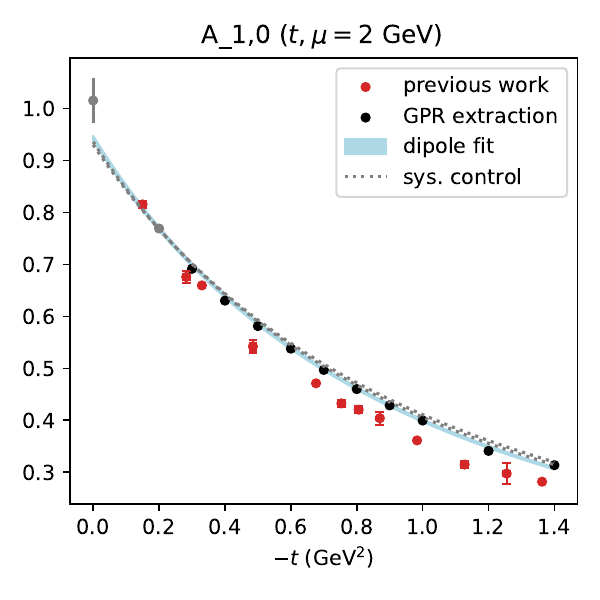}\includegraphics[width=0.25\linewidth]{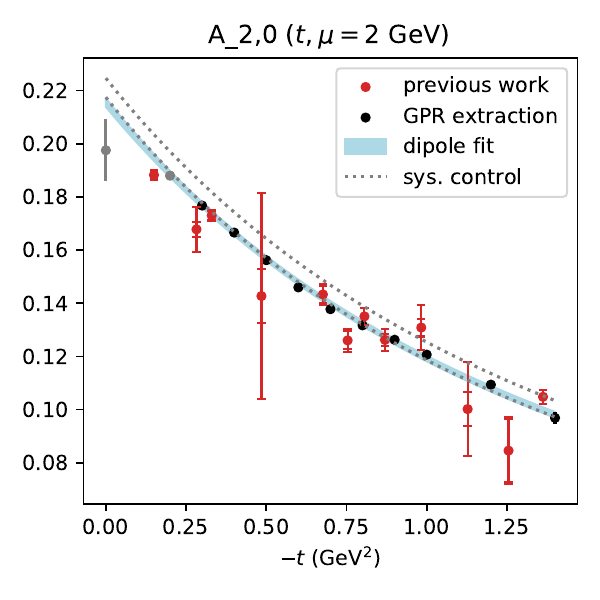}\includegraphics[width=0.25\linewidth]{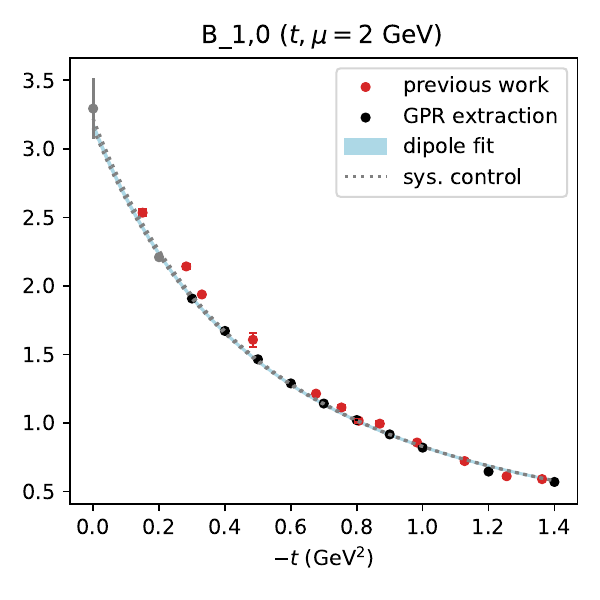}\includegraphics[width=0.25\linewidth]{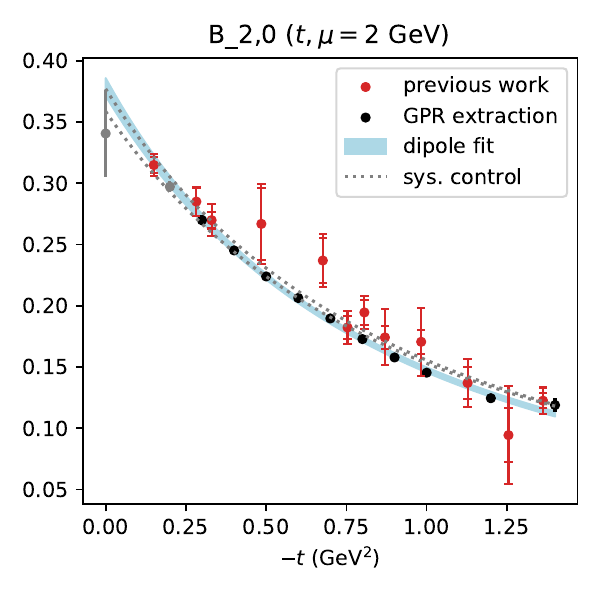}

\includegraphics[width=0.25\linewidth]{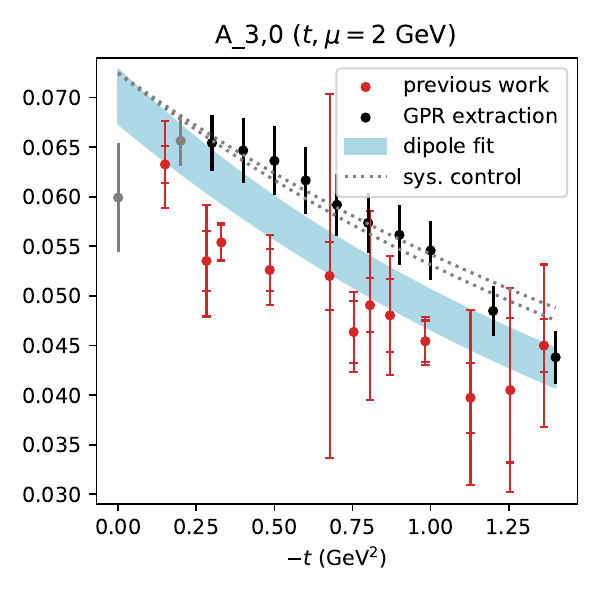}\includegraphics[width=0.25\linewidth]{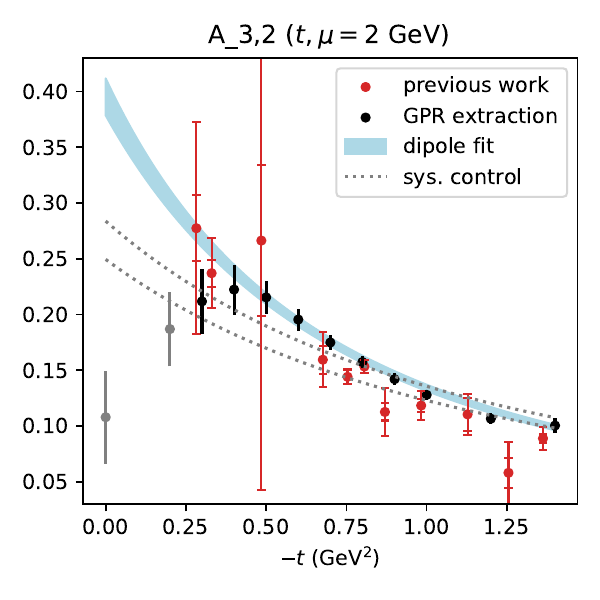}\includegraphics[width=0.25\linewidth]{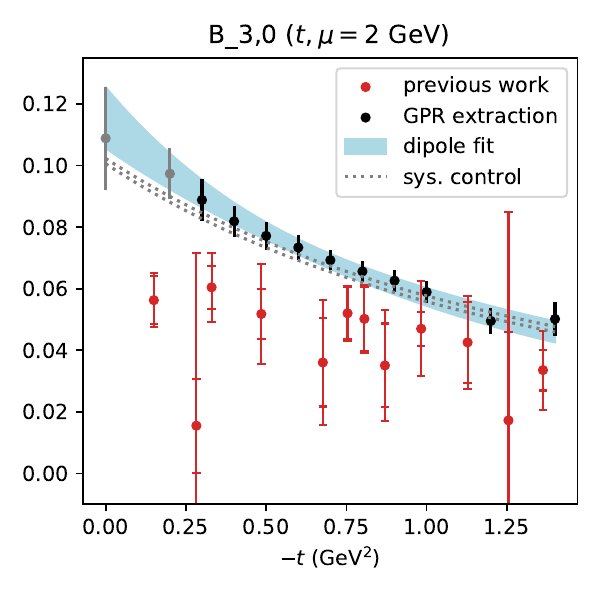}\includegraphics[width=0.25\linewidth]{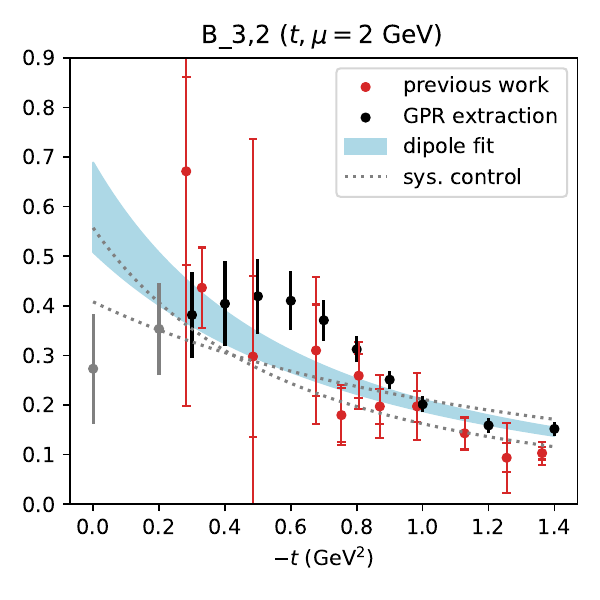}

\includegraphics[width=0.25\linewidth]{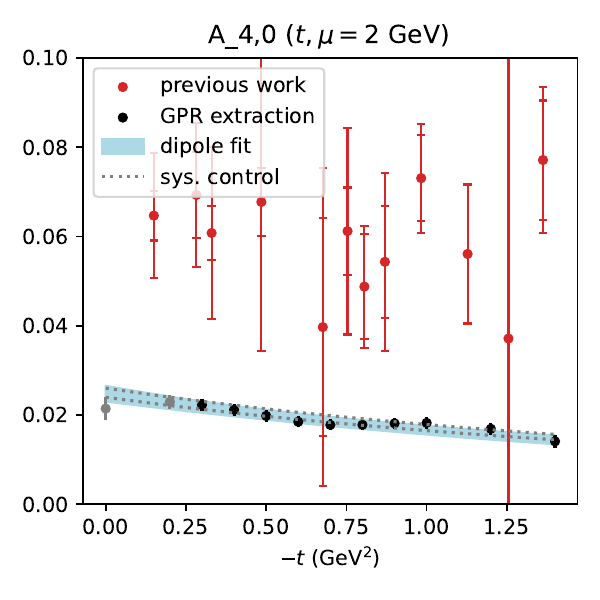}\includegraphics[width=0.25\linewidth]{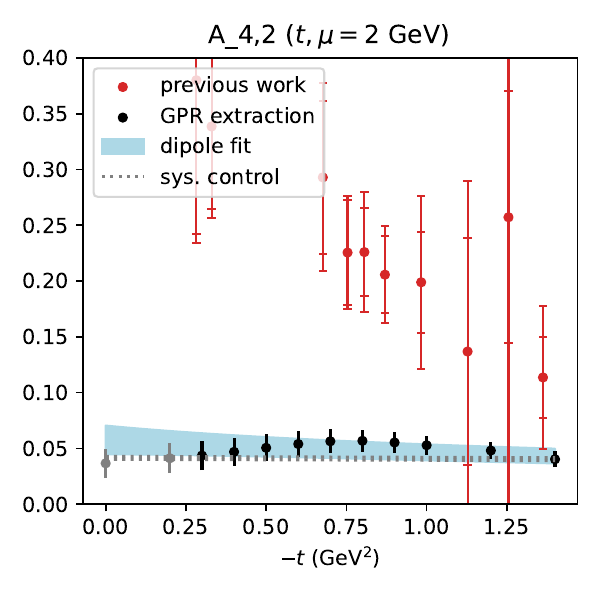}\includegraphics[width=0.25\linewidth]{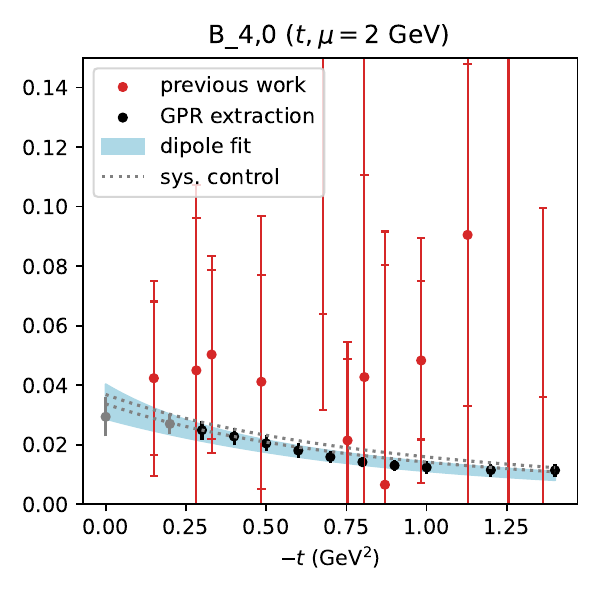}\includegraphics[width=0.25\linewidth]{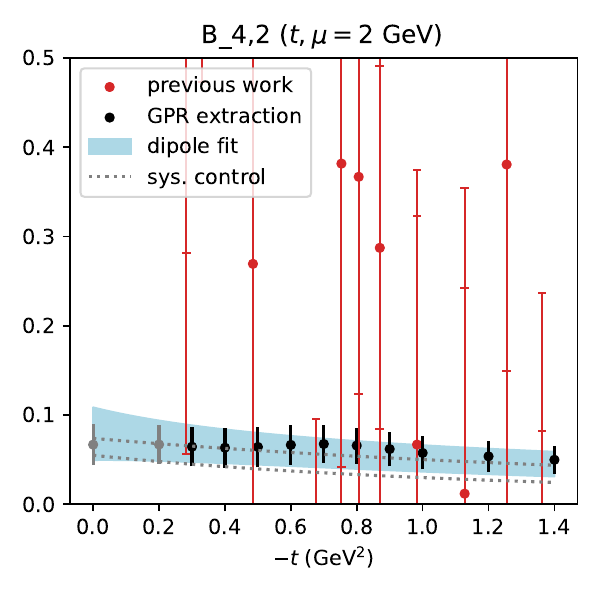}

\includegraphics[width=0.25\linewidth]{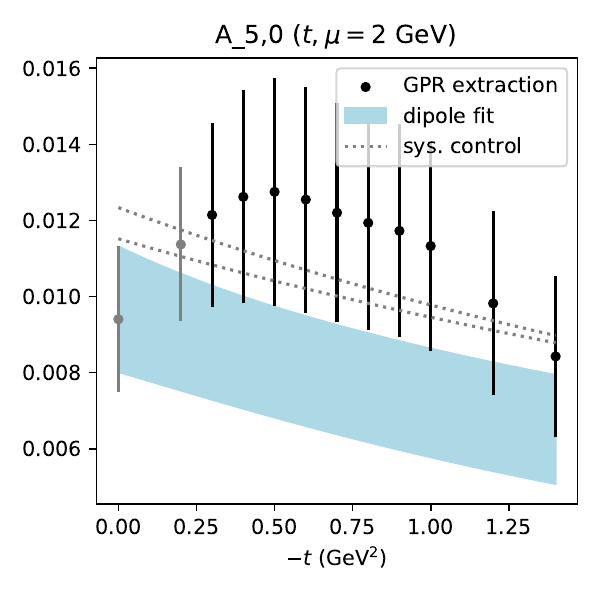}\includegraphics[width=0.25\linewidth]{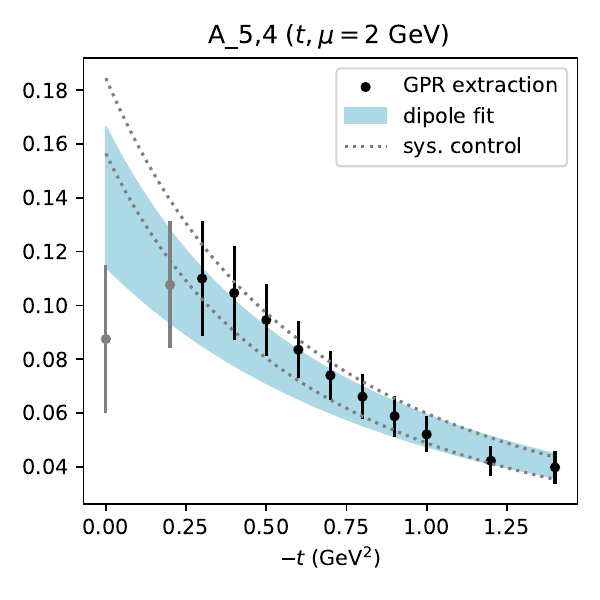}\includegraphics[width=0.25\linewidth]{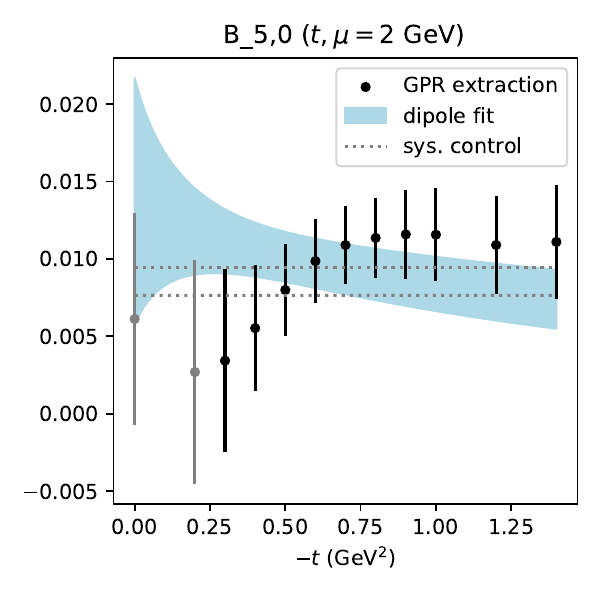}\includegraphics[width=0.25\linewidth]{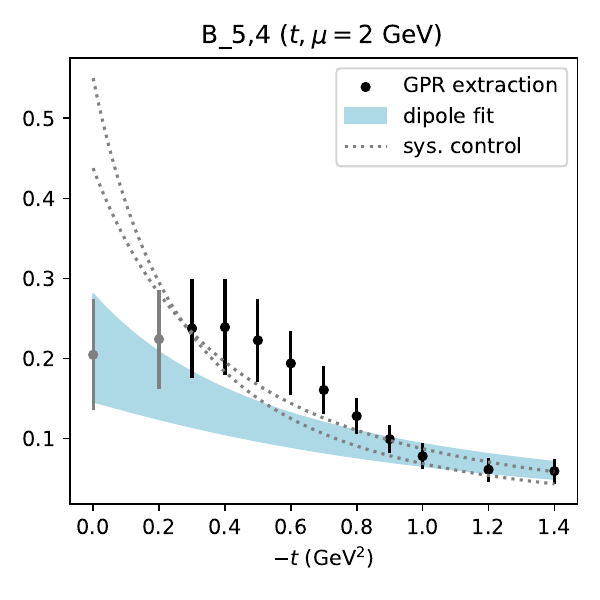}

\includegraphics[width=0.25\linewidth]{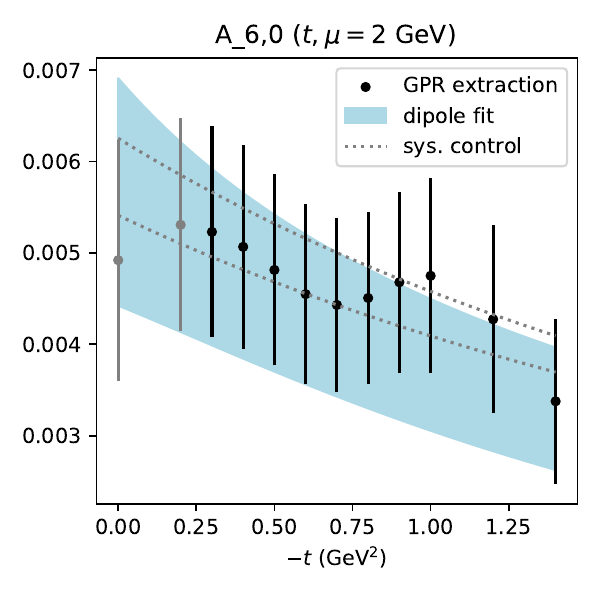}\includegraphics[width=0.25\linewidth]{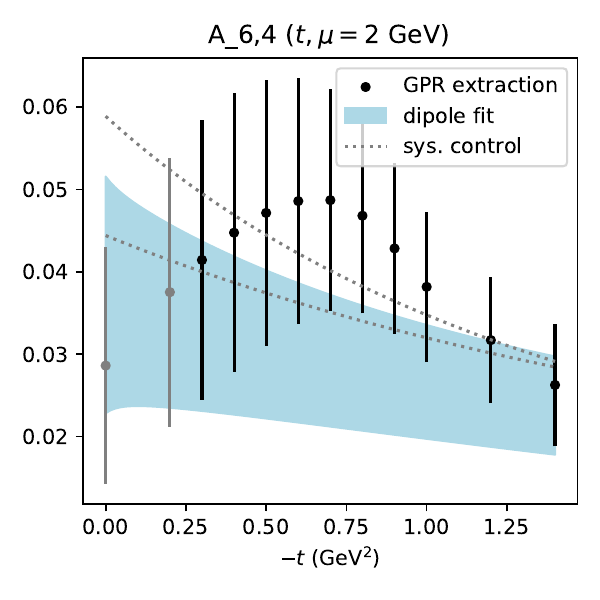}\includegraphics[width=0.25\linewidth]{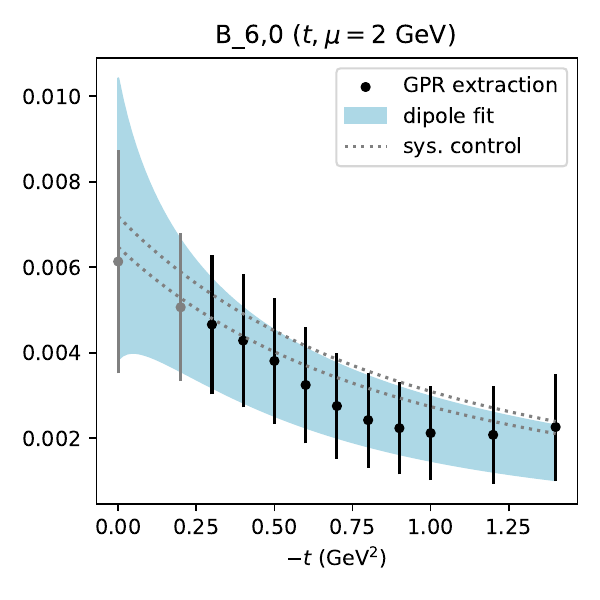}\includegraphics[width=0.25\linewidth]{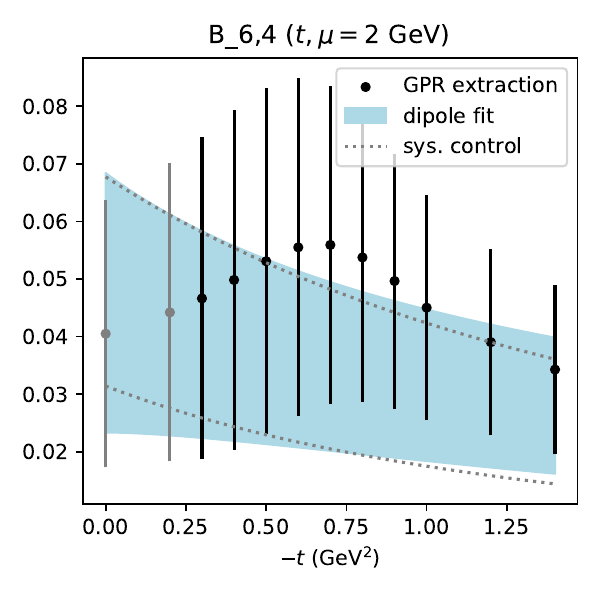}

  \caption{$t$-dependence of generalized form factors computed from the GPR reconstruction (black points, $z = 0.47$ fm, $t_s = 3$) compared with the extraction of our previous work \cite{HadStruc:2024rix, dutrieux_2024_13504271} (red points - inner error corresponding to $t_s = 3$, outer error with addition in quadrature of the mean difference with $t_s = 4$). A dipole fit of the black points is shown, excluding the two leftmost points. The dotted grey lines show dipole fits of the GPR reconstruction with $z = 0.47$ fm, $t_s = 4$ and $z = 0.56$ fm, $t_s = 3$ to evaluate additional systematic uncertainty.} \label{fig:GFFs}
\end{figure*}

\begin{figure*}[t]
    \centering
\includegraphics[width=0.48\linewidth]{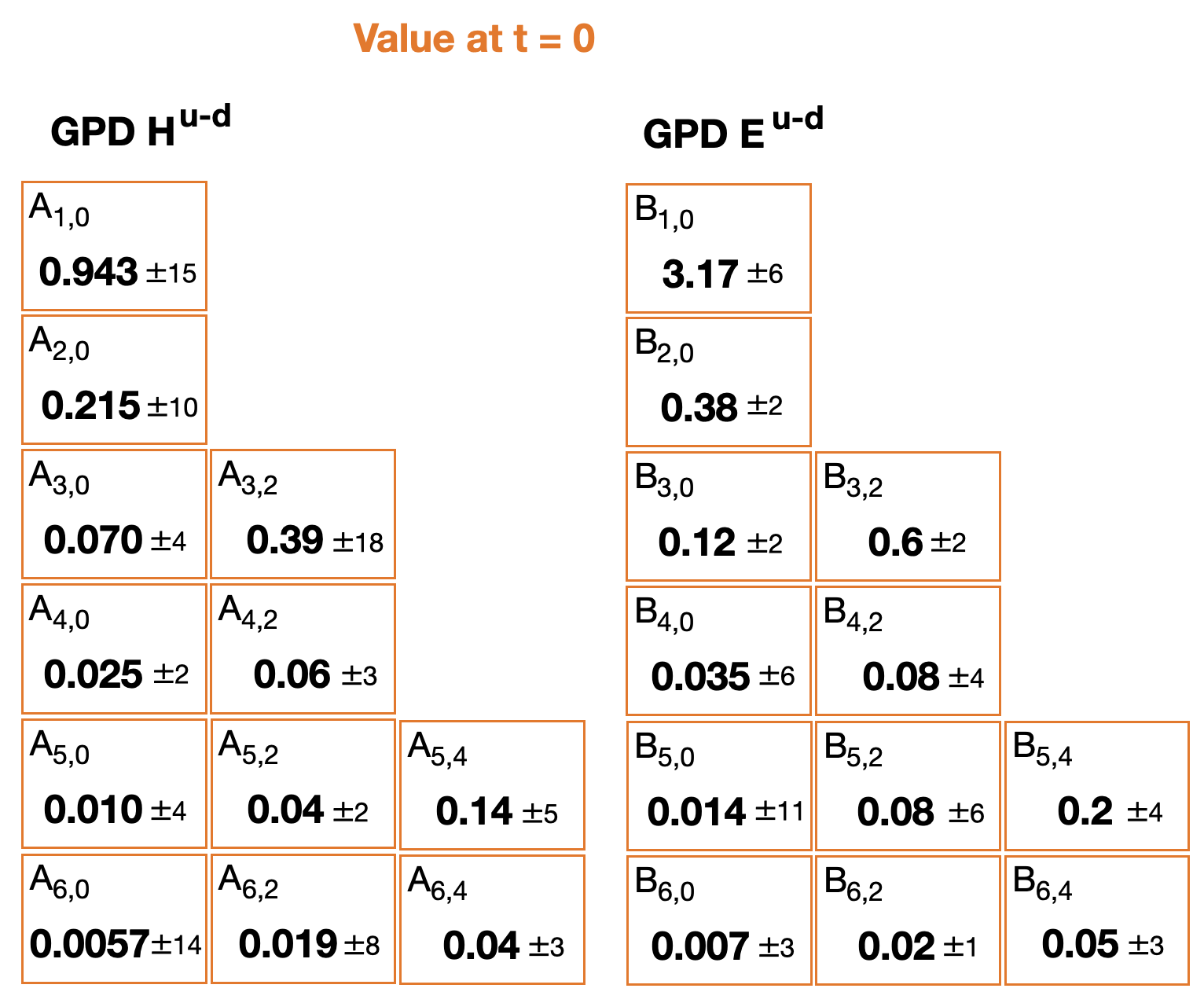}\ \ \includegraphics[width=0.48\linewidth]{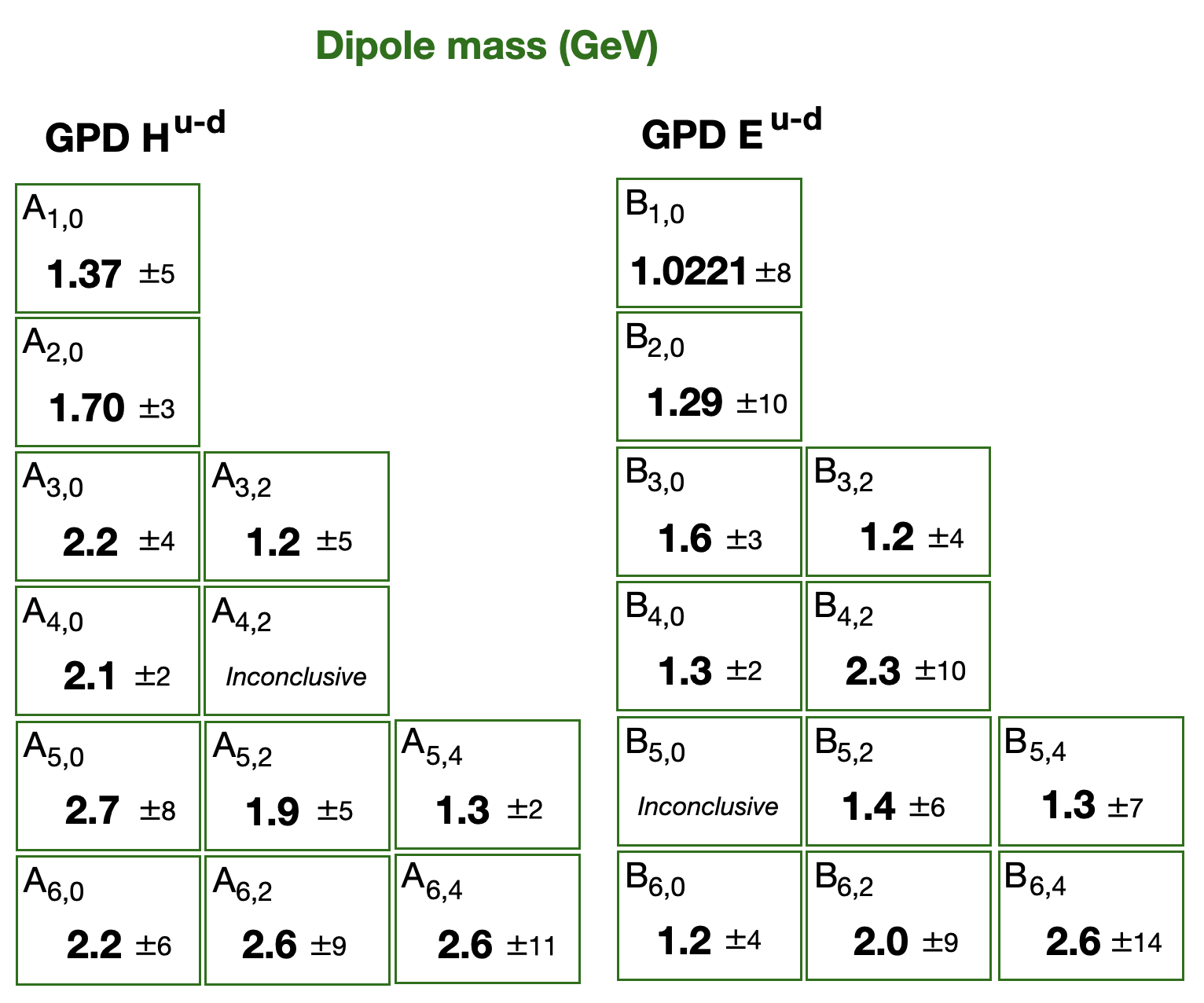}

\caption{Result of dipole fits on the GPR extraction of generalized form factors. For $A_{4,2}$ and $B_{5,0}$, the dipole mass is virtually unbounded, showing no fittable $t$-dependence in the data.} \label{fig:dipol_fit}
\end{figure*}

We can use our new GPR results to extract GFFs, by simply computing the Mellin moments of our final GPD reconstructions. The new GFF extraction will differ from the previous one both in the range in Ioffe time of the data and the systematic uncertainty linked to the analysis pipeline. We show in Fig.~\ref{fig:GFFs} our new GFF extraction compared, where possible, to the previous extraction. The black points show our new GFFs using the $z = 0.47$ fm, $t_s = 3$ data. The blue fit represents a dipole model of this data:
\begin{equation}
A_{n,k}(t) = A_{n,k}(t=0) \left(1-\frac{t}{\Lambda_{n,k}^2}\right)^{-2}\,.
\end{equation}
The dotted lines give a sense of additional systematic uncertainty by using a different $t_s$ or $z$ value, while the red points show our previous extraction of \cite{HadStruc:2024rix, dutrieux_2024_13504271}. The results of the new dipole fits with added systematic uncertainty are shown in Fig.~\ref{fig:dipol_fit}. Several comments are in order. 

\paragraph{Elastic form factors} One of the striking lessons of our previous analysis \cite{HadStruc:2024rix} was that we found a significantly different result when we extracted EFFs from non-local matrix elements compared to local matrix elements. One can constrain EFFs from data with $z \neq 0$ using either hadron momenta $p_{i,z} = p_{f,z} = 0$ so that $\nu = \bar\nu = 0$, or simply, as we have done again in this paper, by a fit of data at $\nu \neq 0$ and an extrapolation to $\nu = 0$. Both methods were studied in our previous paper and consistently showed a dependence of the EFF on the value of $z$. As EFFs are unaffected by perturbative matching, we concluded that this effect likely showed sizeable discretization errors or higher-twist contamination, without being able to investigate the question further with a single lattice ensemble.

Since our current GPR extraction is based on data at $z = 5a$ and $6a$, it suffers from a different EFF $z$-dependent shift compared to our previous work which used fits of the full, highly correlated, range in $z$. It is very visible particularly for $A_{1,0}$ in Fig.~\ref{fig:GFFs}. Note that the newly added kinematics, which exist at a large value of $\nu$, have little to no effect on the EFF at $\nu = 0$. The discrepancy, that we choose not to try and reduce to remind the reader of unaccounted systematic uncertainty in our extraction, is solely a result of previously discussed features of the data in \cite{HadStruc:2024rix}. In particular, the issue of consistency of the new extraction of $F_2$ at small $-t$ compared to our previous data which was investigated in Sec.~\ref{sec:effs} does not seem to play a major role in this discussion.

Finally, an obvious systematic uncertainty is introduced by the model-dependence inherent to a dipole fit. We have shown on the plots the extrapolation performed by the GPR at $t = 0$ with grey points, not included in the fit, with a considerably larger uncertainty compared to the dipole fit. For instance, the result of the dipole fit at $A_{1,0}$ misses badly the expected value of unity, while the GPR extrapolation reaches the correct value within its much larger uncertainty.

\paragraph{GFFs with $n = 2$ and $3$} For $n = 2$, the agreement between our new extraction and the previous one is good, while it remains fair for $n = 3$ with the largest tension being observed for $B_{3,0}$. We note that our previous extraction used a simple binning in $t$, meaning that red points in the plots are weakly correlated. On the other hand, our GPR reconstruction is a global reconstruction of the $t$-dependence, and shows therefore large correlations between $t$-values. This correlation information, absent from the plot, makes a strict comparison of the respective standard deviation of red and black points or eye-gauging fit results misleading.

Since the GFF at $n=2$ corresponds to the slope at small $\nu$ of the imaginary part of the data, and $n = 3$ to the quadratic curvature at small $\nu$ of the real part, the newly added kinematics at larger $\nu$ play a very small role here again. Therefore, the difference between our new and previous results is once again mostly driven by the different systematic uncertainty of our analysis pipeline.

\paragraph{GFFs with $n \geq 4$} This time, the new data plays a role in the GFF extraction. For $n = 4$, it is crucial to be able to locate the first extremum of the imaginary part. As our previous data barely extended to $\nu = 3.5$, the position of this extremum was uncertain. The polynomial fit of the small $(\nu,\bar\nu)$ expansion, which introduced a systematic uncertainty through the choice of order truncation, resulted in an overestimation of the GFFs for $n = 4$. This fit resulted in an extremum of the imaginary part at smaller $\nu$ than the new data, which extends up to $\nu = 7$, indicates. As for $n > 4$, we could not extract any signal in the previous analysis. Notice that our extraction remains very imprecise for large $n$ GFFs, and that the $t$-dependence is poorly resolved by the data, leading occasionally to unreliable dipole fits.

\paragraph{General observations} As $n$ increases and $k$ remains fixed, the magnitude of moments decreases quickly, which is the expected behavior for a distribution which vanishes at $x = 1$. The $t$-dependence becomes milder, or equivalently is fittable by a larger dipole mass. This is in line with the perturbative estimate of Eq.~\eqref{eq:largexGPD}, according to which GPDs are independent of $t$ at large $x$. 

When $k$ increases at fixed $n$, the magnitude of moments increases notably, and their $t$-dependence is much steeper. Since the fitted dipole mass is clearly statistically resolved to be different for several GFFs, we invalidate, as in our previous study, a factorized Ansatz for the $t$-dependence.

The ratio $A_{2,0}(t) / A_{1,0}(t)$ was recently analyzed in soft-collinear effective theory (SCET) at large $-t$ in \cite{Hatta:2025xuf}. Using some approximations, the authors find the ratio to be dominated by 1 plus perturbative corrections in $\alpha_s$ and $\alpha_s \ln(-t / \mu^2_{UV})$. We note that if our dipole Ansatz were to be trusted in the large $-t$ regime, which we have no data-driven evidence to support, the ratio of GFFs would tend to a constant at large $-t$ given by
\begin{equation}
\lim_{-t \rightarrow \infty} \frac{A_{2,0}(t)}{A_{1,0}(t)} = \frac{A_{2,0}(t=0) \Lambda_{2,0}^4}{A_{1,0}(t=0)\Lambda_{1,0}^4}\,.
\end{equation}
Using numerical values from our previous study \cite{HadStruc:2024rix} gives 0.96, while values from our GPR reconstruction give 0.54. The main culprit for this large difference is the systematic error in the evaluation of the dipole parameters of the EFF, demonstrating the necessity to tackle the question of discretization errors and higher-twist contamination for a precise phenomenology of GPDs.
 
\section{Conclusion}

For the first time in the literature, we have extracted double distributions directly from lattice data, producing a reconstruction of the full kinematic dependence in $(x, \xi, t)$ of the proton isovector unpolarized GPDs in agreement with important theoretical properties of GPDs. Our results, obtained from data at $m_\pi = 358$ MeV and lattice spacing $a = 0.09$~fm, are displayed in Fig.~\ref{fig:final}. They show a broad qualitative agreement with the Goloskokov-Kroll phenomenological model \cite{Goloskokov:2005sd,Goloskokov:2007nt,Goloskokov:2009ia} 
at $t = 0$
based on Radyushkin's double distribution ansatz  \cite{Radyushkin:1998es,Radyushkin:1998bz,Musatov:1999xp}. We make our results available to the community in the public database \cite{dutrieux_2026_19695472}. 

We extended the framework of multidimensional Gaussian process regression to produce a flexible regularization of the inverse problem in the pseudo-distribution formalism \cite{Radyushkin:2017cyf}, enforce correct physical properties on our reconstruction, and perform a partial assessment of the model dependence. We used one-loop matching \cite{Radyushkin:2019owq} with renormalization-group improvement, data up to $z = 0.56$ fm and proton momenta up to 2.7 GeV. The treatment of perturbative uncertainties at such large values of $z$ remains summary, and we cannot assess the size of lattice discretization errors and higher-twist contamination. Evidence of their presence mostly affects the elastic form factors due to their good statistical precision. Further refinements on lattice systematic uncertainties and closer to physical pion masses are important to fully benefit from the impact of this research.

Regarding the phenomenological impact, it is remarkable that, for the first time, one accesses a GPD from first-principles calculation that fulfils both the polynomiality properties (by construction) and the positivity one (see sec. \ref{sec:pheno}), a longstanding goal of the phenomenological studies \cite{Chouika:2017dhe,Chouika:2017rzs,Mezrag:2022pqk,Dutrieux:2021wll}.
Our work provides the first necessary steps to compute deep exclusive processes amplitudes (see e.g the case of charged vector meson production \cite{Mankiewicz:1997aa}) based on pure lattice and perturbative QCD inputs, following what has been done for one observable of Deep Virtual Compton Scattering in \cite{Martinez-Fernandez:2025jvk} using generalized Form Factors.
Such a target would however include the necessary steps of extrapolating our results both to the physical pion mass and to the continuum limit.
It will also require the computation of the $\rho$ meson distribution amplitude, which is accessible through the short-distance factorisation formalism.
These points will be the aims of forthcoming publications.

\section{Acknowledgements}

We would like to thank Huey-Wen Lin for useful comments on the matching procedure. This project was supported by the U.S.~Department of Energy, Office of Science, Contract No.~DE-AC05-06OR23177, under which Jefferson Science Associates, LLC operates Jefferson Lab. SZ, C\'edric M. and HD acknowledge support in part from the Agence Nationale de la Recherche
(ANR) under Project No. ANR-23-CE31-0019. Christopher M.~is supported in part by U.S.~DOE Grant \mbox{\#DE-SC0025908}. KO is supported in part by U.S.~DOE Grant \mbox{\#DE-FG02-04ER41302}. The work of DR was conducted in
part under the Laboratory Directed Research and Development Program at
Thomas Jefferson National Accelerator Facility for the U.S.\ Department of Energy. AR acknowledges support by U.S.~DOE Grant \mbox{\#DE-FG02-97ER41028}. 

 This work has benefited from the collaboration enabled by the Quark-Gluon Tomography (QGT) Topical Collaboration, U.S.~DOE Award \mbox{\#DE-SC0023646}. Computations for this work were carried out in part on facilities of the USQCD Collaboration, which are funded by the Office of Science of the U.S.~Department of Energy. The authors acknowledge William \& Mary Research Computing for providing computational resources and/or technical support, which were provided by contributions from the National Science Foundation (MRI grant PHY-1626177), and the Commonwealth of Virginia Equipment Trust Fund. In addition, this work used resources at NERSC, a DOE Office of Science User Facility supported by the Office of Science of the U.S. Department of Energy under Contract \#DE-AC02-05CH11231, as well as resources of the Oak Ridge Leadership Computing Facility at the Oak Ridge National Laboratory, which is supported by the Office of Science of the U.S. Department of Energy under Contract No. \mbox{\#DE-AC05-00OR22725}. 
The authors acknowledge support as well as computing and storage resources by GENCI on Adastra (CINES), Jean-Zay (IDRIS) under project (2020-2024)-A0080511504.
The software codes {\tt Chroma} \cite{Edwards:2004sx}, {\tt QUDA} \cite{Clark:2009wm, Babich:2010mu}, {\tt QPhiX} \cite{QPhiX2}, and {\tt Redstar} \cite{Chen:2023zyy} were used in our work. The authors acknowledge support from the U.S. Department of Energy, Office of Science, Office of Advanced Scientific Computing Research and Office of Nuclear Physics, Scientific Discovery through Advanced Computing (SciDAC) program, and of the U.S. Department of Energy Exascale Computing Project (ECP). The authors also acknowledge the Texas Advanced Computing Center (TACC) at The University of Texas at Austin for providing HPC resources, like Frontera computing system~\cite{frontera} that has contributed to the research results reported within this paper.

\begin{appendix}

\section{Evolution of DDs at small $\beta$ \label{sec:appA}}

The goal of this Appendix is to study the singularities generated by one-loop perturbative evolution on DDs. We already know from the inspection of Radon integration lines in Section \ref{sec:DDrep} that some divergence must occur at small $\beta$, but we would like to characterize it better analytically.

A convenient tool to study parton distributions at small $x$ or $\beta$ is the behavior of their lowest Mellin moments. For instance, if the distribution converges to a constant value $\lambda$ at small $x$, the Mellin moments will diverge when $n \rightarrow 0$ as $\lambda/n$. If the small $x$ behavior is a power divergence $x^{-\lambda}$, the Mellin moments will diverge when $n \rightarrow \lambda$, etc. 

Using the notations of Eq.~\eqref{eq:polrelHE}, the GPD moments are related to moments of the DD through:
\begin{equation}
A_{n,k}(t) = \begin{pmatrix}
    n-1 \\ k
\end{pmatrix} \int_\Omega \mathrm{d}\beta\mathrm{d}\alpha\,\beta^{n-1-k}\alpha^k f^q(\beta, \alpha, t)\,.\label{eq:momDD}
\end{equation}
If we allow ourselves to consider indices $n \in \mathbb{R}$, the small-$\beta$ behavior for a DD with no power divergence is described by the limit $n \rightarrow k$, where the integral in Eq.~\eqref{eq:momDD} is dominated by small $\beta$ contributions. Due to the parity in $\alpha$ of the DD, only $k$ even will be considered. For $k = 0$,
\begin{equation}
A_{n,0}(t) = \int_\Omega \mathrm{d}\beta\mathrm{d}\alpha \,\beta^{n-1} f^q(\beta, \alpha,t)\,,\label{eq:dewfvr}
\end{equation}
so the limit of $A_{n,0}$ when $n \rightarrow 0$ is reflective of the DD behavior at small $\beta$ integrated over the entire $\alpha$ dependence.

For $k \geq 2$,
\begin{align}
\int_\Omega \mathrm{d}\beta\mathrm{d}\alpha\,\beta^{n-1-k}\alpha^k f^q(\beta, \alpha,t) &= \frac{k!}{(n-1)...(n-k)} A_{n,k}(t) \,,\\
&\stackrel{n \rightarrow k}{\sim} \frac{k}{n-k} A_{n,k}(t)\,.\label{eq:dfvg}
\end{align}
The factor $1/(n-k)$ diverges when $n \rightarrow k$, but the behavior of $A_{n,k}$ remains to be clarified.

\subsection{Integrated behavior over $\alpha$}

Let us start with the simple case $k = 0$, where the $\alpha$-dependence is integrated over. From Eq.~\eqref{eq:dewfvr}, the DD moment is simply $A_{n,0}(t)$, which in turn is a Mellin moment of the GPD at $\xi = 0$. $A_{n,0}(t)$ behaves with respect to evolution exactly like a PDF moment. In particular, one has:
\begin{equation}
A_{n,0}(t,\mu_1^2) = A_{n,0}(t,\mu_0^2) \left(\frac{\alpha_s(\mu_0^2)}{\alpha_s(\mu_1^2)}\right) ^{\gamma_n/(2\pi\beta_0)}\,,
\label{eq:evolPDF}\end{equation}
where $\gamma_n$ are the Mellin moments of the DGLAP evolution operator:
\begin{widetext}
\begin{equation}
\gamma_n = C_F \int_0^1 \mathrm{d}\alpha\,\frac{1+\alpha^2}{1-\alpha}\,(\alpha^{n-1} -1) = C_F\left[-2\psi^{(0)}(n)-2\gamma_E+\frac{3}{2}-\frac{1}{n}-\frac{1}{n+1}\right]\,,
\end{equation}
\end{widetext}
with $\psi^{(0)}(n) = \mathrm{d} \ln\Gamma(n) / \mathrm{d}n$ the digamma function and $\gamma_E = -\psi^{(0)}(1)$ the Euler-Mascheroni constant. $\beta_0 = (33-2n_f)/(12\pi) \approx 0.716$.

In Section 3.3 of \cite{Dutrieux:2023zpy}, we studied the evolution operator in the limit $n \rightarrow 0$:
\begin{equation}
\left(\frac{\alpha_s(\mu_0^2)}{\alpha_s(\mu_1^2)}\right) ^{\gamma_n/(2\pi\beta_0)} \stackrel{n \rightarrow 0}{\sim} \left(\frac{\alpha_s(\mu_0^2)}{\alpha_s(\mu_1^2)}\right)^{C_F/(2\pi\beta_0) [1/n + 1/2]}\,. \label{eq:rrevr}
\end{equation}
The Mellin moments diverge when $n \rightarrow 0$ in an exponential fashion, demonstrating a small-$x$ divergence which is smaller than any power law. The inverse Mellin transform of the right hand side of Eq.~\eqref{eq:rrevr} gave the behavior of the PDF evolution operator at small $x$. Let us recall the result here: if we write the evolution of the PDF $q(x, \mu^2)$ from $\mu_0^2$ to $\mu_1^2$ as
\begin{equation}
q(x, \mu_1^2) = \int_x^1 \frac{\mathrm{d}y}{y} \,\mathcal{E}(y; \mu_0^2, \mu_1^2) q\left(\frac{x}{y}, \mu_0^2\right)\,,
\end{equation}
we find the behavior of the evolution operator $\mathcal{E}$ at small $y$ to be dominated at one-loop by:
\begin{equation}
\mathcal{E}(y; \mu_0^2, \mu_1^2)\ \  \stackrel{y \rightarrow 0}{\sim}\ \  A\  _0F_1\left(;2;-B\ln(y)\right)\,, \label{eq:divPDF}
\end{equation}
where $A$ and $B$ depend on the scales, but not on $y$, and $_0F_1\left(;2;x\right)$ is the confluent hypergeometric limit function defined by its series expansion:
\begin{equation}
    _0F_1(;2;x) = \sum_{n = 0}^\infty \frac{x^n}{n!(n+1)!}\,.
\end{equation}
Eq.~\eqref{eq:divPDF} informs on the rate of divergence at small $x$ induced solely by evolution.

Since the DD moment for $k = 0$ is just $A_{n,0}$, 
we can conclude that, \textbf{when averaged over the \mbox{$\alpha$-dependence,} the small-$\beta$ behavior of DDs under evolution is the same as the small-$x$ behavior of PDFs}, described at one-loop by Eq.~\eqref{eq:divPDF}.

\subsection{Understanding the $\alpha$-dependence at small $\beta$}

To go beyond the previous observation, we need to address the skewness dependence of the GPD evolution. It is well-known that conformal moments, defined as\begin{widetext}
\begin{equation}
O_n(\xi, t, \mu^2) = \frac{\sqrt{\pi}\Gamma(n)}{2^n\Gamma(n+1/2)}\xi^{n-1}\, \int_{-1}^1 \mathrm{d}x\,C_{n-1}^{(3/2)}\left(\frac{x}{\xi}\right) H^q(x, \xi, t, \mu^2)\,,\label{eq:defGeg}
\end{equation}
diagonalize the leading-order evolution of GPDs, such that they obey a similar evolution equation to Eq.~\eqref{eq:evolPDF}:
\begin{equation}
O_n(\xi, t, \mu_1^2) = O_n(\xi, t, \mu_0^2) \left(\frac{\alpha_s(\mu_0^2)}{\alpha_s(\mu_1^2)}\right) ^{\gamma_n/(2\pi\beta_0)}\,.
\label{eq:evolconf}\end{equation}
Let us write the explicit decomposition of Gegenbauer polynomials as:
\begin{equation}
C_n^{(3/2)}(x) = \sum_{\substack{k=0\\k \equiv n[2]}}^n c_{n,k}x^k\,.
\end{equation}
The initial factor in the definition of Eq.~\eqref{eq:defGeg} ensures that $c_{n,n} = 1$. Then the conformal moments write:
\begin{align}
O_n(\xi) = [A_{n,0}&+A_{n,2}\xi^2 + ...] + \xi^2 c_{n-1,n-3} [A_{n-2,0}+A_{n-2,2}\xi^2+...] + ...
\end{align}
Therefore, one finds:
\begin{equation}
A_{n,0} = O_n(\xi = 0)\,.
\end{equation}
This just shows that conformal moments reduce to simple Mellin moments in the limit $\xi = 0$, which we have analyzed in the previous section. Keeping on,
\begin{equation}
\frac{\mathrm{d}^2}{\mathrm{d}\xi^2} O_n(\xi = 0) = 2 [ A_{n,2} + c_{n-1,n-3}A_{n-2,0}]\,,\label{eq:dweehtrge}
\end{equation}
so
\begin{align}
A_{n,2} = \frac{1}{2}\frac{\mathrm{d}^2}{\mathrm{d}\xi^2} O_n(\xi = 0) + \frac{(n-1)(n-2)}{2(2n-1)} A_{n-2,0}\,, \label{eq:efrhbre}
\end{align}
where we have used the actual value of $c_{n-1,n-3}$. Now introducing the scale dependence, we find:
\begin{align}
A_{n,2}(\mu_1^2) = \frac{1}{2}\frac{\mathrm{d}^2}{\mathrm{d}\xi^2}&O_n(\xi = 0, \mu_0^2) \left(\frac{\alpha_s(\mu_0^2)}{\alpha_s(\mu_1^2)}\right) ^{\gamma_n/(2\pi\beta_0)} \nonumber \\
&+ \frac{(n-1)(n-2)}{2(2n-1)} A_{n-2,0}(\mu_0^2) \left(\frac{\alpha_s(\mu_0^2)}{\alpha_s(\mu_1^2)}\right) ^{\gamma_{n-2}/(2\pi\beta_0)}\,. \label{eq:rf3trb}
\end{align}
We are interested in the limit $n \rightarrow 2$, which will give us insight on the small-$\beta$ behavior of the moment $\langle \alpha^2\rangle$ of the DD. The first term in the right hand side of Eq.~\eqref{eq:rf3trb} is simply finite when $n \rightarrow 2$. On the other hand, $A_{n-2,0}$ typically diverges as $\mathcal{O}(1/(n-2))$ if the GPD has a finite non-zero value at $x = \xi = 0$, and more sharply if the GPD is itself divergent. The scale dependent factor of the second term has already been studied in the previous section and diverges as well. Therefore the first term can be neglected. Reminding ourselves of Eq.~\eqref{eq:dfvg},
\begin{align}
\int_\Omega \mathrm{d}\beta\mathrm{d}\alpha\,\beta^{n-3}\alpha^2 f^q(\beta, \alpha,t, \mu_1^2) \stackrel{n \rightarrow 2}{\sim} \frac{1}{3} A_{n-2,0}(\mu_0^2) \left(\frac{\alpha_s(\mu_0^2)}{\alpha_s(\mu_1^2)}\right) ^{\gamma_{n-2}/(2\pi\beta_0)}\,.\label{eq:drbvdcs}
\end{align}
\end{widetext}
Up to the factor $1/3$, this is exactly the result we found for $k = 0$. Therefore, it appears that the scale dependence of $\langle \alpha^0 \rangle$ and $\langle \alpha^2 \rangle$ is similar at small-$\beta$, and reflects the $x$-evolution of ordinary PDFs. Notice that while it was exactly the case for $\langle \alpha^0 \rangle$, it is now just an approximation at small $\beta$ (or small $n$ in Eq.~\eqref{eq:rf3trb}). The term which we have neglected, which was constant, would correspond to a non-divergent contribution to the small-$\beta$ region, because $\gamma_2$ is finite whereas $\gamma_n \rightarrow \infty$ when $n \rightarrow 0$.

This argument can be easily generalized beyond $k=2$. Proceeding in a similar fashion as in Eq.~\eqref{eq:dweehtrge}, we can isolate:
\begin{equation}
A_{n,k} = \sum_{i = 0 \textrm{ even}}^k d_{n,i,k} \frac{\mathrm{d}^i}{\mathrm{d}\xi^i}O_{n-k+i}(\xi = 0)\,,
\end{equation}
where $d_{n,i,k}$ is a combination of coefficients of Gegenbauer polynomials. The only divergent contribution in the limit $n \rightarrow k$ comes again from the term $i = 0$, which corresponds to $A_{n-k,0}(\mu_0^2)$ and the anomalous dimension $\gamma_{n-k}$. Not only does it mean that every DD moment $\langle \alpha^k \rangle$ has the same asymptotic scale evolution at $\beta \rightarrow 0$, but it only depends on the $\langle \alpha^0 \rangle$ moment at initial scale. In other words, evolution at very small $\beta$ tends towards a universal profile in $\alpha$ independent of the shape at starting scale.

To compute this universal profile, we start by deriving the general expression of the coefficients of the Gegenbauer polynomials:
\begin{align}
c_{n-1,n-1-2i} = \frac{(-1)^i}{2^i i!} \frac{(n-1)(n-2)...(n-2i)}{(2n-1)(2n-3)...(2n-2i+1)}\,.
\end{align}
It is easy to verify that for $i = 0$, $c_{n-1,n-1} = 1$ and for $i = 1$, $c_{n-1,n-3} = -(n-1)(n-2)/(2(2n-1))$, a formula we used in Eq. \eqref{eq:efrhbre}.

Then we notice that, keeping only the dominant terms:
\begin{equation}
A_{n,k} \stackrel{n \rightarrow k}{\sim} -\sum_{i=2 \textrm{ even}}^k c_{n-1,n-i-1} A_{n-i, k-i}\,.
\end{equation}
This recurrence relation relates ultimately $A_{n,k}$ to $A_{n-k,0}$ times a relatively simple but combinatorially growing combination of Gegenbauer coefficients. Performing all simplifications, we end up with the relationship:
\begin{widetext}
\begin{equation}
A_{n,2k} \stackrel{n \rightarrow 2k}{\sim} A_{n-2k,0} \times \frac{(n-1)(n-2)...(n-2k)}{2^k k!(2n-2k+1)(2n-2k-1)...(2n-4k+3)}\,.
\end{equation}
Finally, combining with the $k / (n-k)$ factor from Eq.~\eqref{eq:dfvg}, we obtain the remarkably simple:
\begin{align}
\int_\Omega \mathrm{d}\beta\mathrm{d}\alpha\,\beta^{n-1-k}\alpha^k f^q(\beta, \alpha,t, \mu_1^2) \stackrel{n \rightarrow k}{\sim} \frac{1}{k+1} A_{n-k,0}(\mu_0^2) \left(\frac{\alpha_s(\mu_0^2)}{\alpha_s(\mu_1^2)}\right) ^{\gamma_{n-k}/(2\pi\beta_0)}\,.
\end{align}
It is immediate to cross-check that if $k = 0$, the coefficient should be 1, and that we have already computed the coefficient of $k = 2$ to be $1/3$ in Eq.~\eqref{eq:drbvdcs}. The profile towards which evolution makes the DD tend at very small $\beta$ is therefore one such that $\int_{-1}^1 \mathrm{d}\alpha\,\alpha^k f(\alpha) \propto \frac{1}{k+1}$, which is trivially found to be a constant in $\alpha$.

Of course, even if the divergence created by the perturbative evolution lifts up the very small $\beta$ region towards a uniform distribution in $\alpha$, it remains a divergence smaller than any power law. If the DD presents a non-flat profile in $\alpha$ at initial scale with a power-law divergence in $\beta$, the equalization of the $\alpha$ profile through the effect of evolution would be extremely slow. These observations open new insights on the modelling of the skewness dependence of GPDs at small $x$, where several approaches based on the asymptotic effect of perturbative kernels have been already considered (see \cite{Dutrieux:2023qnz} for more details).

\end{widetext}

\section{Performance of interpolation and integration procedures \label{app:perf}}

\begin{figure*}
    \centering
    \includegraphics[width=0.4\linewidth]{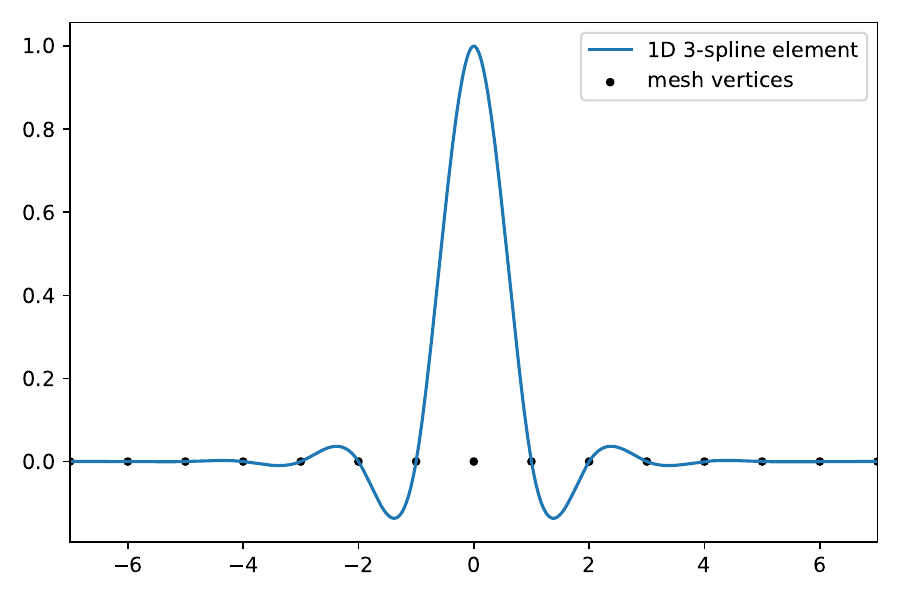} \includegraphics[width=0.4\linewidth]{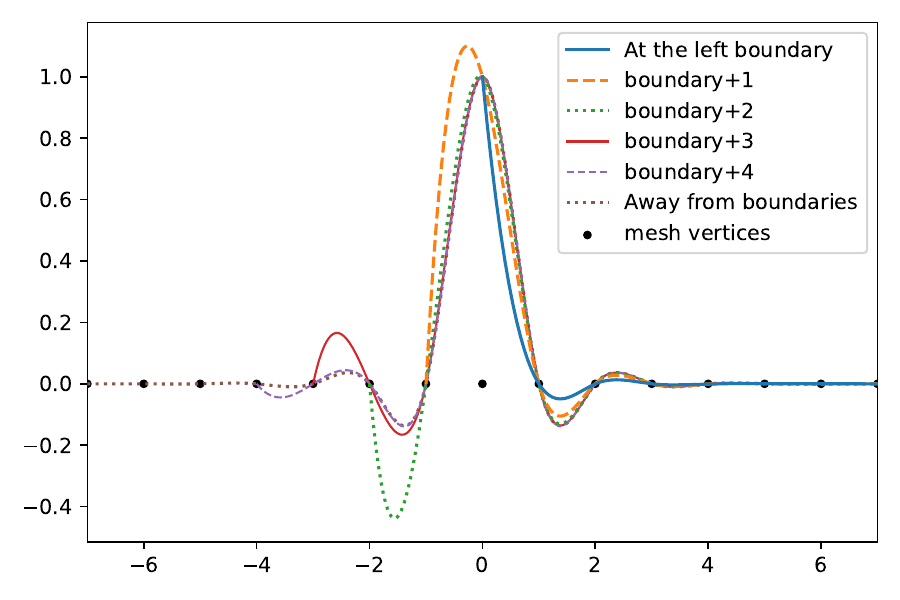}
    \caption{(left) An ECS in one dimension, at large distance from the boundaries of the domain. (right) Effect of boundaries on ECSs. The ECS is always centered on the middle vertex of the plot, which is either at the left boundary (blue), or at an increasing distance from the left boundary.}
    \label{fig:1D3spline}
\end{figure*}

Following the presentation of Section~\ref{sec:GPRDD}, the main computational task for our use of GPR is the determination of the matrix $B$, which represents the 2D Fourier transform of the cubic spline interpolation basis for each kinematic in the data. We call the elements of the basis \textit{elementary cubic splines} (ECSs). They are 3390 oscillating functions centered at each vertex of the mesh such as the one represented in 1D on Fig.~\ref{fig:1D3spline}. The 249 values of $(\vec{p}_f, \vec{p}_i)$ of our lattice dataset correspond to 160 kinematic sets of $(|\nu|, |\bar\nu|, t)$. We need therefore to perform about 540000 double integrals of oscillating functions for each value of $z^2$. While the computational effort of GPR in 1D was essentially negligible, it is no longer the case.

To measure the difficulty of a naive evaluation of $B$, for each of the 160 kinematic values, we pick at random two vertices in the mesh and compute the 2D cosine transform associated to those ECSs. Since we only picked 2 vertices out of 3390, we only carry out $\sim 0.06\%$ of the required computation. 

    On a simple Jupyter notebook running on one core of an M4 Pro laptop computer, using 

\verb|scipy.integrate.dblquad|

\verb|scipy.interpolate.RegularGridInterpolator|

\hspace{150pt}\verb|(method="cubic")| 

\noindent takes 486s, so about 8 minutes. Extrapolating to the full task with 3390 vertices would give about 10 days of calculation for each value of $z^2$.

A first observation is that each double integral is entirely independent of the others, so the task is embarrassingly parallel. A factor 10 improvement is easy to obtain on a simple laptop by using all cores, bringing down the computation to about 1 day.

\paragraph{First optimization: restricting the support of ECSs}

\begin{figure*}
    \centering
    \includegraphics[width=0.4\linewidth]{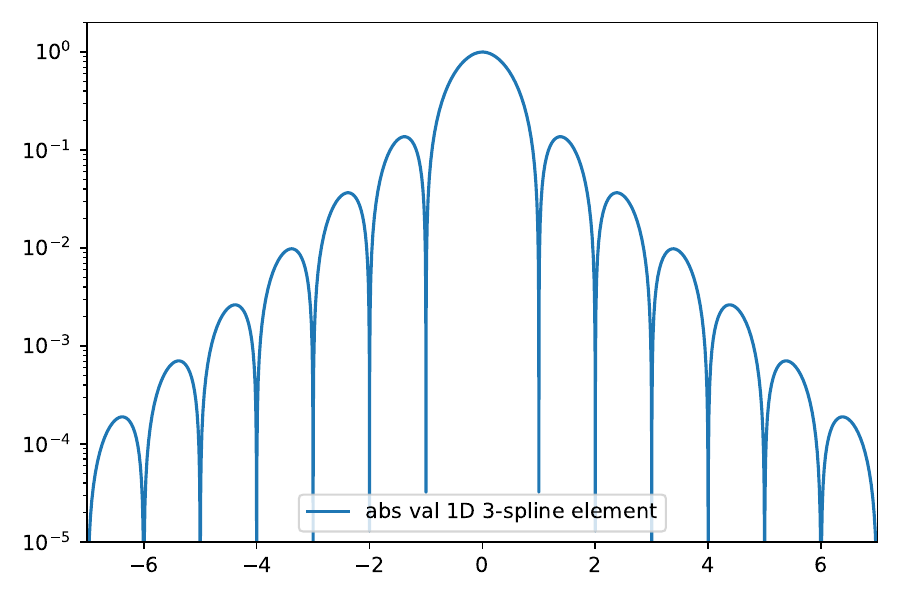}
    \includegraphics[width=0.4\linewidth]{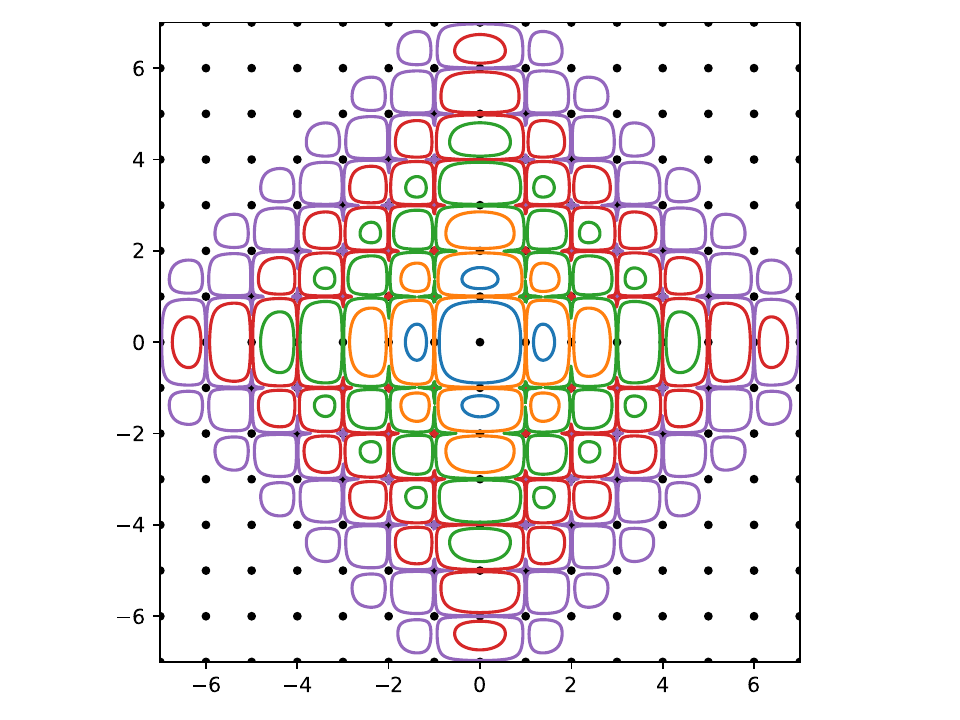}
    \caption{The absolute value of an ECS in 1D (left) and 2D (right), at large distance from the boundaries of the domain. On the right plot, the colors correspond to the levels $\{10^{-5}, 10^{-4}, 10^{-3}, 10^{-2}, 10^{-1}\}$ from the outside to the inside of the plot. Along the horizontal or vertical axis going through 0, the levels correspond to those of the left plot. It appears that the exponential decrease depends on the $L^1$ distance.}
    \label{fig:2D3spline}
\end{figure*}

A first optimization stems from observing that ECSs are mostly local. Although they have formally a support that extends to the entire mesh, one can observe from Fig.~\ref{fig:2D3spline} that the magnitude of an ECS decreases exponentially with the number of vertices to the center of the spline. Note that this observation is only true for a regular mesh: if vertices are distributed logarithmically in one direction, the locality of the ECSs is greatly reduced.

\begin{figure}[]
    \centering

\includegraphics[width=0.7\linewidth]{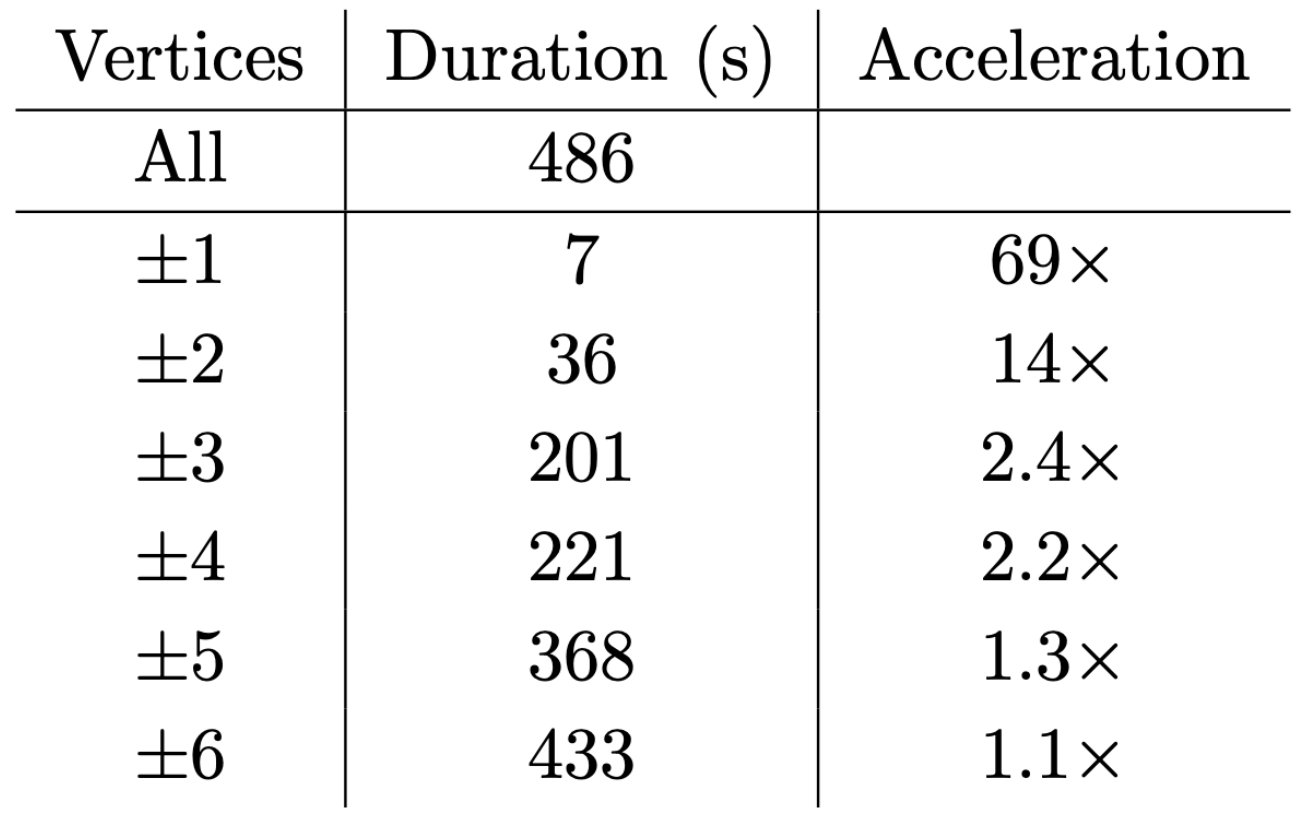} \includegraphics[width=0.85\linewidth]{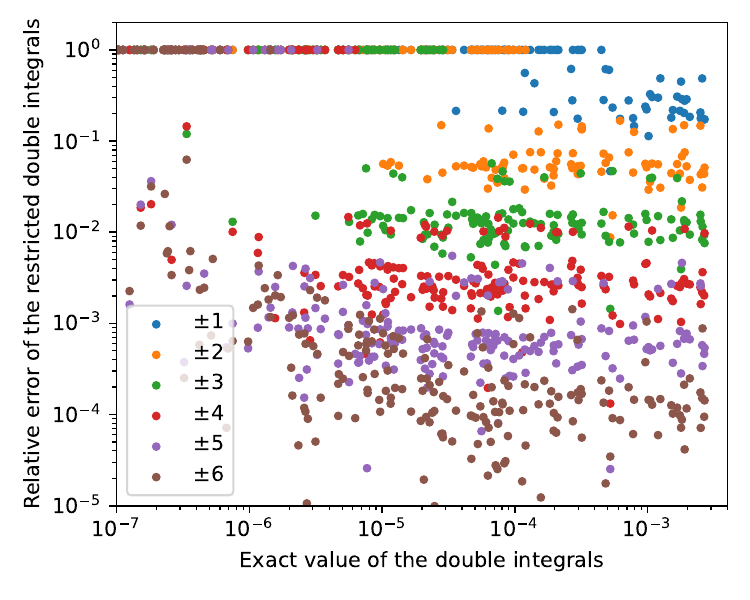}
    \caption{(top) When the double integrals of ECSs are restricted to the neighboring $\pm n$ vertices, the computing time, presented here for 320 randomly chosen double integrals, can be reduced significantly. -- (bottom) Relative error of restricted double integrals compared to their exact value. When the value of the kinematic $t$ is too far from the ECS central vertex, the restricted integral is 0, yielding a 100\% relative error.}
    \label{tab:reducedspline}
\end{figure}

Therefore, a first approximation is to neglect the contribution of an ECS more than $n$ vertices away from its center in each of the three directions. The support of an ECS becomes contained in a rectangular cuboid with $2n a_i$ side length in each direction where $a_i$ is the mesh spacing in direction $i = \{\beta, \alpha, t\}$, or less if the boundary is too close. For instance, using $n=\pm 1$ means only considering the central peak and no ``ripples'', which is obviously a fairly bad approximation that violates significantly the continuity of the first and second derivatives that cubic splines are intended for. The reduction of computation time when restricting the splines to only the closest $n$ vertices is shown on the top table of Fig.~\ref{tab:reducedspline}. $n = \pm 1$ brings a massive acceleration of 69 times, $n = \pm 3$ and 4 are about the same with a factor $\sim$ 2.3 of acceleration. 

The error introduced by the restriction is shown on the bottom panel of the same figure. It decreases exponentially with $n$. As expected, $n = \pm 1$ is bad with a relative error typically in excess of 20\%. $n=\pm 4$ appears as a natural candidate, both since it is computationally about as cheap as $n = \pm 3$ and offers a relative error typically below 1\%. The error for $n = \pm 4$ (red points on Fig.~\ref{tab:reducedspline}) is distinctively split into two groups: one corresponds to a precision of about 1\%, while a much larger group has a precision significantly below one percent. We observe that the less precise group corresponds exactly to ECSs centered at vertices which are in 6-th position from 0 either in $\alpha$ or $\beta$, showing that boundary effects have a slight influence on the precision of our approximation.

Using $n=\pm 4$, the full computation for each value of $z^2$ on a laptop takes about 10 hours.

\paragraph{Second optimization: using approximate translation invariance} The exponential locality of ECSs has another consequence: ECSs are exponentially similar to one another as the distance -- in number of vertices -- to the boundaries of the domain increases, as can be observed in 1D on the right panel of Fig.~\ref{fig:1D3spline}. If we name an ECS centered on a vertex $(\beta_0, \alpha_0, t_0)$ away from the boundaries as
$\mathcal{I}_{\beta_0,\alpha_0,t_0}(\beta, \alpha, t)$,
we can approximate $\mathcal{I}$ from a universal profile function $\mathcal{E}$ such that:
\begin{equation}
\mathcal{I}_{\beta_0,\alpha_0,t_0}(\beta, \alpha, t) \approx \mathcal{E}(\beta-\beta_0,\alpha-\alpha_0,t-t_0)\,.
\end{equation}
Then the double integral becomes
\begin{align}
\int_{\Omega^+} &\mathrm{d}\beta\mathrm{d}\alpha\,\cos(\beta\nu)\cos(\alpha\bar\nu) \mathcal{I}_{\beta_0,\alpha_0,t_0}(\beta, \alpha, t) \nonumber\\&\approx \theta(|t-t_0|-na_t)\int_{\mathcal{V}_n(\beta_0, \alpha_0)} \mathrm{d}\beta\mathrm{d}\alpha\,
\nonumber \\ & \times 
\cos(\beta\nu)\cos(\alpha\bar\nu) \mathcal{E}(\beta-\beta_0,\alpha-\alpha_0,t-t_0)\,,
\end{align}
where we have restricted the ECS support to a rectangular cuboid neighborhood of $(\alpha_0, \beta_0, t_0)$ of size controlled by $n$ as in the first optimization. Then performing a change of variables, expanding the resulting trigonometric functions and removing the terms which vanish due to parity reasons,
\begin{align}
\int_{\Omega^+} &\mathrm{d}\beta\mathrm{d}\alpha\,\cos(\beta\nu)\cos(\alpha\bar\nu) \mathcal{I}_{\beta_0,\alpha_0,t_0}(\beta, \alpha, t) \nonumber\\&\approx \cos(\beta_0\nu)\cos(\alpha_0\bar\nu) \theta(|t-t_0|-na_t)
\nonumber \\ & \times \int_{\mathcal{V}_n(0,0)} \mathrm{d}\beta\mathrm{d}\alpha\,\cos(\beta\nu)\cos(\alpha\bar\nu) \mathcal{E}(\beta, \alpha, t-t_0)\,.\label{eq:reducedset}
\end{align}

This expression has trivialized the $(\beta_0, \alpha_0)$ dependence of the double integral. If it were not for the remaining $t_0$-dependence, we would only compute for each kinematic $(|\nu|, |\bar\nu|)$ a double integral for one vertex centered at 0 and distant from the edges, and it would give us an excellent approximation for all vertices far from the boundaries. This would produce a speed-up proportional to the mesh volume, \textit{a priori} far more advantageous than the previously discussed constant speed-up.

To solve the issue of the remaining $t_0$-dependence, we can simply reconstruct the $t$-dependence by interpolation: for each of the 20 different $(|\nu|, |\bar\nu|)$ in the dataset, we compute the integral
\begin{equation}
\mathcal{D}_k(\nu, \bar\nu) = \int_{\mathcal{V}_n(0,0)} \mathrm{d}\beta\mathrm{d}\alpha\,\cos(\beta\nu)\cos(\alpha\bar\nu) \mathcal{E}(\beta, \alpha, ka_t)
\end{equation}
for $0 \leq k \leq n$. The result is very economical, with only 100 double integrals for $n = 4$ providing results for 160 times every vertex away from the boundaries of the mesh.

However, the applicability of this method in our case is limited by the relatively small size of our mesh. Only 55\% of our vertices are at least one vertex away from the boundaries. The proportion drops to 26\% if we require two vertices from the boundaries, 9.6\% with 3 vertices and a mere 1.8\% for 4 vertices. Tab.~\ref{tab:vertxbound} shows the distribution of vertices close to the boundaries depending on the number of vertices required to be considered as away from a boundary.

\begin{table}[]
    \centering
\begin{tabular}{c|c|c|c|c}
    & Away from  &  Except one & Except two  & Except three \\
    $n$ & all boundaries & boundary & boundaries & boundaries \\\hline
    1 & 1872 & 1263 & 241 & 14 \\
    2 & 891 & 1666 & 741 & 92 \\
    3 & 324 & 1467 & 1293 & 306 \\
    4 & 63 & 968 & 1647 & 712
\end{tabular}
    \caption{Number of vertices out of 3390 which are away from all boundaries, or close to only one, two or three boundaries. A vertex is close to a boundary if moving by $n$ vertices in any direction puts it out of the mesh. $n=1$ corresponds to points that are exactly on the boundary.}
    \label{tab:vertxbound}
\end{table}

For a finer mesh, where a large fraction of the vertices are away from the boundaries, this optimization could prove crucial. Let us note that similar improvements can be performed for terms which are close to a boundary. However, the increasing difficulty of implementation dissuaded us from pursuing such developments on our relatively small mesh.

\paragraph{Third optimization: getting rid of double integrals} A final optimization strategy that we pursued is to replace the general purpose double integration routine \verb|scipy.integrate.dblquad| by a series of one-dimensional integrations. There are several ways to organize this calculation. 

The most straightforward is to first integrate over the $\beta$-variable at fixed values of $\alpha$ on a regular grid of $m$ points spanning $\alpha \in [0,1]$. Then we perform a cubic interpolation of the dependence in $\alpha$ and compute the second integral. The parameter controlling the precision of the approximation and its cost is the number $m$ of points.

\begin{figure}[h]
    \centering

\includegraphics[width=0.8\linewidth]{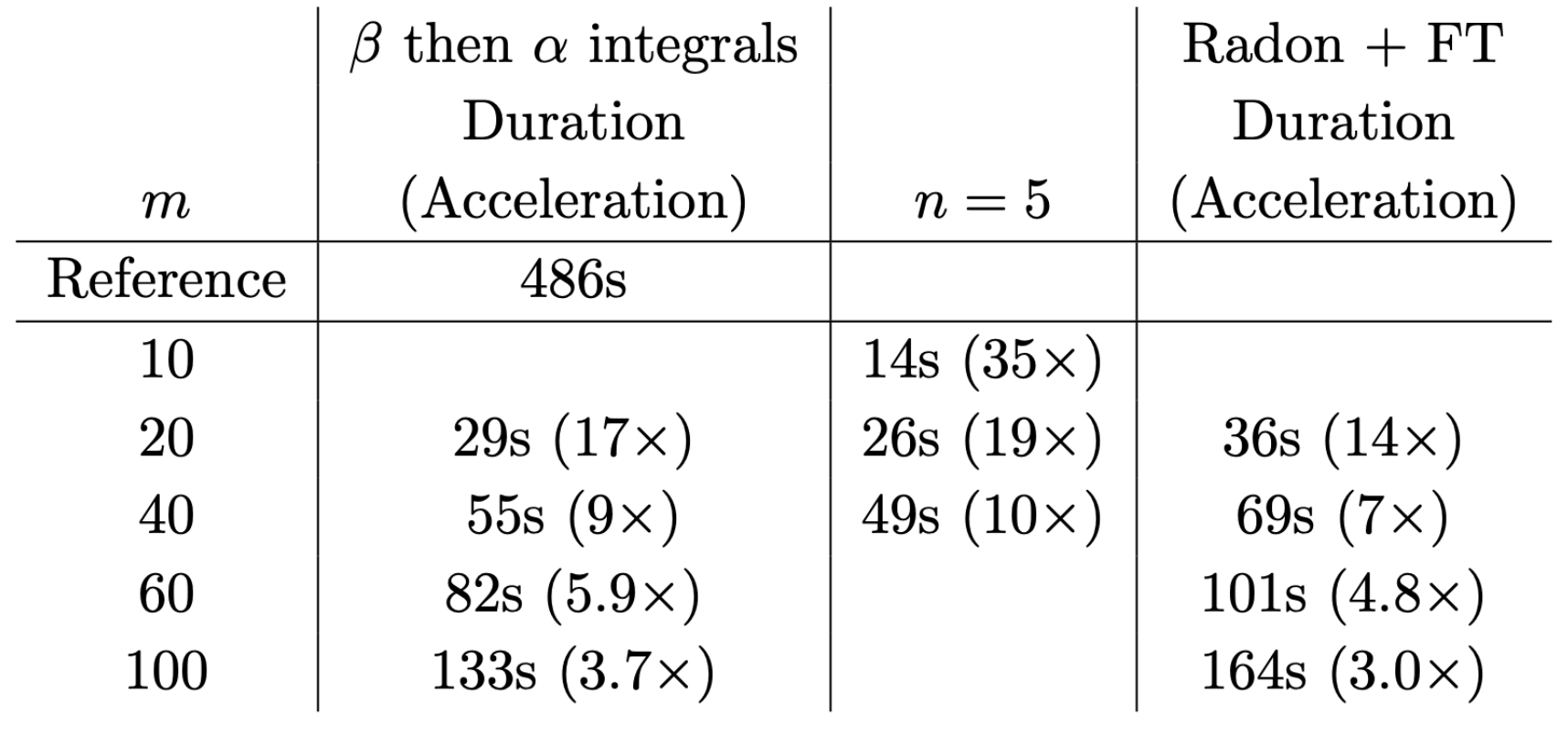}
\includegraphics[width=0.7\linewidth]{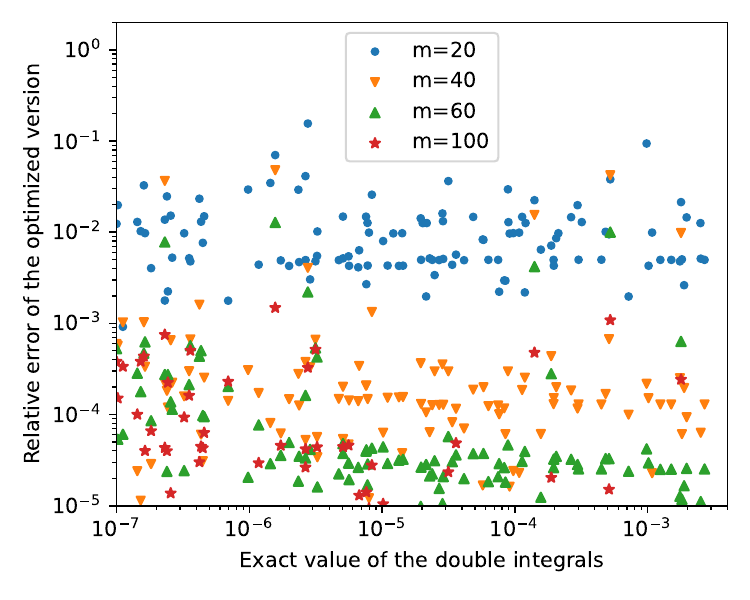}
\includegraphics[width=0.7\linewidth]{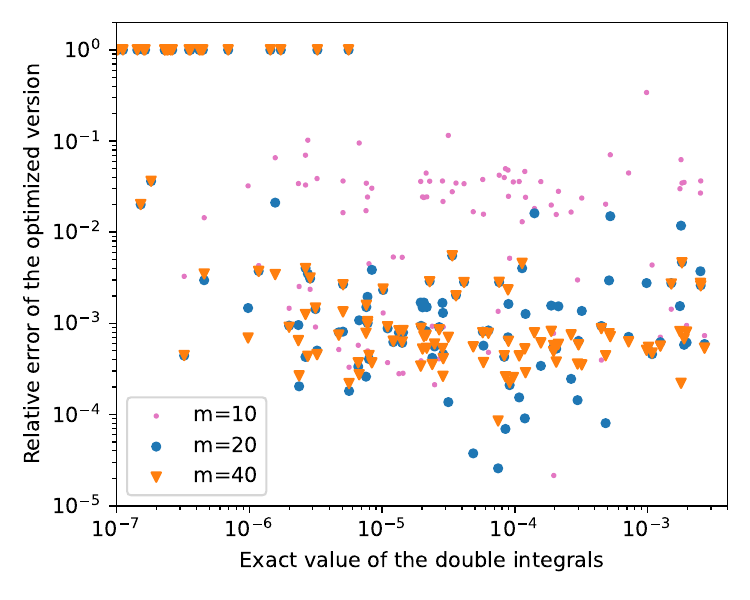}
\includegraphics[width=0.7\linewidth]{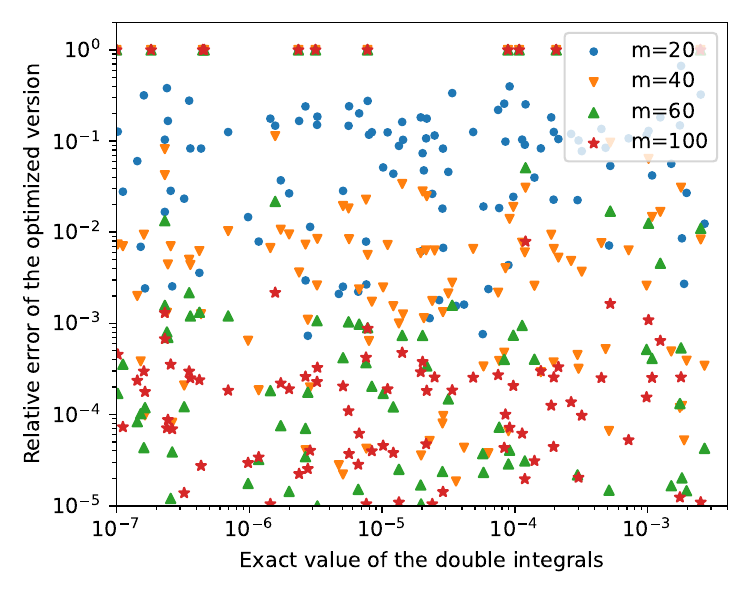}
    \caption{(top) Acceleration of the computation time on 320 double integrals when replacing scipy.integrate.dblquad by one-dimensional integrals and interpolation on $m$ points, using either successive integrals over the $\beta$ and $\alpha$ variables (second column), the same strategy but restricting the support of ECS to $n = \pm 5$ (third column), or performing Radon transforms followed by a Fourier transform (fourth column). (top right) Relative error when performing $\beta$ then $\alpha$ integrals. (bottom left) Relative error when performing $\beta$ then $\alpha$ integrals with restricted ECSs to $n = \pm 5$. When the value of the kinematic $t$ is too far from the ECS central vertex $t_0$, the restricted integral is 0, yielding a 100\% relative error. (bottom right) Relative error when using Radon transform followed by a Fourier transform. The kinematics with $\nu = 0$ cannot be computed with this method, and are represented by points with 100\% relative error.}
    \label{tab:succ1Dint}
\end{figure}

The second column of the table on the top panel of Fig.~\ref{tab:succ1Dint} shows the acceleration of computing time depending on the value of $m$. The second plot shows the achieved precision. Already $m=20$ gives a satisfactory approximation at the percent level in a lot of cases, for a 17 times acceleration. However, some points are poorly reproduced with errors in excess of 10\%. It is particularly clear with the cases $m = 40$ and 60 that a few points have a much worse relative error than the rest. They correspond to vertices which are very close to the edges of the domain, at either very large $\alpha$ or $\beta$. 

A way to understand this lack of precision is to notice that evenly probing $\alpha \in [0,1]$ with $m$ points is a waste of resources, and leads to imprecise results if $\alpha_0$ is large or small. A better use of resources is to probe the region in the vicinity of $(\beta_0, \alpha_0, t_0)$, which is exactly the spirit of our first optimization. Therefore, we limit ourselves to $n = \pm 5$ vertices away from $(\beta_0, \alpha_0, t_0)$. The result is shown on the third panel of Fig.~\ref{tab:succ1Dint}. The precision for $m=20$ is significantly increased, and in general way below percent accuracy. It also slightly improves the computing time as seen in the third column of the table.

From the purple dots in Fig.~\ref{tab:reducedspline}, we know that restricting the support of ECSs to $n = \pm 5$ imposes a cap on the relative error of the order of $10^{-3} - 10^{-4}$. Therefore, it is not useful to increase $m$ once this level of accuracy is achieved. In fact, $m = 20$ essentially saturates the bound, and seems perfectly satisfactory for our use. We will use this method to compute the matrix $B$, with a factor $\sim 19$ of acceleration compared to a naive computation, reducing the time for each value of $z^2$ to barely more than an hour on a M4 Pro laptop computer.

There are other ways to organize the calculation. One arises naturally from the physics of the problem: we first perform a 1D Radon transform, and then a 1D Fourier transform, that is for $\nu \neq 0$:
\begin{align}
\int_{\Omega^+} &\mathrm{d}\beta\mathrm{d}\alpha\,\cos(\beta\nu)\cos(\alpha\bar\nu) \mathcal{I}_{\beta_0,\alpha_0,t_0}(\beta, \alpha, t) \nonumber\\&\hspace{-30pt}= \frac{1}{2} \int_0^{\max(\xi, 1)} \mathrm{d}x\,\cos(x\nu)  \nonumber\\&\int_{\Omega}\mathrm{d}\beta\mathrm{d}\alpha\,\delta(x-\beta-\alpha\xi)\mathcal{I}_{\beta_0,\alpha_0,t_0}(\beta, \alpha, t)\,,
\end{align}
where $\xi = \bar\nu / \nu$. The inner integral on the right hand side computes a GPD value at $(x, \xi, t)$. We remind that $\xi$ may exceed 1, but remains finite since $\nu \neq 0$. Then inspecting the Radon integration lines on Fig.~\ref{fig:DD_Radon}, we see that the support in $x$ also exceeds 1, but $x \leq \xi$. Notice that the inner integral is expressed on $\Omega$, but must be converted to an integral on $\Omega^+$ by adequately reflecting the initial Radon line on the boundaries of the triangular domain using the appropriate parity constraints.

This time, for each ECS centered at $(\beta_0, \alpha_0, t_0)$ and each kinematic $(|\nu|, |\bar\nu|, t)$ with $\nu \neq 0$, we compute the GPD values on a grid of $m$ values of $x \in [0, \max(\xi, 1)]$, then interpolate it to compute the outer integral. The method is slightly more involved, and both less efficient computationally and less precise. We will therefore not use it to compute the matrix $B$. However, we will use it when it comes to studying the $\nu$- and $x$-dependence of our reconstructions at fixed skewness. It is therefore worthwhile to know the cost and precision scaling with $m$ of this alternative method, which are respectively depicted in the last column of the top table in Fig.~\ref{tab:succ1Dint} and the plot in the bottom panel. Notice that at a given $m$, this method is not tremendously lengthier than the previous, but the accuracy is much worse. Instead of $m=20$, we need at least $m=60$ to reach approximate percent accuracy, which takes 4 times longer. Restricting the ECS support to $n = \pm 5$ does not change the accuracy this time, but generally reduces the computing time by a mere 1.5 factor.

\FloatBarrier

\begin{widetext}

\section{Notice on how to read online datafiles of GPD results\label{app:notice}}

The results are stored at the address \cite{dutrieux_2026_19695472} in four datafiles in the HDF5 file format of 32 MB each: \verb|H_umd_m.h5| contains the $H^{u-d(-)}$ data, \verb|H_umd_p.h5| for $H^{u-d(+)}$, \verb|E_umd_m.h5| for $E^{u-d(-)}$ and \verb|E_umd_p.h5| for $E^{u-d(+)}$.
Each file contains the following hierarchical group structure: 
\begin{enumerate}
    \item Value of $t$ spanning $\{0, -0.2, -0.4, -0.6, -0.8, -1.0, -1.2, -1.4\}$ GeV$^2$ in
    
    \verb|['t0', 't-20', 't-40', 't-60', 't-80', 't-100', 't-120', 't-140']|
    \item Value of $\xi$ spanning $\{0, 0.25, 0.5, 0.75, 1\}$ in
    
    \verb|['xi0', 'xi25', 'xi50', 'xi75', 'xi100']|
    \item Choice between \verb|['samples', 'z5ts4', 'z6ts3']|

    \verb|'samples/data'| is an array of shape $(1000, 100)$ containing 1000 samples of the value of the GPD at the chosen $(\xi, t)$ values on a regular grid of 100 points in $x \in [0,1]$. This is the result of the fit with hyperparameter variation on the data $(z = 5a = 0.47 \textrm{ fm}, t_s = 3)$.

    \verb|'z5ts4/data'| is a vector of length $100$ containing the central value on a regular grid of 100 points in $x \in [0, 1]$ of the fit on the data $(z = 5a = 0.47 \textrm{ fm}, t_s = 4)$.

    \verb|'z6ts3/data'| is a vector of length $100$ containing the central value on a regular grid of 100 points in $x \in [0, 1]$ of the fit on the data $(z = 6a = 0.56 \textrm{ fm}, t_s = 3)$.
\end{enumerate}
\end{widetext}

\textbf{WARNING 1} Please note that the 1000 samples are made of 100 batches of 10 samples. Each batch of 10 samples is drawn from the posterior corresponding to one choice of prior hyperparameters. If one does not wish to use all 1000 samples, it is advised to draw a random subset of samples rather than pick the first, say 30, as they may not reflect the full uncertainty of the distribution.

\textbf{WARNING 2} Each of the four GPDs was reconstructed with its own set of hyperparameters. There is no expectation of correlations between samples of different GPDs.

A simple exemplary code to plot the value of $H^{u-d(-)}$ at $t = -0.8$ GeV$^2$ and $\xi = 0.5$ is shown below, and produces the result of Fig.~\ref{fig:database}.

\begin{widetext}
\begin{verbatim}
import numpy as np
import matplotlib.pyplot as plt
import h5py as h5

# Collect the data

file = h5.File('H_umd_m.h5', 'r')
data_1 = file['t-80/xi50/samples/data'][()]
data_2 = file['t-80/xi50/z5ts4/data'][()]
data_3 = file['t-80/xi50/z6ts3/data'][()]
file.close()

# Compute the mean of the samples and add systematic uncertainty in quadrature

m = np.mean(data_1, axis=0)
s = np.std(data_1, axis=0, ddof=1)
s = np.sqrt(s**2 + (data_2-m)**2 + (data_3-m)**2)

# Plot result

x_ = np.linspace(0, 1, 100)
plt.fill_between(x_, m-s, m+s, alpha=0.3)
plt.plot([0,1], [0,0], "--", color="grey")
plt.show()
\end{verbatim}
\end{widetext}

\begin{figure}
    \centering
    \includegraphics[width=0.8\linewidth]{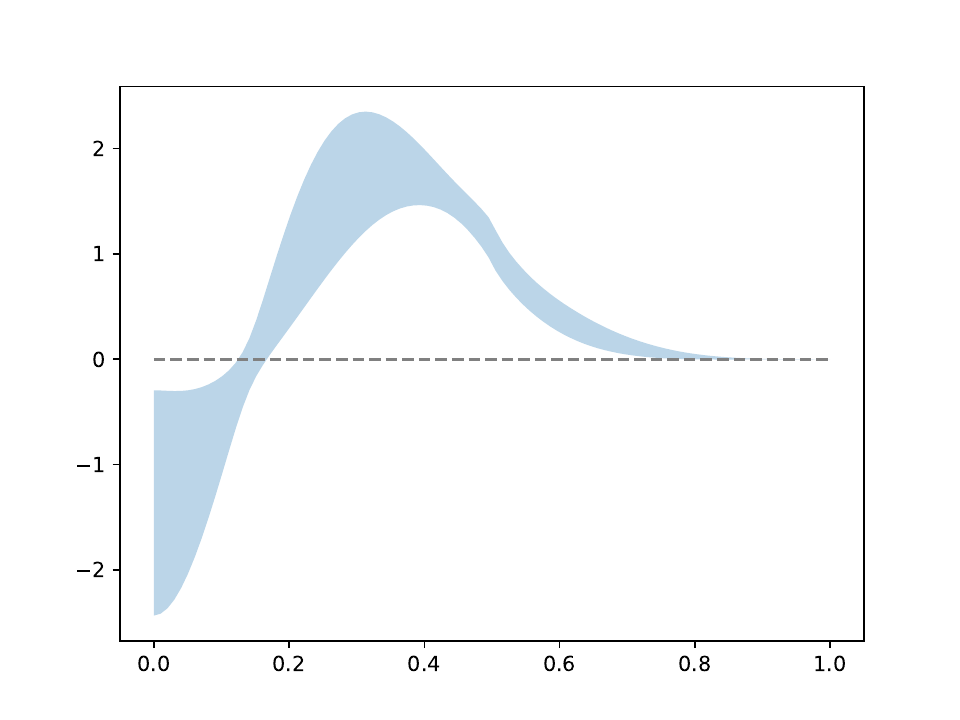}
    \caption{Result from running the exemplary code of the database.}
    \label{fig:database}
\end{figure}

\section{Window observables}\label{app:windows}
The problem of extracting GPDs from experiment rests in the limited sensitivity of the integral relating them to the Compton Form Factor. The value of the CFF is heavily dominated by the region of $\xi\sim x$ while the other regions are poorly constrained. In order to minimize model biases from both directions, comparing lattice QCD with experiment will be best done by integrating the lattice calculated GPD in the continuum limit and predicting the CFF. For benchmarking $x$-space GPD phenomenology and lattice QCD, it has been proposed in Ref~\cite{Karpie:2025ikc} to calculate integrals of the distributions which allow for higher precision from lattice QCD. In the case of GPD phenomenology, the narrowest windows in $x$ about $x=\xi$ would be most precise, and model biases would become more present in wide windows. As the windows widen, these model biases could be studied by comparison to future high precision LQCD results in the continuum limit. 

\begin{figure}
    \centering
    \includegraphics[width=0.85\linewidth]{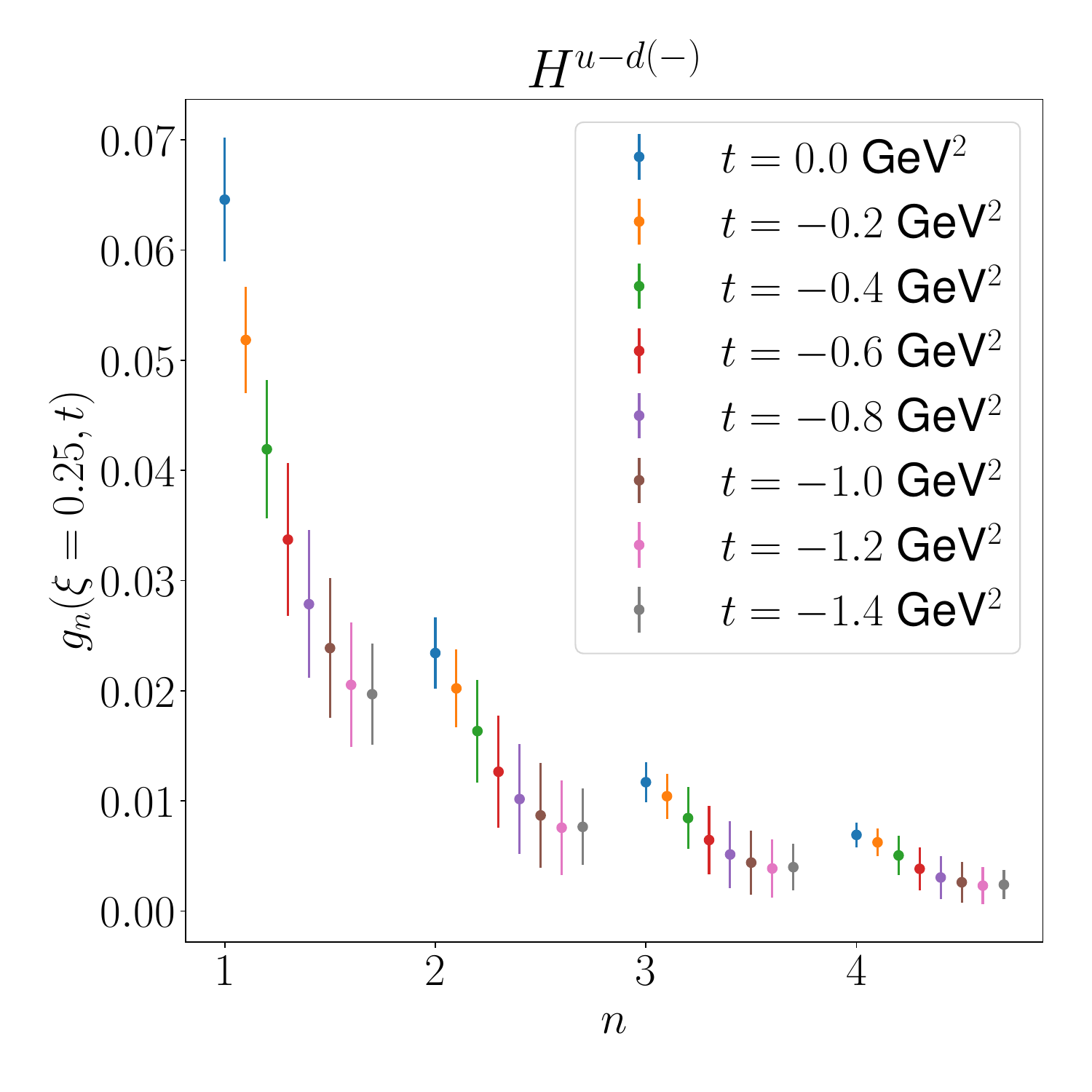}
    \includegraphics[width=0.85\linewidth]{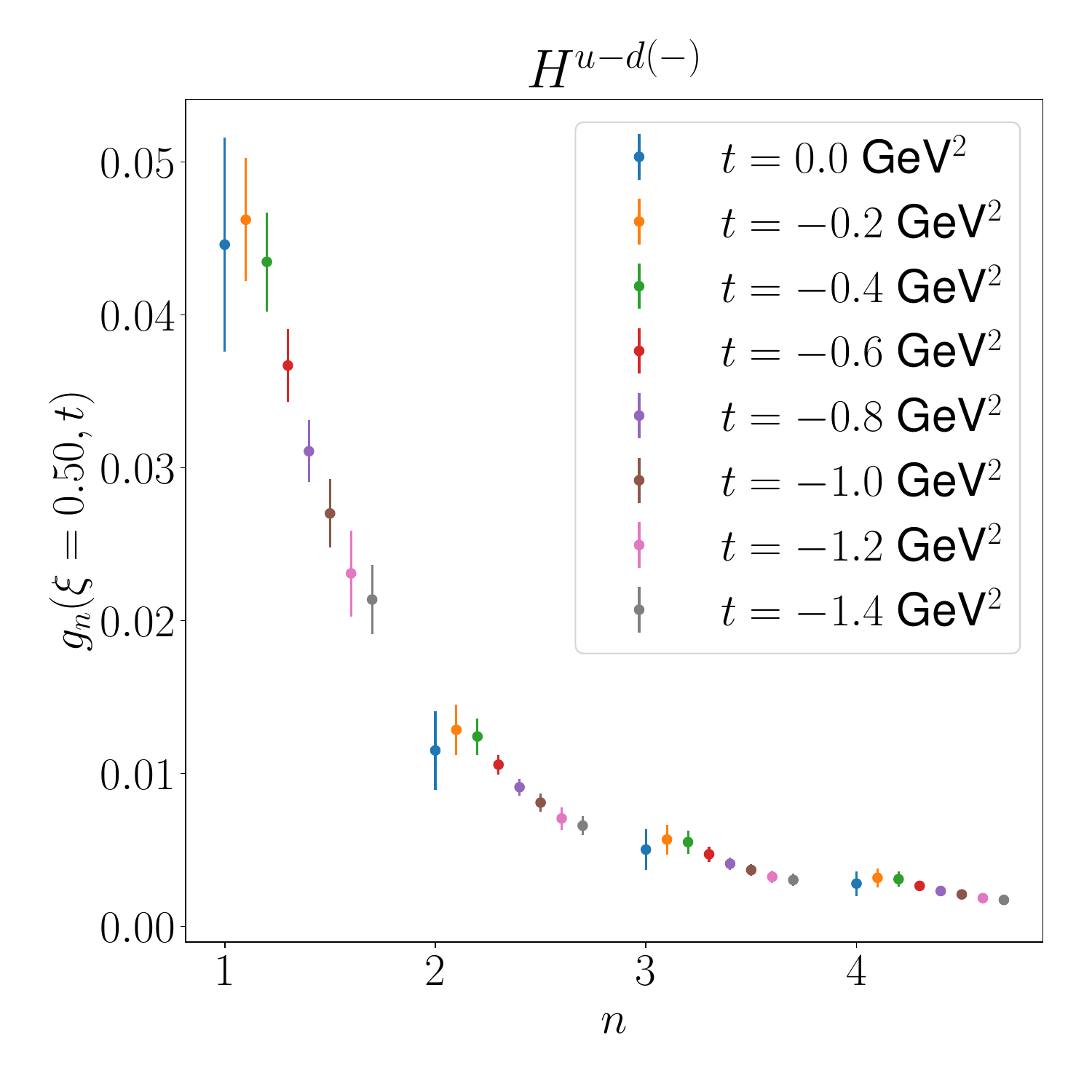}
    \caption{The Gaussian windows of $H$ centered about $x=\xi=0.25$ and 0.5. The different $t$ values are displaced from their $n$ for visibility.}
    \label{fig:window_plot1}
\end{figure}

In that vein, we have calculated the gaussian windows centered on $x=\xi$. In the notation of Ref~\cite{Karpie:2025ikc} we have chosen $x_\pm=\xi\pm 0.125$ and increase $n$ to narrow the windows. For 2 $\xi$ values on a range of $t$, Figs.~\ref{fig:window_plot1} shows the Gaussian windows of $H$. They generally fall off with $t$ as the GPD does. The size of the errors vary but it is reassuring that even as $n$ increases. As continuum, chiral, and leading twist behaviors are controlled, Gaussian windows like this can be used to compare to $x$ space GPD phenomenology.

\end{appendix}

\bibliography{srcs.bib}

\end{document}